\documentclass[a4paper,11pt]{article}
\pdfoutput=1 

\usepackage{jheppub} 
\usepackage{mathtools}
\usepackage{bm}
\usepackage{amsmath,amssymb,amsthm,amscd,graphicx,simpler-wick}
\usepackage{psfrag}
\input epsf.sty
\usepackage{xparse}
\usepackage{tikz}
\usetikzlibrary{tikzmark}
\usepackage{tikz-cd}
\usepackage{needspace}
\usepackage{footmisc}

\usepackage{silence}
\usepackage[missing=]{gitinfo2}

\usepackage{varioref}
\usepackage{subcaption}
\usepackage{textcomp}

\newcommand{\I}{{\mathrm i}}
\newcommand{\e}{{\mathrm e}}
\newcommand{\de}{\mathrm{d}}
\newcommand{\ri}{\text{i}}

\newcommand{\abs}[1]{\lvert#1\rvert}

\newcommand{\cM}{{\mathcal M}}

\newcommand{\cI}{{\mathcal I}}

\newcommand{\cS}{{\mathcal S}}
\newcommand{\cN}{{\mathcal N}}

\newcommand{\cW}{{\mathcal W}}
\newcommand{\cA}{{\mathcal A}}
\newcommand{\cX}{{\mathcal X}}
\newcommand{\bX}{{\mathbb X}}

\newcommand{\tot}{{\mathrm{tot}}}

\newcommand{\E}{\text{e}}

\newcommand{\rd}{\text{d}}

\newcommand{\be}{\begin{equation}}
\newcommand{\ee}{\end{equation}}
\newcommand{\ba}{\begin{aligned}}
\newcommand{\ea}{\end{aligned}}

\DeclareMathOperator{\disc}{disc}

\DeclareMathOperator{\re}{e}

\usetikzlibrary{positioning,arrows,patterns}
\usetikzlibrary{decorations.markings}
\usetikzlibrary{calc}
\usetikzlibrary{shapes}
\usetikzlibrary{topaths}
\usetikzlibrary{decorations}
\usetikzlibrary{decorations.pathmorphing,fit}
\usetikzlibrary{bending}
\usetikzlibrary{calc}

\newcommand*\circled[1]{\tikz[baseline=(char.base)]{
            \node[shape=circle,draw,inner sep=2pt] (char) {#1};}}

\tikzset{
  fermion/.style={draw=black, postaction={decorate},decoration={markings,mark=at position .55 with {\arrow{>}}}},
  vertex/.style={draw,shape=circle,fill=black,minimum size=3pt,inner sep=0pt},
  vertex2/.style={draw,shape=circle,fill=black,minimum size=1pt,inner sep=0pt},
  vertex3/.style={draw,shape=circle,fill=\psiplusColor,minimum size=3pt,inner sep=0pt},
  vertex4/.style={draw,shape=circle,fill=\psiminusColor,minimum size=3pt,inner sep=0pt}  
}

\tikzset{dot/.style={circle, fill, inner sep=1.5pt}}

\newcommand{\branchpointColor}{orange}
\newcommand{\wallColor}{black}
\newcommand{\psiplusColor}{purple!70}
\newcommand{\psiminusColor}{blue!70}
\newcommand{\leashColor}{green!55!black}
\tikzset{snake it/.style={decorate, decoration=snake}}

\newlength{\branchpointmarkersize}
\newcommand{\drawbranchpointmarker}[1]{
  \draw [line width=0.5mm, \branchpointColor] ($(#1)-(\branchpointmarkersize/2,\branchpointmarkersize/2)$) -- ($(#1)+(\branchpointmarkersize/2,\branchpointmarkersize/2)$);
  \draw [line width=0.5mm, \branchpointColor] ($(#1)-(\branchpointmarkersize/2,-\branchpointmarkersize/2)$) -- ($(#1)+(\branchpointmarkersize/2,-\branchpointmarkersize/2)$);
}

\tikzset{
    psipluscontour/.style={thick,\psiplusColor},
    psiminuscontour/.style={thick,\psiminusColor},
    leash/.style={dashed,\leashColor}
  }

\title{\boldmath Exact WKB of solutions by Borel summation and open TBA}
\author[1]{Qianyu Hao}
\affiliation[1]{Section de Math\'{e}matiques, Universit\'{e} de Gen\`{e}ve, 1211 Gen\`{e}ve 4, Switzerland}

\emailAdd{qianyu.hao@unige.ch}

\makeatletter
\gdef\@fpheader{\bigskip}
\makeatother

\begin{document}

\makeatletter
\renewcommand{\sectionautorefname}{\S\@gobble}
\renewcommand{\subsectionautorefname}{\S\@gobble}
\renewcommand{\subsubsectionautorefname}{\S\@gobble}
\makeatother

\abstract{In this paper, we study the exact WKB methods for solutions of the Schr\"{o}dinger equations corresponding to quantum Seiberg-Witten curves in 4d $\cN=2$ theories with surface defects. The tools are Borel summation and Gaiotto-Moore-Neitzke (GMN) open TBA, which is a conjecture that has largely stayed unexplored so far. We study the Borel plane of the solution regularized by a BPS soliton charge and find that its poles on the Borel plane have the physical meaning of BPS soliton central charges and 4d BPS state central charges. We discuss in detail the discontinuities associated with the singularities and propose general expressions for these discontinuities which also apply to higher-order ODEs. We apply the open TBA to genuinely nontrivial examples -- the 4d $\cN=2$ theory with a single flavor BPS state corresponding to the Weber equation and an example of the pure $SU(2)$ theory corresponding to the modified Mathieu equation -- and show how to match it with Borel summation. We provide numerical evidence for the matching of the two exact WKB methods and also for an exact solution to the modified Mathieu equation via open TBA.
}

\maketitle

\flushbottom

\bibliographystyle{JHEP}

\section{Introduction}\label{sec:intro}

In this work, we study solutions to differential equations of Schr\"{o}dinger type
\begin{equation}
\label{dex2}
\left(-\epsilon^2\partial_z^2+P(z,\epsilon)\right)\psi(z)=0, 
\end{equation}
where $\epsilon$ is a small expansion parameter and $P(z, \epsilon)$ is a meromorphic function of $z$ on a Riemann surface $C$ which we will specify later in examples. This type of equation has been studied in depth using exact WKB methods \cite{voros-quartic,bpv,voros-fp,ddpham,reshyper}. Furthermore, the correspondence relating differential equations to 4d $\cN=2$ supersymmetric theories with surface defects in the Nekrasov-Shatashvili phase of the $\Omega$ background\,\cite{Alday:2009aq,Alday:2009fs,Nekrasov:2009rc,Nekrasov:2011bc,Gaiotto:2014bza,Jeong:2018qpc} significantly extended the scope of study -- culminating in what are now known as generalized exact WKB methods. In this correspondence, \eqref{dex2} is regarded as a quantization of the Seiberg-Witten curve $\Sigma$ of the supersymmetric theory.

The main object studied in this work is the exact WKB solution to \eqref{dex2}. Based on the supersymmetric theory/differential equation correspondence, we propose generic expressions for the singularities on the Borel plane and the corresponding discontinuities of the Borel summation, given in terms of physical quantities. This generalization applies to differential equations of arbitrary order within the framework of the supersymmetric theory/differential equation correspondence.

As another important part of this work, we investigate a conjectured exact WKB method for solutions known as the Gaiotto–Moore–Neitzke (GMN) open TBA, a remarkable construction that originates from the study of supersymmetric theories. 
So far, the validity of the open TBA has only been examined in the most elementary example -- the Airy equation -- as briefly discussed in the last two pages of \cite{Gaiotto:2014bza}. In this work, we provide the first solid application of this conjecture to differential equations corresponding to supersymmetric theories with richer gauge and matter contents: namely, the Weber equation arising from a 4d $\mathcal{N}=2$ theory of Argyres-Douglas type with a single 4d flavor BPS state, and the modified Mathieu equation corresponding to the pure $SU(2)$ gauge theory. In this context, we provide numerical evidence for an exact solution to the modified Mathieu equation via open TBA --  which so far has remained unexplored.

\subsection{Borel summation}\label{intro:borel}

We review the basics of Borel summation in \autoref{sec:padeborel}. As an example, we apply Borel summation to the asymptotic series $Y(z,\epsilon)$ defined by the WKB ansatz for a solution of \eqref{dex2}:
\begin{equation}\label{intro:Y}
\psi(z)=\e^{\frac{1}{\epsilon}\int_{z_0}^zY(z,\epsilon)\rd z}=\e^{\frac{1}{\epsilon}\int_{z_0}^z\sum\limits_{n=0}^\infty Y_n(z)\epsilon^n\rd z}.
\end{equation}

The Borel summation $s(Y)(z,\epsilon)$ is defined by two steps, the Borel transform \eqref{boreltr}, followed by the Laplace transform \eqref{laplace}. Singularities of the Borel transform $\hat{Y}(z,\zeta)$ lead to discontinuities of $s(Y)(z,\epsilon)$.

\subsection{The supersymmetric theory/differential equation correspondence}\label{intro:poleY}

Our study of Borel singularities and their associated discontinuities is motivated by previous work on the dictionary between Borel summation and the BPS spectra within the supersymmetric theory/differential equation correspondence \cite{Grassi:2019coc,Grassi:2021wpw}. We start by reviewing results of \cite{Grassi:2019coc,Grassi:2021wpw}.

The central object studied in \cite{Grassi:2019coc,Grassi:2021wpw} is the quantum period 
 \be\label{intro:QPdef} \cX_{\gamma}(\epsilon)=\exp\Bigg(s\Big(\frac{1}{\epsilon}\sum_{n=0}^\infty\oint_{\gamma} Y_n(z) \de z\,\epsilon^n\Big)(\epsilon)\Bigg),\ee
where $\gamma\in H^1(\overline{\Sigma},\mathbb{Z})$, with $\overline{\Sigma}$ obtained by filling in the punctures of $\Sigma$. $x_{\gamma}=\log\cX_\gamma$ quantizes the classical period \(\frac{1}{\epsilon}\oint_\gamma\sqrt{P(z,0)} \de z\). In physics, quantum periods can be interpreted as expectation values of line operators \cite{Gaiotto:2009hg,Gaiotto:2012rg,Hollands:2019wbr}. In \cite{Grassi:2019coc}, it is pointed out that the singularities of $\hat{x}_\gamma(\zeta)$ are at the minus central charges of 4d BPS states. Inspired by this, \cite{Grassi:2021wpw} further identifies the singularities of $\hat{Y}(z,\zeta)$ as the negative central charges of BPS solitons in the coupled 2d-4d  system, which is a generalization of the statement in the exact WKB literature \cite{kawai2005algebraic}.

Although quantum periods are not the main focus of this paper, they are needed for calculating the primary objects of interest -- solutions -- using the GMN open TBA method, which is one of the main topics of this work. Strictly speaking, \eqref{intro:QPdef} is defined by Borel summation but in this paper we use GMN closed TBA  as the equivalent exact WKB method to calculate quantum periods \cite{Gaiotto:2009hg,Gaiotto:2012rg,Hollands:2019wbr,Grassi:2019coc,Grassi:2021wpw}; see \autoref{disccomp}, \autoref{3dN=4} and \autoref{sec:su2closetba} for a particular example in the pure $SU(2)$ theory.
Our choice of method is based on its ability to  produce the values of quantum periods at arbitrary points on the $\epsilon$-plane. For a more thorough review of exact WKB methods for quantum periods, we direct the interested reader to \cite{Grassi:2021wpw}. More related progress can be found in \cite{Gaiotto:2014bza,Gaiotto:2011tf,Iwaki:2014vad,Hollands:2019wbr,Grassi:2019coc,Dumas:2020zoz,Imaizumi:2021cxf,Yan:2020kkb,Ito:2024nlt,Ito:2021sjo,Marino:2024yme,Bridgeland:2016nqw,Fioravanti:2020udo}.

The correspondence between singularities on the Borel plane and BPS spectra has been further generalized to the 3d-5d supersymmetric theory/difference equation  correspondence \cite{Aganagic:2011mi,Mironov:2009uv,Nekrasov:2009rc}. In \cite{Grassi:2022zuk}, it has been proposed that positions of singularities on the Borel plane of the WKB expansion for solution of the difference equation correspond to central charges of 3d-5d BPS KK-modes. And singularities for the Borel transformed series for closed topological string partition functions correspond to central charges of 5d BPS KK-modes. More related developments in exact WKB or resurgence in 5d can be found in \cite{DelMonte:2022kxh,Rella:2022bwn,Gu:2022fss,Gu:2023mgf,Gu:2023wum,Gu:2021ize}.

\subsection{Normalization of solutions, Borel singularities and discontinuities}\label{sec:intro:disc}

The appearance of the point $z_0$ in \eqref{intro:Y} raises some subtle issues. In this paper, we choose $z_0=b_0$, where $b_0$ represents a branch point of the projection map $\Sigma\xrightarrow[]{2:1} C$. We calculate the integral $I^{n,(i)}$ defined in \eqref{Upsilonallorder} of each coefficient $Y^{n}(z)$. However, the integral for $n>1$ is singular at $b_0$, so that it needs regularization. There are actually two independent WKB series of \eqref{dex2} labeled by subscript $i=1,2$ in $\psi^{(i)}(z,\epsilon)$ as explained around \eqref{psidefori}. We choose $\psi^{(1)}(z,\epsilon)$ for clarity. We introduce a regularization \eqref{Upsilonallorder}
where we regard the integral contour on the first sheet of $\Sigma$ from $b_0$ to $z$ to be $\frac12$ of a soliton charge $-\gamma_{21}$ surrounding $b_0$ \cite{Iwaki:2014vad,Marino:2021lne}. Similarly, the regularization for generic $i$ uses $-\gamma_{ji}$ in the set of relative homology classes on the Seiberg-Witten curve $\Sigma$ of 1-chain with boundary at the lifts of $z$ to the $j$th and $i$th sheets of $\Sigma$. $\gamma_{ji}$ has the physical meaning of a soliton central charge; see the discussion around \eqref{solZ}.

\begin{equation}
\text{Reg}(\begin{tikzpicture}[baseline={(0,0)},scale=0.5]
\draw[snake it, orange] (-3, 0) to (0, 0);
\node at (2.2,0) [right] {$z$};
\draw[->] (1,0 ) -- (1.2,0);
\draw (0,0) -- (2,0);
\drawbranchpointmarker{0, 0};
\node at (0,0) [above] {$b_0$};
\node at (1.1,0.1) [above] {$1$};
\filldraw[black] (2,0) circle(3pt); 
\end{tikzpicture})
=
\frac{1}{2}
\begin{tikzpicture}[baseline={(0,0)},scale=0.5]
\draw[snake it, orange] (-3, 0) to (0, 0);
\node at (2.2,0) [right] {$-\gamma_{21}$};
\node at (1,0.1) [above] {$1$};
\node at (1.1,0.1) [below] {$2$};
\draw[->] (1,0.1 ) -- (1.2,0.1);
\draw (0.2,0.1) -- (2,0.1);
\draw[<-] (1.2,-0.1 ) -- (1.4,-0.1);
\draw (0.2,-0.1) -- (2,-0.1);
\draw [domain=26:360-26,line width=0.2mm] plot ({0.2*cos(\x)}, {0.2*sin(\x)});
\drawbranchpointmarker{0, 0};
\node at (0,0) [above] {$b_0$};
\filldraw[black] (2,0) circle(3pt); 
\end{tikzpicture}
=
\frac{1}{2}
\begin{tikzpicture}[baseline={(0,0)},scale=0.5]
\draw[snake it, orange] (-3, 0) to (0, 0);
\node at (2.2,0) [right] {$-\gamma_{21}$};
\draw [domain=0:360,line width=0.2mm] plot ({1.3*cos(\x)+0.7}, {0.4*sin(\x)});
\draw[->] (0.9,0.41 ) -- (1,0.41);
\drawbranchpointmarker{0, 0};
\node at (0,0.2) [above] {$b_0$};
\node at (1,0.3) [above] {$1$};
\filldraw[black] (2,0) circle(3pt); 
\end{tikzpicture}
\end{equation}

The main object to study in this paper -- the exact WKB solution -- is defined by Borel summation of the series of integrals $I^{(i)}$ defined in \eqref{Upsilonallorder},
\begin{equation}\label{intro:psidef}
\Upsilon^{(i)}(z,\epsilon)=s(I^{(i)})(z,\epsilon).
\end{equation}
Similarly we define $\Psi^{(i)}(z,\epsilon)$ as its exponential.
We further introduce the reduced $\Upsilon^{(i)}_{\rm red}(z,\zeta)$ or $\Psi^{(i)}_{\rm red}(z,\epsilon)$ defined by \eqref{defUpsilonred} or \eqref{defPsired} respectively which are obtained by removing the leading order term $I^{0,(i)}(z)$ in the WKB series for matching with the open TBA later.

The singularities on the Borel plane of solutions, along with their associated discontinuities \eqref{eqn:defdisc1}, are among the central objects studied in this work. Our main results in this context are generic expressions for these objects in terms of physical quantities -- they are highlighted in boxes throughout \autoref{sec:intro:disc}. These formulae are also expected to apply to differential equations of other orders within the supersymmetric theory/differential equation correspondence.

Since there are different solitons $-\gamma_{ji}$, we label them by subscripts $n=1,2,\cdots,$ and assume the soliton used for the normalization is $-\gamma_{ji_1}$\footnote{We emphasize $1$ is the subscript of the whole $\gamma_{ji}$ not of $i$.}. We find that there are two types of singularities. Correspondingly, the associated discontinuities are of two types and they are conveniently described by $\disc(\frac{\Upsilon^{(i)}_{\rm red}}{\epsilon})(z,\epsilon)$ or $\disc(\Psi^{(i)}_{\rm red})(z,\epsilon)$.
The two types of singularities and discontinuities include:
\begin{itemize}
\item The first type of singularity is at negative central charges of BPS solitons in the coupled 2d-4d system, given by $-Z_{\gamma_{{ji}_n}}(z)=-\int\limits_{\gamma_{ji_n}}\sqrt{P(z,0)} \rd z$. The positions of those singularities are the same as the ones of $\hat{Y}(z,\epsilon)$ reviewed in \autoref{intro:poleY} \cite{Grassi:2021wpw}. We propose that along the codim-1 locus given by $\arg(\epsilon)=\arg(-Z_{\gamma_{{ji}_n}}(z))$, the corresponding discontinuity is \eqref{sec2:jumpsol2} which we reproduce here for convenience
\begin{equation}\label{intro:jumpsol2gen}
\boxed{
\begin{aligned} 
&{\rm disc}(\Psi^{(i)}_{\rm red})(z,\epsilon)=\mu(\gamma_{{ji}_n})\Psi^{(j)}_{\rm red}(z,\epsilon)\e^{\frac{Z_{\gamma_{ji_n}}(z)}{\epsilon }+\frac{1}{\pi\ri}\sum\limits_{\gamma>0}\langle\gamma_n,\gamma\rangle\Omega(\gamma)\mathcal{I}_{\gamma}(\epsilon)},\\
&\quad\quad\quad\quad\quad\quad\quad\quad\quad\quad\quad\quad\quad\quad\quad\quad\quad\quad\quad\quad\quad\quad\quad\text{at }\arg(\epsilon)=\arg(-Z_{\gamma_{{ji}_n}}(z)),
\end{aligned}}
 \end{equation}
where the quantities $\mathcal{I}_{\gamma}(\epsilon)$, $\gamma_{n}=(\gamma_{{ji}_n}-\gamma_{{ji}_1
})\in H^1(\overline{\Sigma},\mathbb{Z})$, $\mu(\gamma_{{ji}_n})$, $\Omega(\gamma)$ and $\langle-,-\rangle$ are defined in the main text around \eqref{sec2:jumpsol2}. \eqref{intro:jumpsol2gen} generalizes the results of \cite{Voros1983, 10.1007/978-4-431-68170-0_1, Iwaki:2014vad} to all second-order ODEs and their higher-order analogues within the supersymmetric theory/differential equation correspondence, leveraging the relevant physical quantities.

When $n=1$, for those familiar with the subject, \eqref{intro:jumpsol2gen} is consistent with the intuitive picture offered by the spectral network framework~\cite{Gaiotto:2012rg}. A non-exhaustive list of applications includes \cite{Hollands:2016kgm,Longhi:2016wtv,Hao:2019ryd,Neitzke:2020jik,Hao:2024vlg}.

Note that the condition
\begin{equation}\label{stokescond}
\arg(\epsilon)=\arg(-Z_{\gamma_{{ji}_n}}(z))
\end{equation} of \eqref{intro:jumpsol2gen} imposes a constraint that depends on both $\epsilon$ and $z$. So \eqref{intro:jumpsol2gen} controls discontinuities of $s(\Psi^{(i)}_{\rm red})(z,\epsilon)$ as a function of both $\epsilon$ and $z$.  For the purpose of comparing with the open TBA, we typically fix $z$ and study the discontinuities with respect to $\epsilon$ so we represent $Z_{\gamma_{{ji}_n}}(z)$ by $Z_{\gamma_{{ji}_n}}$ for simplicity in this paper. Fixing \( \epsilon \) and varying \( z \), on the other hand, produces the Stokes graph (also known as the spectral network) where the condition \eqref{stokescond} defines codim-1 walls on the Riemann surface $C$.

\item We propose that there is a second type of singularity at minus 4d BPS state central charges $-Z_{\gamma}$ such that the intersection number $\langle\gamma_{ji_1},\gamma\rangle\neq 0$, and the corresponding discontinuity is \eqref{sec2:jumppi} which we reproduce here
\begin{equation}\label{intro:jumppi} \boxed{
{\rm disc}(\frac{\Upsilon^{(i)}_{\rm red}}{\epsilon})(z,\epsilon)= \frac{\langle-\gamma_{ji_1},\gamma\rangle\Omega(\gamma)}{2}\log(1-\sigma(\gamma)\cX_{\gamma}), \text{ at }\arg(\epsilon)=\arg(-Z_{\gamma}).}
\end{equation}
We propose that the discontinuity \eqref{intro:jumppi} purely results from the normalization at $b_0$ regularized by $-\gamma_{ji_1}$. In fact, this discontinuity is $\frac{1}{2}$  of the Kontsevich-Soibelman transformation. The quantum open path integral
$
X_{-\gamma_{ji_1}}(z,\epsilon)=\exp\Bigg(s\Big(\frac{1}{\epsilon}\sum\limits_{n=0}^\infty\int\limits_{-\gamma_{ji_1}} Y_n(z) \de z\,\epsilon^n\Big)(\epsilon)\Bigg) 
$
is analogous to the Voros symbol for an open path \cite{Iwaki:2014vad}, thus we expect it to be subject to the Kontsevich-Soibelman transformation. The  factor $\frac{1}{2}$ appears because the regularized integral $\text{Reg}(\int_{b_0}^{z}\cdots)$ contour is $\frac{1}{2}$ of the charge $-\gamma_{ji_1}$.  

The statement of this type of singularity also exists in the resurgence literature as fixed singularities\footnote{We thank Marcos Marino for his clarification on this.} \cite{delabaere:hal-01886535,Takei2008} and the discontinuity \eqref{intro:jumppi} is discussed in \cite{Takei2008} for the example of the Weber equation. As a generalization, \eqref{intro:jumppi} applies to all second-order ODEs and their higher-order analogues which correspond to supersymmetric theories. In fact \eqref{intro:jumppi} is crucial in the matching with the GMN open TBA conjecture.

\end{itemize}

We discuss the details of the singularities in the Borel plane and the corresponding dicontinuities in \autoref{padeborelsol}, \autoref{sec:borel4d} and \autoref{sec:disc}.  Certain special cases of \eqref{intro:jumpsol2gen} and \eqref{intro:jumppi} have been rigorously proven in the literature \cite{Voros1983, 10.1007/978-4-431-68170-0_1, Iwaki:2014vad, Takei2008}. We also provide additional numerical evidence through examples: the Weber equation in \autoref{sec:weber}, and the modified Mathieu equation at $u=0$  in \autoref{sec:su2}.

\subsection{Gaiotto-Moore-Neitzke open TBA equations}

In this work, we investigate a conjectural TBA-based exact WKB method known as the GMN open TBA \cite{Gaiotto:2011tf, Gaiotto:2014bza} which applies to  differential equations of arbitrary order within the context of the supersymmetric theory/differential equation correspondence. Although this approach was briefly tested on the Airy equation in the concluding pages of~\cite{Gaiotto:2014bza}, its application to genuinely nontrivial examples has remained largely unexplored. Here, we present the first detailed study of this method in more intricate settings: the Weber equation and the modified Mathieu equation, both of which have rich physical and mathematical structure. 

In \autoref{3dN=4} and \autoref{sec:conflimit}, we provide a brief review of the GMN open TBA equations. These equations are formulated as integral equations in the variable \(\epsilon\). For clarity, we suppress the dependence on \(z\) and other parameters, writing, for example, \(g_i(\epsilon) \equiv g_i(z, \epsilon)\) and \(x_{\gamma_{ji_n}}(\epsilon) = x_{\gamma_{ji_n}}(z, \epsilon)\). The open TBA reads
\begin{equation} 
g_i(\epsilon) = g_i^+ + \frac{1}{2 \pi \I}\sum_{n} \mu(\gamma_{j i_n})
\int_{\ell_{\gamma_{j i_n }}} \frac{\de \epsilon'}{\epsilon'}
\frac{\epsilon }{\epsilon' - \epsilon} g_j (\epsilon') X_{\gamma_{j i_n}}(\epsilon').
\end{equation}
Definitions for the relevant quantities can be found near equation \eqref{xijconformal}.
Since the examples we study involve second-order differential equations, we can use the symmetry
$g_2(\epsilon) = \ri\, g_1(-\epsilon)$,
to simplify the analysis by focusing on an integral equation of a single function $g_1(\epsilon)$, as given in \eqref{eq:int-22}, which we reproduce here:
\begin{equation} \label{intro:TBAopen}
g_1(\epsilon) = \frac{1}{\left(-\sqrt{P(z,0)}\right)^{1/2}} + \sum_{n} \mu(\gamma_{21_n})
\frac{\epsilon}{2\pi} \int\limits_{\ell_{\gamma_{21_n}}} \frac{d\epsilon'}{\epsilon'} \,
\frac{1}{\epsilon' - \epsilon} \, g_1(-\epsilon') \, X_{\gamma_{21_n}}(\epsilon').
\end{equation}

The GMN open TBA solves a Riemann-Hilbert problem encoding the asymptotics $g(z,\epsilon)\sim\frac{1}{\left(-\sqrt{P(z,0)}\right)^{\frac{1}{2}}} $ at $\epsilon\rightarrow 0$ and the discontinuities 
\begin{equation}
 {\rm disc}(g_1)(z,\epsilon)=\ri\, g_1(z,-\epsilon)\e^{\frac{Z_{\gamma_{21_n}}}{\epsilon }+
\frac{1}{\pi \ri}\sum\limits_{\gamma>0}  \langle\gamma_{21_n},\gamma\rangle\Omega(\gamma)
\mathcal{I}_{\gamma}(\epsilon)}
 \end{equation}
across the codim-1 locus $\arg(\epsilon)=\arg(-Z_{\gamma_{21_n}})$. While this formulation differs from the Riemann-Hilbert problem described in \eqref{sec:intro:disc}, the two can be reconciled with a careful treatment; see the next subsection.
 
\subsection{Matching Borel summation and open TBA}
We motivate the matching between Borel summation and the GMN open TBA by examining the Riemann-Hilbert problem for $\Psi^{(1)}_{\rm red}(\epsilon)$.  From the perspective of Borel summation, we obtain the same asymptotic behavior: $\Psi^{(1)}_{\rm red}(z,\epsilon)\sim\frac{1}{\left(-\sqrt{P(z,0)}\right)^{\frac{1}{2}}} $.  However, as discussed in \autoref{sec:intro:disc},  $\Psi^{(1)}_{\rm red}(z,\epsilon)$ exhibits different discontinuities, which are captured by the following TBA equation:
\begin{equation}\label{intro:natTBA}
\begin{aligned}
&\quad \ \Psi^{(1)}_{\rm red}(\epsilon) \\
&= \e^{-\frac{1}{2\pi\ri}\sum\limits_{\gamma>0}\langle\gamma_{21_1},\gamma\rangle\Omega(\gamma)\cI_\gamma(\epsilon)}\Bigg(\frac{1}{\left(-\sqrt{P(z,0)}\right)^{\frac{1}{2}}} \\
&\quad\quad + \sum_{n} \mu(\gamma_{21_n})
\frac{\epsilon}{2 \pi } \int\limits_{\ell_{\gamma_{21_n}}} \frac{\de \epsilon'}{\epsilon'}
\frac{1}{\epsilon' - \epsilon} \Psi^{(1)}_{\rm red}(-\epsilon')\e^{\frac{Z_{\gamma_{21_n}}}{\epsilon }+\frac{1}{\pi\ri}\sum\limits_{\gamma>0}\langle\gamma_n,\gamma\rangle\Omega(\gamma)\cI_\gamma(\epsilon')}\e^{\frac{1}{2\pi\ri}\sum\limits_{\gamma>0}\langle\gamma_{21_1},\gamma\rangle\Omega(\gamma)\cI_\gamma(\epsilon')}\Bigg).\\
\end{aligned}
\end{equation}

It is straight forward to see that \eqref{intro:natTBA} and \eqref{intro:TBAopen} are identical up to a factor
\begin{equation}\label{intro:relateTBAborel}
\boxed{
g_1(z,\epsilon)=\Psi_{\rm red}(z,\epsilon)N_{\rm GMN}(\epsilon)\equiv \Psi_{\rm red}(z,\epsilon)\e^{\frac{1}{2\pi \ri}
\sum\limits_{\gamma>0}  \langle\gamma_{21_1},\gamma\rangle\Omega(\gamma)
\cI_\gamma(\epsilon)}.}
\end{equation}
We present a more concrete derivation of \eqref{intro:relateTBAborel} in \autoref{sec:matchTBAborel} based on a conjecture. Note that the factor $N_{\rm GMN}(\epsilon)$ is a function of $\epsilon$ alone. In fact, the GMN open TBA solution corresponds to a specific normalization that symmetrizes the contributions of all solitons $\gamma_{21_n}$. Remarkably, this normalization also eliminates the discontinuities arising from 4d BPS states. Using \eqref{intro:relateTBAborel}, the numerical checks of the Borel summation discontinuities presented in \autoref{sec:weber} and \autoref{sec:su2} can be regarded as preliminary evidence supporting the conjectured GMN open TBA equations.

\subsection{Numerical comparison of Borel summation and open TBA}

Although Borel summability is a well-developed topic, there have been few detailed numerical investigations into its application for computing actual solutions. Similarly, the GMN TBA is firmly grounded in the rich physical and geometric structures of the 4d $\mathcal{N}=2$ theories and the GMN closed TBA has been successfully tested numerically across a range of examples \cite{Grassi:2019coc,Grassi:2021wpw,Ito:2024nlt,Ito:2021sjo};  however, the GMN open TBA has received comparatively little attention in the literature and remains largely unexplored even numerically.

We test numerically both the Pad\'{e}-Borel method and open TBA method in the examples of Argyres-Douglas-type with only a flavor BPS state corresponding to the Weber equation in \autoref{sec:weber} and of the pure $SU(2)$ theory corresponding to the modified Mathieu equation at $u=0$ in \autoref{sec:su2}. The numerical results \autoref{openTBA3}, \autoref{openTBA2i}, \autoref{openTBAsu2} and \autoref{openTBAsu22} provide further support for the consistency and agreement between the two methods.

\subsection{Future directions}
We comment on a few interesting points and future directions to explore:
\begin{itemize}
\item In the GMN TBA method, we have focused on the strong coupling region where the number of coupled integral equations is finite. It would be nice if we can solve the GMN TBA equations also in the weak coupling region where there are infinite number of coupled integral equations.
\item Another useful exact WKB method for the solution is the instanton counting method \cite{Jeong:2017mfh,Jeong:2018qpc,Jeong:2017pai,Jeong:2023qdr}. This method has the advantage that the Nekrasov partition function is an analytic special function. It would be interesting to compare the two exact WKB methods in this paper with the instanton counting method in some example.
\item In this paper, we work with differential equations and the corresponding 4d $\cN=2$ supersymmetric field theories. In particular, the open TBA equations give a new solution to the modified Mathieu equation corresponding to the pure $SU(2)$ theory. It would be interesting to generalize the GMN open TBA to the difference equations which correspond to 5d $\cN=1$ theories to produce exact solutions. It would also be interesting to compare it with other exact results obtained in \cite{Marino:2016rsq,Francois:2025wwd}. The exponential networks \cite{Eager:2016yxd,Banerjee:2018syt,Banerjee:2020moh,Grassi:2022zuk,Gupta:2024ics,Gupta:2024wos} and BPS states of local CY threefolds with 3d defects \cite{Douglas:2000qw,Longhi:2021qvz,Banerjee:2018syt,Closset:2019juk,Bonelli:2020dcp} should be important ingredients; see also \cite{Grassi:2014zfa,Codesido:2015dia,Marino:2016rsq,Gu:2022fss, DelMonte:2022kxh,DelMonte:2024dcr} for the exact WKB of quantum periods in 5d.

\item To discuss the uniqueness of solutions to the TBA equations, it is important to specify the boundary behavior as $\epsilon\rightarrow\infty$. In \cite{Grassi:2019coc}, a boundary condition for closed TBA in the limit $\epsilon\rightarrow -\infty$ is obtained from the identification of a quantum period with the Fredholm determinant in certain region of the parameter space and the instanton counting expression of the later. An analogous boundary condition for the GMN open TBA might be useful\footnote{We thank Alba Grassi for bringing this to the author's attention.}.

\end{itemize}

\section*{Acknowledgement}
We thank Ines Aniceto, Matijn Francois, Pavlo Gavrylenko, Jie Gu, Alba Grassi, Kohei Iwaki, Cristoforo Iossa, Pietro Longhi, Marcos Marino, Andrew Neitzke and Maximilian Schwick for helpful discussions. We especially thank Jie Gu for sharing the code for computing quantum periods in the pure $SU(2)$ theory. The work of QH is supported by the
Swiss National Science Foundation Grant No. 185723.

\section{Borel summation}\label{sec:padeborel}
In this section, we review the background of Borel summation and some known facts about it. We also discuss our new proposals on the singularities in the Borel plane and the corresponding discontinuities partially in \autoref{padeborelsol} and also in \autoref{sec:borel4d} and \autoref{sec:disc}.
\subsection{All-orders WKB}
\label{allorder}
The all-orders WKB analysis of \begin{equation}
\label{maindex2}
\left(-\epsilon^2\partial_z^2+P(z,\epsilon)\right)\psi(z)=0,
\end{equation}
 uses the following ansatz for a solution of 
\eqref{maindex2}:
\begin{equation}
\label{ansatzo}
\psi(z)=\exp\left(\frac{1}{\epsilon}\int_{z_0}^zY(z,\epsilon) \de z\right).
\end{equation}
Plugging in \eqref{maindex2}, $Y(x,\epsilon)$ satisfies the Ricatti equation
\begin{equation}\label{ricattix}
-\left(Y(z,\epsilon)\right)^2-\epsilon {\frac{\rd}{\rd z}} Y(z,\epsilon)+P(z,\epsilon)=0\, .
\end{equation}
$Y(z,\epsilon)$ can be solved formally as a power series in $\epsilon$
\begin{equation}\label{wkbY}
Y(z,\epsilon) = \sum_{n=0}^\infty Y^{n}(z)\epsilon^{n}\, ,
\end{equation}
where $Y^{0}=\pm\sqrt{P(z,0)}$.
We denote the two choices of sign by $Y^{0,(1)}$ and $Y^{0,(2)}$.
Note that the choice of $i=1,2,$ in $Y^{0,(i)}$ determine all the higher-order terms, so we denote \eqref{ansatzo} by the same label 
\begin{equation}\label{psidefori} \psi^{(i)}(z)=\exp\left(\frac{1}{\epsilon}\int_{z_0}^zY^{(i)}(z,\epsilon) \de z\right),
\end{equation}
with the relation
\begin{equation}\label{symsol}
\psi^{(1)}(z,\epsilon)=-\ri \psi^{(2)}(z,-\epsilon).
\end{equation}
The label $i$ can be identified with different sheets of $\Sigma$. In this paper we mainly study the sheet $i=1$. The point $z_0$ in \eqref{psidefori} is actually subtle. In this paper, we choose the normalization with $z_0=b_0$, where $b_0$ is a branch point of the projection map $\Sigma\xrightarrow[]{2:1} C$. We define
\begin{equation}\label{Upsilonallorder}
I^{(i)}(z,\epsilon)=\sum_{n=0}^\infty I^{n,(i)}(z)\epsilon^n=\sum_{n=0}^\infty \text{Reg}\left(\int_{b_0}^z Y^{n,(i)}(z)\rd z\right)\epsilon^{n}=\sum_{n=0}^\infty(\frac12\int\limits_{-\gamma_{ji}} Y^{n,(i)})\epsilon^n,
\end{equation}
where $\text{Reg}$ means the regularization of the integral, which is needed because $\int_{b_0}^z Y^{n,(i)}(z)\rd z$ for $n>1$ diverges at $b_0$ and $\gamma_{ji}\in H_1(\overline{\Sigma},\{z^{(j)},z^{(i)}\},\mathbb{Z})$ is the charge of BPS soliton. 
\begin{equation}\label{reggamma}
\text{Reg}(\begin{tikzpicture}[baseline={(0,0)},scale=0.5]
\draw[snake it, orange] (-3, 0) to (0, 0);
\node at (2.2,0) [right] {$z$};
\draw[->] (1,0 ) -- (1.2,0);
\node at (1.1,0) [above] {$i$};
\draw (0,0) -- (2,0);
\drawbranchpointmarker{0, 0};
\node at (0,0) [above] {$b_0$};
\filldraw[black] (2,0) circle(3pt); 
\end{tikzpicture})
=
\frac{1}{2}
\begin{tikzpicture}[baseline={(0,0)},scale=0.5]
\draw[snake it, orange] (-3, 0) to (0, 0);
\node at (2.2,0) [right] {$-\gamma_{ji}$};
\draw[->] (1,0.1 ) -- (1.2,0.1);
\draw (0.2,0.1) -- (2,0.1);
\draw[<-] (1.2,-0.1 ) -- (1.4,-0.1);
\draw (0.2,-0.1) -- (2,-0.1);
\draw [domain=26:360-26,line width=0.2mm] plot ({0.2*cos(\x)}, {0.2*sin(\x)});
\drawbranchpointmarker{0, 0};
\node at (0,0) [above] {$b_0$};
\node at (1.1,0.1) [above] {$i$};
\filldraw[black] (2,0) circle(3pt); 
\end{tikzpicture}
=
\frac{1}{2}
\begin{tikzpicture}[baseline={(0,0)},scale=0.5]
\draw[snake it, orange] (-3, 0) to (0, 0);
\node at (2.2,0) [right] {$-\gamma_{ji}$};
\draw [domain=0:360,line width=0.2mm] plot ({1.3*cos(\x)+0.7}, {0.4*sin(\x)});
\draw[->] (0.9,0.41 ) -- (1,0.41);
\drawbranchpointmarker{0, 0};
\node at (0,0.2) [above] {$b_0$};
\node at (1.1,0.3) [above] {$i$};
\filldraw[black] (2,0) circle(3pt); 
\end{tikzpicture}
\end{equation}
We now review briefly what a BPS soliton is. In the coupled 2d-4d system corresponding to the differential equation, there is a type of BPS state living on the surface defect supported at point $z\in C$. They are very similar to the BPS solitons living in the 2d $\cN=(2,2)$ supersymmetric theory, so they are also referred to as BPS solitons. A BPS soliton interpolating between $i$th and $j$th vacua is specified by the datum of the BPS soliton charge $\gamma_{ij}$, which is in the set of homology classes of open paths on $\Sigma$, running from the lift $z^{(j)}$ of $z$ to the lift $z^{(i)}$, both on $\Sigma$. The BPS solitons are counted by the BPS soliton degeneracy $\mu(\gamma_{ij})$, which is a piecewise invariant of the coupled 2d-4d  system. And the central charge of the BPS soliton is given by
\begin{equation}\label{solZ}
Z_{\gamma_{ij}}(z)=\int_{\gamma_{ij}}\sqrt{P(z,0)}\rd z=\int_{b_0}^z Y^{0,(i)}-Y^{0,(j)}\rd z.
\end{equation}
In this work, we usually fix a $z$ and vary $\epsilon$. So for convenience, we will represent $Z_{\gamma_{ij}}(z)$ simply by $Z_{\gamma_{ij}}$.

In fact, there are multiple solitons $-\gamma_{ji}$ in a coupled 2d-4d  system, thus we label the different solitons by subscripts $n=1,2,\cdots,$ as $-\gamma_{ji_n}$. Particularly, we pick a soliton for the regularization \eqref{Upsilonallorder} and denote it by $-\gamma_{ji_1}$ in this section for clarity.

\subsection{Borel summation}
The all-orders WKB series \eqref{wkbY} or \eqref{Upsilonallorder} are formal, factorially divergent power series with zero radius of convergence. 
Borel summation gives a way to convert this type of asymptotic series into an actual analytic function (for $\epsilon$ lying in some half-plane). We use the formal series \eqref{wkbY} as an example to introduce Borel summation. The Borel summation $s(Y)$ contains two steps. The first one is to take Borel transform of $Y(z,\epsilon)$, which is defined by
 \begin{equation}
 \label{boreltr}
 \hat{Y}(z,\zeta)=\sum_{n=0}^\infty \frac{Y^n(z)}{(n)!}\zeta^{n}.
 \end{equation} 
The Borel transform $ \hat{Y}(z,\zeta)$ is a convergent series for sufficiently small $\abs{\zeta}$.
The second step is to take a Laplace transform 
 \begin{equation}\label{laplace}
  s(Y)(z,\epsilon)=\frac{1}{|\epsilon|}\int_0^{\infty} \hat{Y}(z,{\rm e}^{\ri \arg(\epsilon)}\zeta)\re^{-\zeta/|\epsilon|}d\zeta. 
 \end{equation}
 Note that we need to analytically continue $\hat{Y}(z,\zeta)$ along the integration ray of \eqref{laplace},
beyond the convergence radius of \eqref{boreltr}. It has been discussed for example in \cite{MR3706198,nikolaev2020exact} that such 
an analytic continuation exists and the integral \eqref{laplace} converges 
for generic choices of $\arg(\epsilon)$. 
At certain special $\arg(\epsilon)$ such that a singularity of $\hat{Y}(z,{\rm e}^{\ri \arg(\epsilon)}\zeta)$ is on the integral contour, we say that $Y(z,\epsilon)$ is not Borel summable. In the case when $Y(z,\epsilon)$ is not Borel summable, we define the lateral Borel summation
 \begin{equation}\label{sdeflat}
  s_{\pm}(Y)(z,\epsilon)=\frac{1}{|\epsilon|}\int_0^{{\rm e}^{\ri 0^{\pm}}\infty} \hat{Y}(z,{\rm e}^{\ri \arg(\epsilon)}\zeta)\re^{-\zeta/|\epsilon|}d\zeta. 
 \end{equation}
Such singularities actually contain important information about the BPS spectra of the theory which will be described in \autoref{padeborelsol} and \autoref{sec:borel4d}. 

In numerical computations, we use a Pad\'{e} approximant of finitely many terms in the series \eqref{boreltr} to approximate the desired 
analytic continuation $\hat{Y}(z,\zeta)$. Note that this Pad\'{e} approximant
gives a rational function, meromorphic on the whole $\zeta$-plane. In our study, the poles always form some series. Only the first pole in each series of poles approximates the actual singularity. The Borel summation with this approximation is known as Pad\'{e}-Borel summation.

In this work, we study solutions of \eqref{maindex2} defined by
\begin{equation}\label{psidef}
\Psi^{(i)}(z,\epsilon)=\e^{\frac{\Upsilon^{(i)}(z,\epsilon)}{\epsilon}},
\end{equation}
where
\begin{equation}\label{Upsilondef}
\Upsilon^{(i)}(z,\epsilon)=s(I^{(i)})(z,\epsilon).
\end{equation}
For comparison of Borel summation and open TBA in the later sections, we also define the reduced solution by
\begin{equation} \label{defPsired}
\Psi^{(i)}_{\rm red}(z,\epsilon)=\e^{\frac{\Upsilon^{(i)}_{\rm red}(z,\epsilon)}{\epsilon}},
\end{equation}
where
\begin{equation}\label{defUpsilonred}
\Upsilon^{(i)}_{\rm red}(z,\epsilon)=\Upsilon^{(i)}(z,\epsilon)-\Upsilon^{0,(i)}(z,\epsilon).
\end{equation}
Since $\Psi^{(i)}(z,\epsilon)$ inherits the same relation
\begin{equation}\label{symsolPsi}
\Psi^{(1)}(z,\epsilon)=-\ri \Psi^{(2)}(z,-\epsilon),
\end{equation}
as the one of $\psi^{(i)}(z,\epsilon)$ in \eqref{symsol}, we only work with $\Psi^{(1)}(z,\epsilon)$ in this work.
\subsection{Stokes graphs and singularities at minus soliton central charges}
\label{padeborelsol}
In this chapter, we review some known facts about Borel summation of the all-orders WKB series \eqref{wkbY} and our new findings about the solutions. It is convenient to first introduce a tool, the Stokes graph $\mathcal{W}$, which is also known as the spectral network.

A Stokes graph $\mathcal{W}$ consists of Stokes curves on the Riemann surface $C$. Each Stokes curve of type $ij$ (where $i,j = 1,2$ and $i\neq j$) 
is an oriented trajectory starting from a branch point $b_0$ of $\sqrt{P(z,0)}$, along which 
$\arg(\epsilon)=\arg\left((Y^{0,(i)}-Y^{0,(j)}) \de z\right)$. This condition is equivalent to $\arg(\epsilon)=\arg(-Z_{\gamma_{ji}}(z))$ in terms of  \eqref{solZ}. At a Stokes curve of type $ij$, due to the Stokes phenomenon, the solution $\Psi^{(i)}$ has a discontinuity while the other solution $\Psi^{(j)}$ stays the same. In this work, Stokes curve labels are occasionally omitted for simplicity. The Stokes curve is also referred to as the wall in the language of the spectral network. 

These discontinuities across the Stokes curves are reflected by singularities on the Borel plane. It is known that
$\hat Y^{(i)}(z,\zeta)$ have singularities along the rays $\zeta \in (0,\re^{\I \arg(\epsilon)} \infty)$ 
if $z$ lies on a Stokes curve of type $ij$ \cite{MR3706198,nikolaev2020exact}. 
In \cite{Grassi:2021wpw}, it is further proposed that if $z$ lies on a Stokes curve of type $ij$, then $\hat Y^{(i)}(z,\zeta)$ has a singularity at 
\begin{equation} \label{polesSol}\zeta = -Z_{\gamma_{ji}}=\int_{b_0}^z Y^{0,(i)}-Y^{0,(j)}\rd z,
\end{equation}
where $b_0$ is the branch point where the Stokes curve starts. 
An example of the soliton charge $\gamma_{12}=-\gamma_{21}$ is shown in \eqref{reggamma}. Graphically, fixing a $z$ and using \eqref{polesSol} to determine the singularity $\zeta$, we define a cutoff version of the Stokes graph at $\arg(\epsilon) = \arg(\zeta)$, obtained by plotting the Stokes curves only up to $|2\int_{b_0}^{z} \lambda| = \abs{\zeta}$.  Then, there is a cutoff Stokes curve of $\cW$ that runs exactly up to the point $z$ -- such a cutoff Stokes curve corresponds to the charge of a soliton $\gamma_{ij}=-\gamma_{ji}$. This can be visualized as follows:

\begin{center}
\begin{tikzpicture}
  \begin{scope}
    \coordinate (center) at (0,0);

    \draw [thick,\wallColor] (center) -- ++(120:1.5cm); 
    \draw [thick,\wallColor] (center) -- ++(240:1.5cm); 
    \draw [red,thick] (center) -- ++(0:1.5cm);   

    \drawbranchpointmarker{center};
   \draw [red,domain=0:360,line width=0.2mm] plot ({0.9*cos(\x)+0.6}, {0.4*sin(\x)});
    \draw[->,red] (0.6,0.4 ) -- (0.61,0.4);
       \node[scale=0.6] at (0.6,0.4) [above] {$\color{red}i$};
        \node[scale=0.6] at (1.2,0.3) [above] {$\color{red}-\gamma_{ji}$};
   \filldraw[black] (1.5,0) circle(1pt); 
   \node at (1.5,0) [right] {$z$};
   \node[scale=0.6] at (0.8,0) [above] {$\color{red}ij$};
   \draw[->] (-0.45,-0.45*1.732 ) -- (-0.451,-0.451*1.732);
      \draw[->] (-0.45,0.45*1.732 ) -- (-0.451,0.451*1.732);
         \node[scale=0.6] at (-0.45,0.45*1.732 ) [left] {$ji$};
         \node[scale=0.6] at (-0.45,-0.45*1.732 ) [left] {$ji$};
    \draw[->,red] (0.8,0 ) -- (0.9,0);
   \draw[snake it, orange] (-3, 0) to (0, 0);
  \end{scope}
\end{tikzpicture}
\end{center}

We find that the same type of pole also appears on the Borel plane of $\hat{I}^{(1)}(z,\epsilon)$
 and we propose the same physical explanation for them in terms of minus soliton central charges.

\subsection{Singularities at minus 4d BPS central charges}\label{sec:borel4d}

We propose that there is also a second type of singularity appearing on the Borel plane of $\hat{I}^{(1)}(\zeta,\epsilon)$ normalized by $-\gamma_{ji_1}$. Those singularities are at the minus central charges of 4d BPS states $-Z_{\gamma}$, where $\gamma\in H^1(\overline{\Sigma},\mathbb{Z})$, with $\overline{\Sigma}$ obtained by filling in the punctures of $\Sigma$, such that $\langle\gamma,-\gamma_{ji_1}\rangle\neq 0$, where $\langle-,-\rangle$ is the intersection number of $\gamma$ and the soliton charge $-\gamma_{ji_1}$.

\begin{center}
\begin{tikzpicture}[baseline={(0,0)},scale=0.5]
\draw[snake it, orange] (-3, 0) to (0, 0);
\node at (1.2,-0.5) [below] {$-\gamma_{ji_1}$};
\node at (1.1,0.5) [above] {$i$};
\draw [domain=0:360,line width=0.2mm] plot ({1.3*cos(\x)+0.7}, {0.4*sin(\x)});
\draw[->] (0.9,0.41 ) -- (1,0.41);
\drawbranchpointmarker{0, 0};
\filldraw[black] (2,0) circle(3pt); 
\node at (2.2,0) [right] {$z$};
\draw [domain=0:180,line width=0.2mm] plot ({1*cos(\x)}, {3*sin(\x)-2});
\node at (-2.5,-1) [right] {$\gamma$};
\draw[<-] (-0.1,0.98 ) -- (0.1,0.98);
\node at (-0.2,0) [below] {$b_0$};
\end{tikzpicture}
\end{center}

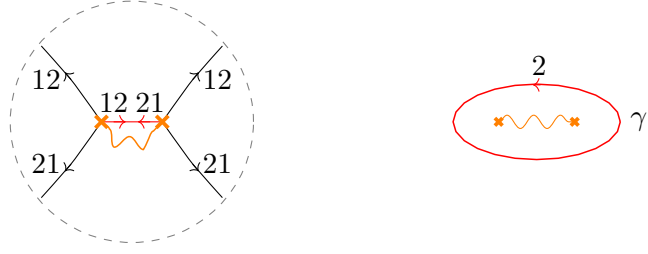
\begin{figure}
\centering
\begin{tikzpicture}[scale=0.8,baseline={(0,0)}]
\draw plot [smooth] coordinates {(0.5, 0)  (0.93592, -0.615053)  (1.18603, -0.913777) (1.38783, -1.14376)  (1.5622, -1.33749)  (1.71812, -1.50792) };
\draw plot [smooth] coordinates {(0.5, 0)  (0.93592, 0.615053)  (1.18603, 0.913777)  (1.38783, 1.14376)  (1.5622, 1.33749)  (1.71812, 1.50792) };
\draw plot [smooth] coordinates {(-0.5, 2.16506*10^-9) (-0.93592, 0.615053) (-1.18603, 0.913777) (-1.38783, 1.14376) (-1.5622, 1.33749) (-1.71812, 1.50792) };
\draw plot [smooth] coordinates {(-0.5, -2.16506*10^-9) (-0.93592, -0.615053) (-1.18603, -0.913777) (-1.38783, -1.14376) (-1.5622, -1.33749) };
\draw[red] (-1/2, 0) to (1/2, 0);
\draw [line width=0.4cm, white] (0,0) circle (2.2cm);
\draw[dashed,gray] (0,0) circle (2cm);
\drawbranchpointmarker{-1/2, 0};
\drawbranchpointmarker{1/2, 0};
\node at (0.3,0) [above] {$21$};
\node at (-0.3,0) [above] {$12$};
\node at (1,0.7) [right] {$12$};
\node at (-1,0.7) [left] {$12$};
\node at (1,-0.7) [right] {$21$};
\node at (-1,-0.7) [left] {$21$};
\draw[<-,red] (0.1,0 ) -- (0.3,0);
\draw[->,red] (-0.3,0 ) -- (-0.1,0);
\draw[->] (1,-0.7 ) -- (1.1,-0.82);
\draw[->] (1,0.7 ) -- (1.1,0.82);
\draw[->] (-1,-0.7 ) -- (-1.1,-0.82);
\draw[->] (-1,0.7 ) -- (-1.1,0.82);
\draw [domain=180:360,line width=0.2mm,orange,snake it] plot ({1/2*cos(\x)}, {0.2*sin(\x)});
\end{tikzpicture}
\hspace{2cm}
\begin{tikzpicture}[scale=0.5,baseline={(0,0)}]
\drawbranchpointmarker{-1, 0};
\drawbranchpointmarker{1, 0};
\draw [domain=0:360,line width=0.2mm,red] plot ({2.2*cos(\x)}, {1*sin(\x)});
\node at (2.2,0) [right] {$\gamma$};
\node at (-0.4,1.5) [right] {$2$};
\draw[<-,red] (-0.1,0.98 ) -- (0.1,0.98);
\draw[snake it, orange] (-1, 0) to (1, 0);
\end{tikzpicture}
\caption{Saddle connection in a degenerate spectral network and the corresponding 4d BPS state charge $\gamma$.}
\label{saddle}
\end{figure}

The 4d BPS states can be detected by Stokes graphs or spectral networks of degenerate type. The examples in this paper correspond to degenerate Stokes curves of the saddle connection type which consists of two curves going head to head on top of each other and connect two branch points. An example is shown in \autoref{saddle} with the saddle connection depicted in red. Such a degenerate wall can be deformed into a 1-loop $\gamma$ surrounding the two branch points at the end.

\subsection{Discontinuities of Borel summation}\label{sec:disc}
There are 2 types of discontinuities corresponding to the two types of singularities 
described in \autoref{padeborelsol} and \autoref{sec:borel4d} respectively.
They are conveniently described by 
\begin{equation}\label{eqn:defdisc1}
{\rm disc}(\frac{\Upsilon^{(i)}_{\rm red}}{\epsilon})(z,\epsilon)= s_+(\frac{I^{(i)}_{\rm red}}{\epsilon})(z,\epsilon)-s_-(\frac{I^{(i)}_{\rm red}}{\epsilon})(z,\epsilon),
\end{equation}
\begin{equation}\label{eqn:defdisc2}
{\rm disc}(\Psi^{(i)}_{\rm red})(z,\epsilon)= \E^{s_+(\frac{I^{(i)}_{\rm red}}{\epsilon})}(z,\epsilon)-\E^{s_-(\frac{I^{(i)}_{\rm red}}{\epsilon})}(z,\epsilon).
\end{equation}
where we use the lateral Borel summation \eqref{sdeflat}.

\subsubsection{The discontinuity corresponding to \texorpdfstring{$-Z_{\gamma_{ji_n}}$}{-Z[gammajin]}}

\label{disccomp}

We propose that the discontinuity of $\Psi^{(i)}_{\rm red}(z,\epsilon)$ corresponding to a minus BPS soliton central charge $-Z_{\gamma_{ji_n}}$ supported at $z$ is

\begin{equation}\label{sec2:jumpsol2}
\boxed{
\begin{aligned} 
&{\rm disc}(\Psi^{(i)}_{\rm red})(z,\epsilon)=\mu(\gamma_{{ji}_n})\Psi^{(j)}_{\rm red}(z,\epsilon)\e^{\frac{Z_{\gamma_{ji_n}}}{\epsilon }+\frac{1}{\pi\ri}\sum\limits_{\gamma>0}\langle\gamma_n,\gamma\rangle\Omega(\gamma)\mathcal{I}_{\gamma}(\epsilon)},\\
&\quad\quad\quad\quad\quad\quad\quad\quad\quad\quad\quad\quad\quad\quad\quad\quad\quad\quad\quad\quad\quad\quad\quad\text{at }\arg(\epsilon)=\arg(-Z_{\gamma_{{ji}_n}}(z)),
\end{aligned}}
 \end{equation}
 where 
 $\mu({\gamma_{ji_n}})$ is the BPS soliton degeneracy,  $\Omega(\gamma)$ is the 4d BPS degeneracy (also known as 4d BPS index), $\langle-,-\rangle$ is the intersection number,
 \begin{equation}\label{sec2:gamma4d}
 \gamma_n=\gamma_{\gamma_{ji_n}}-\gamma_{\gamma_{ji_1}}
 \end{equation} is obtained by concatenation of the two soliton charges,
\begin{equation}
\label{Idefm} \mathcal{I}_{\gamma}(\epsilon)=\int_{\ell_\gamma}d \epsilon' \frac{\epsilon}{(\epsilon')^2- \epsilon^2}
\log(1 - \sigma(\gamma)\cX_{\gamma}(\epsilon'))\, ,
\end{equation}
\begin{equation}\label{deflgamma}
\ell_{\gamma}=\{\epsilon':\frac{Z_{\gamma}}{\epsilon'}\in\mathbb{R}_-\},
\end{equation}
and $\sigma(\gamma)$ is quadratic refinement.
Beyond second-order ODE \eqref{maindex2}, \eqref{sec2:jumpsol2} should also apply to solutions of generic differential equations within the framework of the supersymmetric theory/ differential equation correspondence.

In the simplest case when $n=1$, the discontinuity  
\begin{equation}\label{sec2:jumpsol}
 {\rm disc}(\Psi^{(i)}_{\rm red})(z,\epsilon)=\mu({\gamma_{21_1}})\Psi^{(j)}_{\rm red}(z,\epsilon)\e^{\frac{Z_{{\gamma_{ji_1}}}}{\epsilon }}
 \end{equation}
at $\arg(\epsilon)=\arg(-Z_{{\gamma_{ji_1}}})$ has been proven for $\arg(\epsilon)\neq\arg(Z_{\gamma})$, $\forall \gamma$ and $\mu({\gamma_{ji_1}})=1$ \cite{Voros1983, 10.1007/978-4-431-68170-0_1,Iwaki:2014vad}.

To get \eqref{sec2:jumpsol2} for generic $n$, we pick another normalization $\Psi^{(i)}_{{\rm red},n}(z,\epsilon)$ regularized by $-{\gamma_{ji_n}}$. Then, according to \eqref{sec2:jumpsol} its discontinuity corresponding to $-Z_{\gamma_{ji_n}}$ is simplified to
\begin{equation}\label{dispsip}
 {\rm disc}(\Psi^{(i)}_{{\rm red},n})(z,\epsilon)=\mu({\gamma_{ji_n}})\Psi^{(j)}_{{\rm red},n}(z,\epsilon)\e^{\frac{Z_{{\gamma_{ji_n}}}}{\epsilon }}.
 \end{equation}
 We also know that $\Psi^{(i)}_{{\rm red},n}(z,\epsilon)$ and $\Psi^{(i)}_{\rm red}(z,\epsilon)$ are related by
 \begin{equation}\label{rationnorm}
 \frac{\Psi^{(i)}_{\rm red}(z,\epsilon)}{\Psi^{(i)}_{{\rm red},i}(z,\epsilon)}=(\cX_{\gamma_n}(\epsilon))^{\frac12}\e^{-\frac{ Z_{\gamma_n}}{2\epsilon}},
 \end{equation}
 where 
 \begin{equation}\label{QPdef} 
 \cX_{\gamma_n}(\epsilon)=\exp\Bigg(s\Big(\frac{1}{\epsilon}\sum_{m=0}^\infty\oint\limits_{\gamma_n} Y_m(z) \de z\,\epsilon^m\Big)(\epsilon)\Bigg)
 \end{equation}
is the quantum period for the charge $\gamma_n$. Furthermore, we use an equivalent exact WKB method to the Borel summation -- the conjectured GMN closed TBA to  substitute $\cX_{\gamma_n}(\epsilon)$ in \eqref{rationnorm} by
\begin{equation} \label{sec2:closetbe}
\log\cX_{\gamma_n}(\epsilon) = \frac{Z_{\gamma_n}}{\epsilon} +
\frac{1}{\pi \ri}\sum_{\gamma>0} \Omega(\gamma) \langle\gamma_n,\gamma\rangle
\mathcal{I}_{\gamma}(\epsilon).
\end{equation}
In total, \eqref{sec2:jumpsol2} is obtained by combining \eqref{dispsip}, \eqref{rationnorm} and \eqref{sec2:closetbe}.

\subsubsection{The discontinuity for \texorpdfstring{$-Z_{\gamma}$}{-Z_gamma}}

For the discontinuity corresponding to a 4d BPS state at $\arg(\epsilon)=\arg(-Z_{\gamma})$, our proposal is
\begin{equation}\label{sec2:jumppi} \boxed{
{\rm disc}(\frac{\Upsilon_{\rm red}^{(i)}}{\epsilon})(z,\epsilon)= \frac{\langle-\gamma_{{ji_1}},\gamma\rangle\Omega(\gamma)}{2}\log(1-\sigma(\gamma)\cX_{\gamma}(\epsilon)), \quad\quad\text{at }\arg(\epsilon)=\arg(-Z_{\gamma}).}
\end{equation}
\begin{center}
\begin{tikzpicture}[baseline={(0,0)},scale=0.5]
\draw[snake it, orange] (-3, 0) to (0, 0);
\node at (1.8,0.2) [above] {$1$ $-{\gamma_{ji_1}}$};
\draw [domain=0:360,line width=0.2mm] plot ({1.3*cos(\x)+0.7}, {0.4*sin(\x)});
\draw[->] (0.9,0.41 ) -- (1,0.41);
\drawbranchpointmarker{0, 0};
\filldraw[black] (2,0) circle(3pt); 
\node at (2.2,0) [right] {$z$};
\draw [domain=0:180,line width=0.2mm] plot ({1*cos(\x)}, {3*sin(\x)-2});
\node at (-2,-1) [right] {$\gamma$};
\draw[<-] (-0.1,0.98 ) -- (0.1,0.98);
\end{tikzpicture}
\end{center}
The discontinuity \eqref{sec2:jumppi} is a result of the normalization at $b_0$ regularized by $-\gamma_{ji_1}$. In fact, this discontinuity is $\frac{1}{2}$  of the Kontsevich-Soibelman transformation. This can be seen by using an alternative expression for \eqref{psidef} as
\begin{equation}
\Psi^{(i)}(z,\epsilon)=(\cX_{-{\gamma_{ji_1}}}(z,\epsilon))^{\frac12},
\end{equation}
where
 \begin{equation}\label{QPdef} 
\cX_{-{\gamma_{ji_1}}}(z,\epsilon)=\exp\Bigg(s\Big(\frac{1}{\epsilon}\sum_{n=0}^\infty\int\limits_{-{\gamma_{ji_1}}} Y_n(z) \de z\,\epsilon^n\Big)(\epsilon)\Bigg)
 \end{equation}
is the quantum open path integral defined by Borel summation. $\cX_{-{\gamma_{ji_1}}}(z,\epsilon)$ satisfies Kontsevich-Soibelman transformation parallel to the Vorus symbol for an open path described in \cite{Iwaki:2014vad}, while the  factor $\frac{1}{2}$ is due to the fact that the regularized integral contour in \eqref{Upsilonallorder} is $\frac{1}{2}$ of the charge $-{\gamma_{ji_1}}$. The statement of this type of singularity exists in the resurgence literature as fixed singularities \cite{delabaere:hal-01886535,Takei2008} and the discontinuity \eqref{sec2:jumppi} is discussed in \cite{Takei2008} for the example of the Weber equation. As a generalization, \eqref{intro:jumppi} applies to all second-order ODEs and their higher-order analogues within the supersymmetric theory/differential equation correspondence.  In fact, \eqref{sec2:jumppi} is crucial in the matching with the GMN open TBA conjecture.

\begin{figure}
\begin{centering}
\begin{tikzpicture}[scale=0.5]
  \def\a{1.5}        
  \def\b{0.5}      
  \def\cOne{0.4}     
  \def\cTwo{1.4}     
  \def\theta{110}   
  \begin{scope}[shift={(\cOne,\cTwo)}, rotate=\theta]
    \draw[purple,thick] (0,0) ellipse ({\a} and {\b});
  \end{scope}
  \filldraw [purple] (0.86,0) circle (1.5pt);
  \node at (-1.2,1) [below] {$\color{purple}-{\gamma_{ji_1}}$};
    \node at (0.86,0) [below] {$\color{purple}z$};
  \draw[->,purple] (0.9,1.4) -- (01,1.2);
    \node at (0.9,1.4)  [left] {$\color{purple}1$};
  \def\ap{2.8}        
  \def\bp{0.5}      
  \def\cOnep{2.1}     
  \def\cTwop{1.4}     
  \def\thetap{150}   
  \begin{scope}[shift={(\cOnep,\cTwop)}, rotate=\thetap]
    \draw[magenta,thick] (0,0) ellipse ({\ap} and {\bp});
  \end{scope}
    \filldraw [magenta] (4.5,0) circle (1.5pt);
\node at (5,2) [below] {$\color{magenta}-{\gamma_{ji_1}}$};
\node at (4.5,0) [below] {$\color{magenta}z'$};
\draw[->,magenta] (3,1.4) -- (3.2,1.3);
\node at (3,1.6) [right] {$\color{magenta}1$};
\draw [domain=90:270,line width=0.2mm] plot ({0.5*cos(\x)}, {0.5*sin(\x)+2.5});
\draw [domain=90:270,line width=0.2mm] plot ({0.5*cos(\x)}, {0.5*sin(\x)-2.5});
\draw [domain=-90:90,line width=0.2mm] plot ({2*cos(\x)}, {2*sin(\x)});
\draw [domain=-90:90,line width=0.2mm] plot ({3*cos(\x)}, {3*sin(\x)});
\draw [domain=-90:90,line width=0.2mm, red] plot ({2.5*cos(\x)}, {2.5*sin(\x)});
\drawbranchpointmarker{0,2.5};
\drawbranchpointmarker{0,-2.5};
\drawbranchpointmarker{0,0};
\draw[snake it, orange] (0,2.5) to (0,0);
\draw[snake it, orange] (0,-2.5) to (0,-4);
\filldraw [blue] (0,0) circle (1.5pt);
\node at (3,-1) [right] {$\gamma$};
\node at (3,-0.2) [right] {$1$};
\draw[->] (3,-0.2) -- (3,0.2);
\end{tikzpicture}
\caption{Wall of marginal stability and different intersection numbers for different soliton charges.}
\label{wallcrossing2d4d}
\end{centering}
\end{figure}
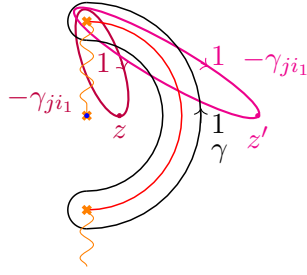

\begin{figure}
     \centering
     \begin{subfigure}[c]{0.47\textwidth}
         \centering
         \includegraphics[width=0.7\textwidth]{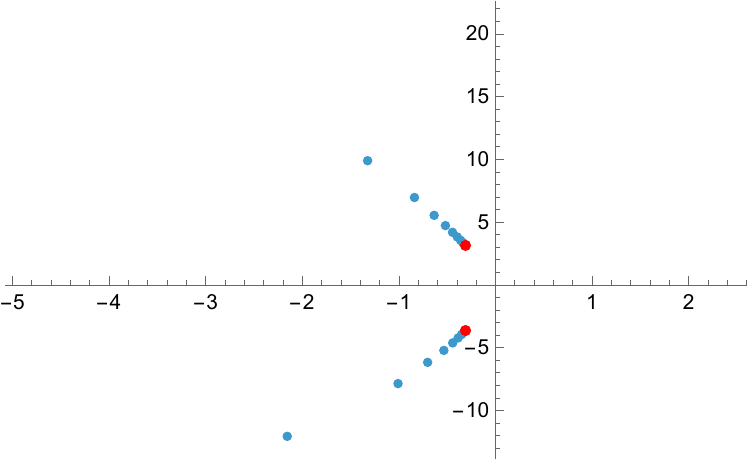}
         \caption{Borel plane of $\Upsilon_{\rm red}^{\text{norm}_1,(1)}(\epsilon)$ at $z=\frac{\ri}{10}+\frac{10}{9}$ with 56 terms. The value of $z$ is chosen such that $\abs{Z_{\gamma_{21_i}}}<\abs{Z_{\gamma_d}}=\abs{Z_{\gamma_m}}$. Thus most of poles appear as series from $- Z_{\gamma_{21_1}}$ and $- Z_{\gamma_{21_2}}$. Only the first pole in each series of poles corresponds to the actual singularity that has physical meaning. The numerical values of the singularities match with $- Z_{\gamma_{21_1}}$ and $- Z_{\gamma_{21_2}}$, which we show as the red dots.}
         \label{BPSrayst}
     \end{subfigure}
     \hfill
          \begin{subfigure}[c]{0.47\textwidth}
         \centering
         \includegraphics[width=0.7\textwidth]{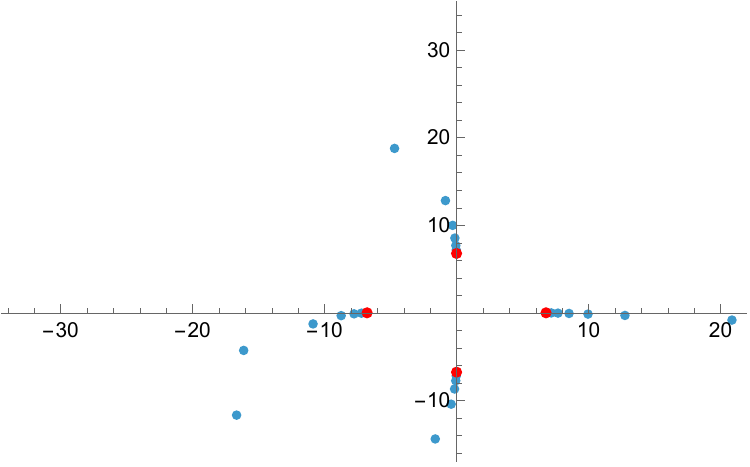}
         \caption{Borel plane of $\Upsilon_{\rm red}^{\text{norm}_1,(1)}$ at $z=8+9\ri$ with 56 terms. The value of $z$ is chosen such that $\abs{Z_{\gamma_{21_i}}}>\abs{Z_{\gamma_d}}=\abs{Z_{\gamma_m}}$. Thus most of poles appear as series from $- Z_{\pm\gamma_{d}}$ and $- Z_{\pm\gamma_{m}}$. Only the first pole in each series of poles corresponds to the actual singularity that has physical meaning.  The numerical values of the singularities match with $-Z_{\pm\gamma_{d}}$ and $- Z_{\pm\gamma_{m}}$, which we show as the red dots.}
         \label{BPSraywk}
     \end{subfigure}
       \\
      \begin{subfigure}[c]{0.45\textwidth}
         \centering
         \includegraphics[width=0.8\textwidth]{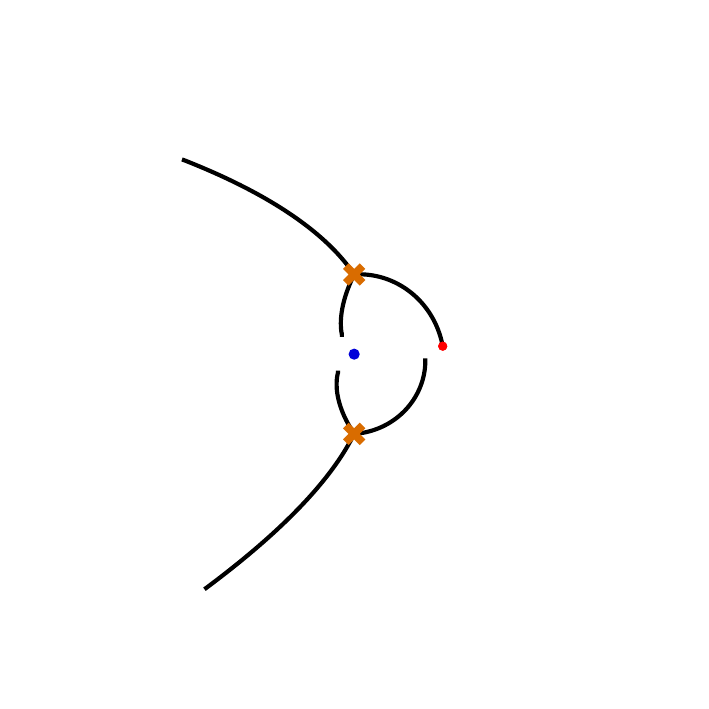}
         \caption{The cutoff version 
     of the Stokes graph with $\arg(\epsilon)= \arg(-Z_{\gamma_{21_1}})$, where we 
     plot the Stokes curves only up
     to $|2\int_{b_0}^{z} \lambda| = \abs{Z_{\gamma_{21_1}}}$.
    $z=\frac{\ri}{10}+\frac{10}{9}$ is plotted as a red dot.}
    \label{su2reg1sol1}
      \end{subfigure}
          \hspace{1cm}
            \begin{subfigure}[c]{0.45\textwidth}
         \centering
         \includegraphics[width=0.8\textwidth]{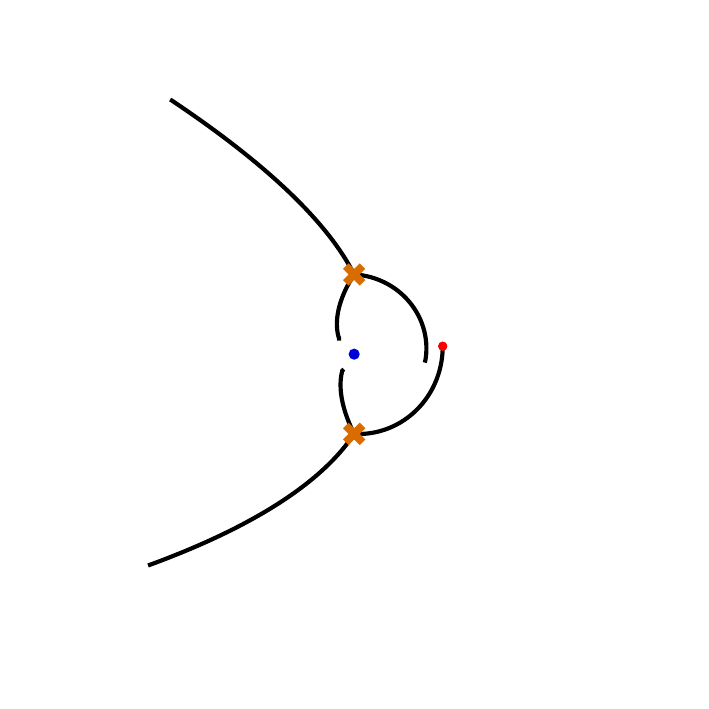}
         \caption{The cutoff version 
     of the Stokes graph with $\arg(\epsilon)= \arg(-Z_{\gamma_{21_2}})$, where we 
     plot the Stokes curves only up
     to $|2\int_{b_0}^{z} \lambda| = \abs{Z_{\gamma_{21_2}}}$.
   $z=\frac{\ri}{10}+\frac{10}{9}$ is plotted as a red dot.}
       \label{su2reg1sol2}
      \end{subfigure}
             \caption{The Borel plane for pure $SU(2)$ theory in region \textcircled{1} and corresponding solitons.}
        \label{4borelsu2r1}
\end{figure}

An important point is that $\langle-{\gamma_{ji_1}},\gamma\rangle$ should be calculated using the $\gamma$ represented by the degenerate wall, for example the red saddle connection in \autoref{wallcrossing2d4d}, and $-{\gamma_{ji_1}}$ represented by the soliton charge, for example the cutoff Stokes curve of $\cW$ running exactly up to the red point in \autoref{su2reg1sol1}. Different $z$ or $z'$ as shown in \autoref{wallcrossing2d4d} contribute to a sign difference in $\langle-{\gamma_{ji_1}},\gamma\rangle$ because the degenerate wall is, in fact, a wall of marginal stability for the coupled 2d-4d system; see, for example, \textsection{}7.2 of \cite{Gaiotto:2011tf}. 

\section{Gaiotto-Moore-Neitzke open TBA equations}\label{sec:openTBA}

GMN open TBA is a conjectural TBA-based exact WKB method originally developed in a type of 4d $\cN=2$ supersymmetric theory known as the theory of class $\cS$ \cite{Gaiotto:2011tf}. The open TBA equations are satisfied by functions which are conjectured to be related to solutions of differential equations of arbitrary order within the context of the supersymmetric theory/differential equation correspondence.  In \autoref{3dN=4} and \autoref{sec:conflimit}, we briefly review the background of open TBA. For more details, derivations and physical explanations, we refer the readers to \cite{Gaiotto:2009hg,Gaiotto:2011tf,MR3115984,Gaiotto:2014bza}. In \autoref{sec:matchTBAborel}, we show how to relate open TBA solutions to the solutions obtained by Borel summation.

\subsection{Coulomb branch of 3d \texorpdfstring{$\mathcal{N}=4$}{N=4} theory}\label{3dN=4}

In the class $\cS$ construction \cite{Gaiotto:2011tf}, a further compactification on $\mathbb{R}^3\times S^1$ results in a $3d$ $\cN=4$ theory. The Coulomb branch $\cM$ of the $3d$ $\cN=4$ theory is a well studied hyperk\"{a}hler manifold  which is also the moduli space of solutions of Hitchin's equations. Label the complex structure by $\zeta$. Given $(A,\Phi)$ solving Hitchin's equations where $A$ is a $G$-connection in a $G$-bundle $E$ on the Riemann surface $C$ and $\Phi\in\Omega^1(\text{End}\ E)$, for each $\zeta$ we can define a flat connection
\begin{equation}
\cA=\frac{\Phi}{\zeta}+A+\zeta\overline{\Phi}.
\end{equation}
In this part, we are mainly focusing on the integral equation with respect to $\zeta$, so we omit the dependence on $z$ and all other parameters. Define a gauge transformation $g_i(\zeta)$ which is a column vector, such that
\begin{equation}
\cA(\zeta) = \sum_i  g_i(\zeta) a_i(\zeta) g_{-i}(\zeta) + g_i(\zeta) \rd g_{-i}(\zeta)
\end{equation}
is diagonalized.
\cite{Gaiotto:2009hg,Gaiotto:2011tf} conjectures that $g_i(\zeta)$ satisfies
\begin{equation} \label{eq:int-2}
g_i(\zeta) = g_i^+ + \frac{1}{2 \pi \I} \sum_{n} \mu(\gamma_{ji_n})
\int\limits_{\ell_{\gamma_{ji_n}}} \frac{\de \zeta'}{\zeta'}
\frac{ \zeta}{\zeta' - \zeta} g_j (\zeta') X_{\gamma_{ji_n}}(\zeta'),
\end{equation}
while 
\begin{equation}
a_{j} (\zeta)= d x_{\gamma_{ij}}(\zeta),
\end{equation} 
where $X_{\gamma_{ji_n}}(\zeta)=\e^{x_{\gamma_{ji_n}}(\zeta)}$ is an auxiliary function defined by
\begin{equation} \label{eq:int-old-x}
 x_{\gamma_{ji_n}}(\zeta) = \frac{Z_{\gamma_{ji_n}}}{\zeta} + \I \theta_{\gamma_{ji_n}} + \bar Z_{\gamma_{ji_n}} \zeta +  \frac{1}{4 \pi \I} \sum_{\gamma} \Omega(\gamma)\langle\gamma_{ji_n},\gamma\rangle \int\limits_{\ell_{\gamma}} \frac{\de \zeta'}{\zeta'} \frac{\zeta' + \zeta}{\zeta' - \zeta} \log(1 - \sigma(\gamma)\cX_{\gamma}(\zeta')),
\end{equation} 
where $\theta_{\gamma_{ji_n}} $ is a constant and we are interested in the case at $\theta_{\gamma_{ji_n}}=0$. The inverse $g_{-k}(\zeta)$ satisfies integral equations analogous to \eqref{eq:int-2}. 
The $\cX_{\gamma}$ or $x_\gamma=\log(\cX_{\gamma})$ showing up in the previous formulae satisfy
\begin{equation} \label{eq:int-old}
x_{\gamma}(\zeta) =\log \cX_{\gamma}(\zeta) = \frac{Z_\gamma}{\zeta} + \I \theta_\gamma + \overline{ Z}_\gamma \zeta +   \frac{1}{4 \pi \I} \sum_{\gamma'} \langle\gamma, \gamma'\rangle \Omega(\gamma')\int_{\ell_{\gamma'}} \frac{\de \zeta'}{\zeta'} \frac{\zeta' + \zeta}{\zeta' - \zeta} \log(1 - \sigma(\gamma')\cX_{\gamma'}(\zeta')),
\end{equation}
where $\theta_\gamma$ is related to the angular coordinate on the fiber and we are interested in the case where $\theta_\gamma=0$. $\cX_{\gamma}$ is related to the quantum period as will be discussed in \autoref{sec:conflimit}.

\subsection{The conformal limit}\label{sec:conflimit}
We are particularly interested in the conformal limit \cite{Gaiotto:2009hg,Gaiotto:2014bza,Dumitrescu:2016ick,Collier:2024uxi}, 
\begin{equation}
\Phi\rightarrow R\Phi ,\quad Z_{\gamma}\rightarrow R Z_{\gamma} ,\quad \zeta\rightarrow 0,\quad R\rightarrow 0,\quad \epsilon=\frac{\zeta}{R}\ \text{ finite},
\end{equation}
which is known to make contact with the Schr\"{o}dinger type differential equation \eqref{maindex2} \cite{Gaiotto:2014bza}. In particular, \eqref{eq:int-old} becomes closed TBA \eqref{sec2:closetbe}, which we reproduce here
\begin{equation} \label{eq:closetbe}
x_{\gamma}(\epsilon) = \frac{Z_{\gamma}}{\epsilon} +
 \frac{1}{\pi \ri}\sum_{\gamma'>0} \langle\gamma,\gamma'\rangle\Omega(\gamma')
\mathcal{I}_{\gamma'}(\epsilon),
\end{equation}
where $\mathcal{I}_{\gamma}(\epsilon)$ is defined in \eqref{Idefm}.
The flat connection 
\begin{equation}
\cA=\sum_i  g_i a_i g_{-i}  + g_i dg_{-i} 
\end{equation}
in this limit consists of 
\begin{equation} 
a_i(\zeta) =\frac{\lambda_i}{\epsilon} +  \frac{1}{\pi \ri}\sum_{\gamma'>0} \Omega(\gamma')\langle\rd\gamma_{ij},\gamma'\rangle 
\cI_{\gamma'}(\epsilon),
\end{equation}
and
\begin{equation} \label{eq:int-2-conf}
g_i(\epsilon) = g_i^+ + \frac{1}{2 \pi \I}\sum_{n} \mu(\gamma_{j i_n})
\int_{\ell_{\gamma_{j i_n }}} \frac{\de \epsilon'}{\epsilon'}
\frac{\epsilon }{\epsilon' - \epsilon} g_j (\epsilon') X_{\gamma_{j i_n}}(\epsilon'),
\end{equation}
which is known as the open TBA where
\begin{equation}\label{xijconformal}
\log(X_{\gamma_{j i_n}}(\epsilon))=x_{\gamma_{ji_n}}(\epsilon) =\frac{ Z_{\gamma_{ji_n}}}{\epsilon} +
\frac{1}{\pi \ri}\sum_{\gamma>0} \langle\gamma_{ji_n},\gamma\rangle\Omega(\gamma) 
\mathcal{I}_{\gamma}(\epsilon).
\end{equation}
The examples studied in this work correspond to $G=SU(2)$. In this case, the integral equations have a symmetry  $g_2(\epsilon) = \I g_1(-\epsilon)$. Therefore we can focus only on $g_1(\epsilon)$ and rewrite the open TBA integral equation as
\begin{equation} \label{eq:int-22}
\begin{aligned}
g_1(\epsilon) &= g_1^+ + \frac{1}{2 \pi }\sum_{n} \mu(\gamma_{21_n})
 \int_{\ell_{\gamma_{21_n}}} \frac{\de \epsilon'}{\epsilon'}
\frac{\epsilon}{\epsilon' - \epsilon} g_1 (-\epsilon') X_{\gamma_{21_n}}(\epsilon')\\
& = g_1^+ - \frac{1}{2 \pi }\sum_{n} \mu(\gamma_{21_n})
 \int_{\ell_{\gamma_{12_n}}} \frac{\de \epsilon'}{\epsilon'}
\frac{\epsilon}{\epsilon' + \epsilon} g_1 (\epsilon') X_{\gamma_{12_n}}(\epsilon'),
\end{aligned}
\end{equation}
such that only $g_1(\epsilon)$ appears in the equation.
The integral equation \eqref{eq:int-22} is defined up to a constant. We choose
\begin{equation}\label{gplus}
g^+_1(z)=\frac{1}{\left(-\sqrt{P(z,0)}\right)^{\frac{1}{2}}}.
\end{equation}

The two elements of the column vector $g_i(z,\epsilon)$ are actually related by the action  of $\Phi$ \cite{Gaiotto:2014bza}. As a consequence, \eqref{eq:int-22} can also be viewed as the integral equation for an element of $g_1(z,\epsilon)$  itself. For simplicity, we loosely refer to the element of  $g_1(z,\epsilon)$ as $g_1(z,\epsilon)$ itself.

\subsection{The match with Borel summation}\label{sec:matchTBAborel}

In the conformal limit \cite{Gaiotto:2014bza,Dumitrescu:2016ick,Collier:2024uxi}, we study the flat section $\varphi(z,\epsilon)$ satisfying
\begin{equation}\label{flatsecdef}
\left(\partial_z+\cA\right)\varphi(z,\epsilon)=0.
\end{equation}

Just as with $g_i(z,\epsilon)$, the two components of the flat section $\varphi_i(z,\epsilon)$ are related, so it suffices to consider only one of them and we also refer to it as $\varphi_i(z,\epsilon)$. The flat sections $\varphi_i(z,\epsilon)$, where $i=1,2$ denotes a basis of the flat section, can be obtained from $g_i(z,\epsilon)$. To see this, we rewrite \eqref{flatsecdef} as
\begin{equation}
\left(g_{-i}(z,\epsilon)\partial_z+g_{-i}(z,\epsilon) \cA\right)g_i(z,\epsilon)\phi_i(z,\epsilon)=0,
\end{equation}
where 
\begin{equation}\label{TBAflatsec}
\varphi_i(z,\epsilon)=g_i(z,\epsilon)\phi_i(z,\epsilon).
\end{equation}
In \cite{Gaiotto:2014bza} it is conjectured that $g_i(z,\epsilon)$ corresponds to a solution of the differential equation \eqref{maindex2}. We further propose that $\varphi_i(z,\epsilon)$ is equivalent to the solution of \eqref{maindex2} obtained by the Borel summation of the WKB ansatz 
\begin{equation}
\varphi_i(z,\epsilon)=\Psi^{(i)}(z,\epsilon),
\end{equation} 
with the same choice of normalization as which will become clear later. More precisely, to obtain $\varphi_i(z,\epsilon)$,
using $g_{-i}g_{j}=\delta_{ij}$, we get
\begin{equation}
\phi_i(z,\epsilon)=\E^{\int_{z_0}^z a_i(z,\epsilon)\rd z}.\end{equation}
 Note that $\phi_i(z,\epsilon)$ is defined up to a normalization factor depending on the end point $z_0$ of the integral. This is the same ambiguity as the $z_0$ in \eqref{psidefori} of the Borel summation method. In order to match the two methods, we pick $a_1$ where the corresponding $\lambda_1=Y^{0,(1)}$ and choose the same normalization as in \eqref{Upsilonallorder}. We define
\begin{equation}\label{TBAphi1}
\begin{aligned}
\phi_1(z,\epsilon)&=\E^{\frac{1}{2}\int\limits_{-\gamma_{21}} a_1(z,\epsilon)\rd z}= \E^{\frac12 x_{\gamma_{12}}(\epsilon) }\\
&=\E^{\frac{1}{2}\frac{ Z_{\gamma_{12}}}{\epsilon} +
\frac{1}{2\pi \ri}\sum\limits_{\gamma>0} \Omega(\gamma)\langle\gamma_{12},\gamma\rangle 
\cI_{\gamma}(\epsilon)}.
\end{aligned}
 \end{equation}

We now go back to the actual flat section $\varphi_1(z,\epsilon)$ in \eqref{TBAflatsec}. The only part of $\phi_1(z,\epsilon)$ depending on $z$ is $\E^{\frac{1}{2}\frac{ Z_{\gamma_{12}}}{\epsilon} }$ and this matches the leading order term of \eqref{psidef}. 
In fact, the part of $\phi_1(z,\epsilon)$ that doesn't depend on $z$ can be regarded as the normalization factor relating \eqref{defPsired} and $g_1(z,\epsilon)$ via
\begin{equation}\label{relateTBAborel}
\boxed{
g_1(z,\epsilon)=\Psi^{(1)}_{\rm red}(z,\epsilon)\e^{
\frac{1}{2\pi \ri}\sum\limits_{\gamma>0} \Omega(\gamma) \langle\gamma_{21},\gamma\rangle
\cI_{\gamma}(\epsilon)}.}
\end{equation}
It is straight forward to check that all the discontinuities of $g_1(z,\epsilon)$ and $\Psi^{(1)}_{\rm red}(z,\epsilon)$ match this way. In fact, it is possible to formulate an equivalent TBA equation for $\Psi^{(1)}_{\rm red}(z,\epsilon)$
\begin{equation}\label{natTBA}
\begin{aligned}
&\quad \ \Psi^{(1)}_{\rm red}(\epsilon) \\
&= \e^{-\frac{1}{2\pi\ri}\sum\limits_{\gamma>0}\Omega(\gamma)\langle\gamma_{{\rm 21}_1},\gamma\rangle\cI_{\gamma}(\epsilon)}\Bigg(\frac{1}{\left(-\sqrt{P(z,0)}\right)^{\frac{1}{2}}} \\
&\quad\quad+ \sum_{i} \mu(\gamma_{{\rm 21}_i})
\frac{\epsilon}{2 \pi } \int\limits_{\ell_{\gamma_{{\rm 21 }_i}}} \frac{\de \epsilon'}{\epsilon'}
\frac{1}{\epsilon' - \epsilon} \Psi^{(1)}_{\rm red}(-\epsilon')\e^{\frac{Z_{\gamma_{{\rm 21}_i}}}{\epsilon }+\frac{1}{\pi\ri}\sum\limits_{\gamma>0}\Omega(\gamma)\langle\gamma_{i},\gamma\rangle\cI_{\gamma}(\epsilon')}\e^{\frac{1}{2\pi\ri}\sum\limits_{\gamma>0}\Omega(\gamma)\langle\gamma_{{\rm 21}_1},\gamma\rangle\cI_{\gamma}(\epsilon')}\Bigg),
\end{aligned}
\end{equation}
where we rename the $\gamma_{21}$ used for the normalization in \eqref{Upsilonallorder} and \eqref{TBAphi1} as $\gamma_{21_1}$ for clarity and all other solitons are denoted by  $\gamma_{21_i}$, $i\neq 1$ and we define $\gamma_{i}=\gamma_{21_i}-\gamma_{21_1}$. Indeed, \eqref{natTBA} directly solves the Riemann-Hilbert problem arising from \autoref{sec:disc}. The term
\[
\e^{\frac{1}{2\pi\ri}\sum\limits_{\gamma>0}\Omega(\gamma)\langle\gamma_{{\rm 21}_1},\gamma\rangle\cI_{\gamma}(\epsilon')}
\]
 in the parenthesis of \eqref{natTBA} precisely cancels the overall factor outside the parenthesis that encodes the discontinuities associated with 4d BPS states, when we focus on the discontinuities arising from the BPS soliton central charges.

\section{Weber equation}\label{sec:weber}
The example of the Airy function is studied in \cite{Gaiotto:2014bza}. The next simplest but more interesting example of a second order ODE is the Weber equation,
\begin{equation}\label{eqn:weber}
\epsilon^2\partial_z^2\psi(z)-(\frac{z^2}{4}-m^2)\psi(z)=0,
\end{equation}
where $m$ is the mass parameter and the residue at $z=\infty$ is $-m^2$. Without loss of generality, we choose $m=1$. In this example
\begin{equation}
P(z)=\frac{z^2}{4}-1.\footnotemark
\end{equation}
\footnotetext{We follow the convention of \cite{Iwaki:2020efz}.}
There are two branch points at $b_0=\pm 2$ respectively. This theory has a 4d flavor BPS hypermultiplet (and its antiparticle) corresponding to a loop $\gamma_f$ ($-\gamma_f$) with central charge 
\begin{equation} Z_{\gamma_f}=\oint_{\gamma_f} Y^{0,(2)} \rd z=- 2\pi \ri,\end{equation}
and  BPS index 
\begin{equation} \Omega(\pm{\gamma_f})=1.\end{equation}

\begin{center}
\begin{tikzpicture}[scale=0.5]
\draw[snake it, orange] (-1, 0) to (1, 0);
\drawbranchpointmarker{-1, 0};
\drawbranchpointmarker{1, 0};
\draw [domain=0:360,line width=0.2mm] plot ({2.2*cos(\x)}, {1*sin(\x)});
\node at (2.2,0) [right] {$\gamma_f$};
\node at (-0.4,1.5) [right] {$2$};
\draw[<-] (-0.1,0.98 ) -- (0.1,0.98);
\end{tikzpicture}
\end{center}
The corresponding quantum period $\cX_{\gamma_f}$ is not subject to quantum corrections and
\begin{equation}\label{Weberhyper}
 \cX_{\gamma_f}(\epsilon)=\e^{\frac{Z_{\gamma_f}}{\epsilon}}=\e^{-\frac{2\pi \ri}{\epsilon}}.
 \end{equation}
In this example, the BPS soliton degeneracies $\mu(\gamma_{21})$ are all 1.

\subsection{Pad\'{e}-Borel summation}
We use Pad\'{e}-Borel summation to get the solutions of the Weber equation. We fix the convention of sheet label by
\begin{equation}
Y^{0,(1)}(z)=-\frac{1}{2} \sqrt{z^2-4}.
\end{equation}
Without loss of generality we focus on  $\Psi^{(1)}_{\rm red}(z,\epsilon)$ or $\Upsilon^{(1)}_{\rm red}(z,\epsilon)$
where we use \eqref{Upsilonallorder} regularized by $-\gamma_{21_{\pm}}$ surrounding either $b_0=2$ or  $b_0=-2$\footnote{The two regularizations are the same for $\Psi^{(1)}_{\rm red}(z,\epsilon)$ in this example, since $\cX_{\gamma_f}$ is not subject to quantum corrections but are different for $\Psi^{(1)}(z,\epsilon)$ by $\cX_{\gamma_f}$.}. The other solution can be obtained using \eqref{symsolPsi}. 
\begin{center}
\begin{tikzpicture}[scale=1]
\def\theta{35}
\draw plot [smooth] coordinates {(0,1) (0.75,0)};
\draw plot [smooth] coordinates {(0,1) (1.1,0.21)};
\draw [domain=170:360] plot ({0.93+0.2*cos(\x)*cos(\theta)-0.4*sin(\x)*sin(\theta)}, {0.1+0.2*cos(\x)*sin(\theta)+0.3*sin(\x)*cos(\theta)});
\def\theta{90+35+20}
\draw plot [smooth] coordinates {(0,1) (-0.75,0)};
\draw plot [smooth] coordinates {(0,1) (-1.1,0.21)};
\draw [domain=-170:-360] plot ({-0.93+0.2*cos(\x)*cos(\theta)-0.4*sin(\x)*sin(\theta)}, {0.1+0.2*cos(\x)*sin(\theta)+0.3*sin(\x)*cos(\theta)});
\draw[snake it, orange] (-1, 0) to (1, 0);
\drawbranchpointmarker{-1, 0};
\drawbranchpointmarker{1, 0};
\node at (0.8,0.8) [right] {$-\gamma_{21_+}$};
\node at (-0.8,0.8) [left] {$-\gamma_{21_-}$};
\node at (0,1) [above] {$z$};
\draw[<-] (-0.48,0.36) --(-0.75,0);
\node at (-0.48,0.36) [right] {$1$};
\draw[<-] (0.87,0.36) --(0.89,0.35);
\node at (0.92,0.36) [right] {$1$};
\end{tikzpicture}
\end{center}

\begin{figure}
     \centering
     \begin{subfigure}[c]{0.47\textwidth}
         \centering
         \includegraphics[width=0.7\textwidth]{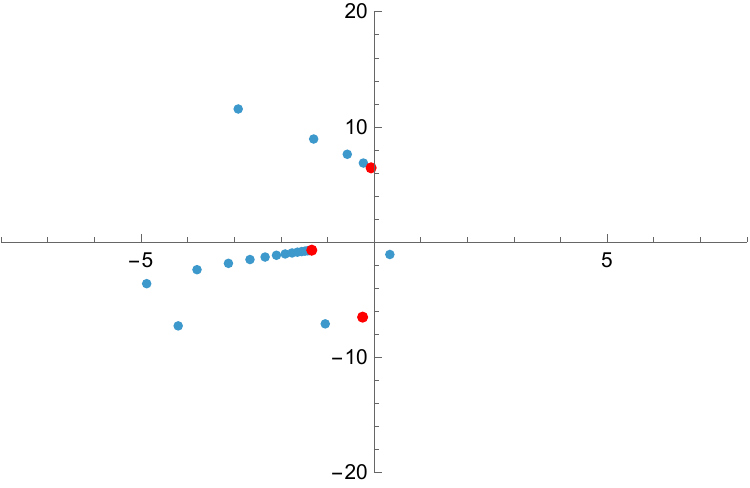}
         \caption{Borel plane of $\Upsilon_{\rm red}^{(1)}(z,\epsilon)$ at $z=3\, \e^{\frac{\ri\pi}{30}}$ with 56 terms.  Only the first pole in each series of poles corresponds to a true singularity of $\hat{\Upsilon}_{\rm red}^{(1)}(z,\zeta)$ that has physical meaning and we show it in red. The numerical values of singularities are at $\zeta=-1.334-0.700\ri, -0.061+6.446\ri, -0.243-6.511\ri$.}
         \label{BPSrayst}
     \end{subfigure}
     \hfill
          \begin{subfigure}[c]{0.47\textwidth}
         \centering
         \vspace{-0.8cm}
         \includegraphics[width=0.7\textwidth]{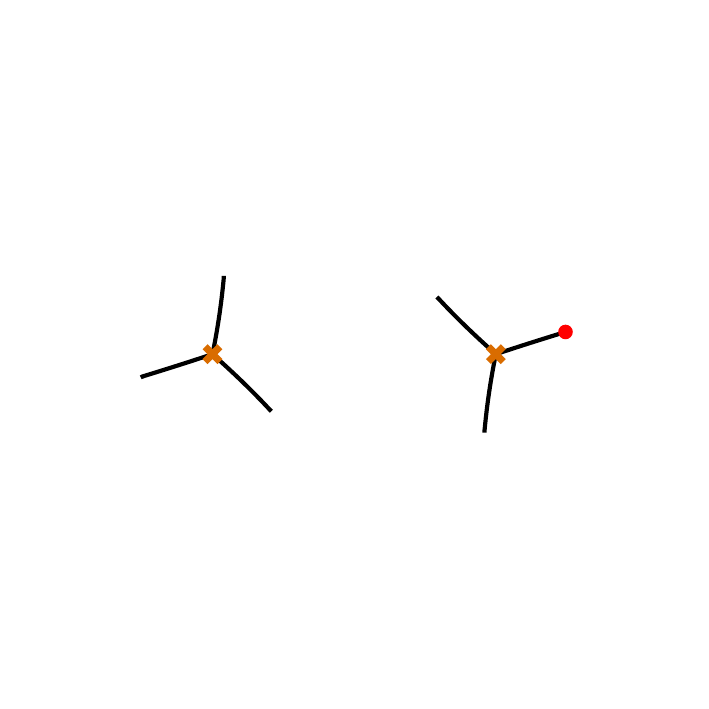}
         \vspace{-1 cm}
         \caption{For fixed $\zeta=-1.334-0.700\ri$ we plot a cutoff version 
     of the Stokes graph at $\arg(\epsilon) = \arg(\zeta)$, 
     plotting the Stokes curves only up
     to $|2\int_{b_0}^{z} \lambda| = \abs{\zeta}$.
    As explained around \eqref{polesSol}, a cutoff Stokes curve of $\cW$ runs exactly up to the point $z=3\, \e^{\frac{\ri\pi}{30}}$, which is plotted as a red dot.}
         \label{BPSraywk}
     \end{subfigure}
          \caption{The Borel plane for $\Upsilon_{\rm red}^{(1)}(z,\epsilon)$ and the corresponding soliton at $z=3\, \e^{\frac{\ri\pi}{30}}$.}
        \label{3borel}
\end{figure}

\begin{figure}
     \centering
     \begin{subfigure}[c]{0.47\textwidth}
         \centering
         \includegraphics[width=0.7\textwidth]{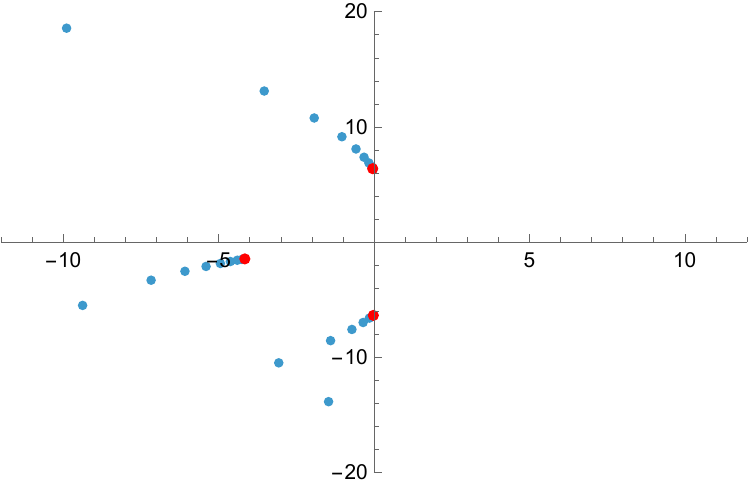}
         \caption{Borel plane of $\Upsilon_{\rm red}^{(1)}(z,\epsilon)$ at $z=4\, \e^{\frac{\ri\pi}{30}}$ with 56 terms. Only the first pole in each series of poles corresponds to a true singularity of $\hat{\Upsilon}_{\rm red}^{(1)}(z,\zeta)$ that has physical meaning and we show it in red. The numerical values of singularities are at $\zeta=-4.155-1.457\ri, -0.041-6.373\ri, -0.02+6.362\ri$.}
         \label{BPSrayst}
     \end{subfigure}
     \hfill
          \begin{subfigure}[c]{0.47\textwidth}
         \centering
         \vspace{-1cm}
         \includegraphics[width=0.7\textwidth]{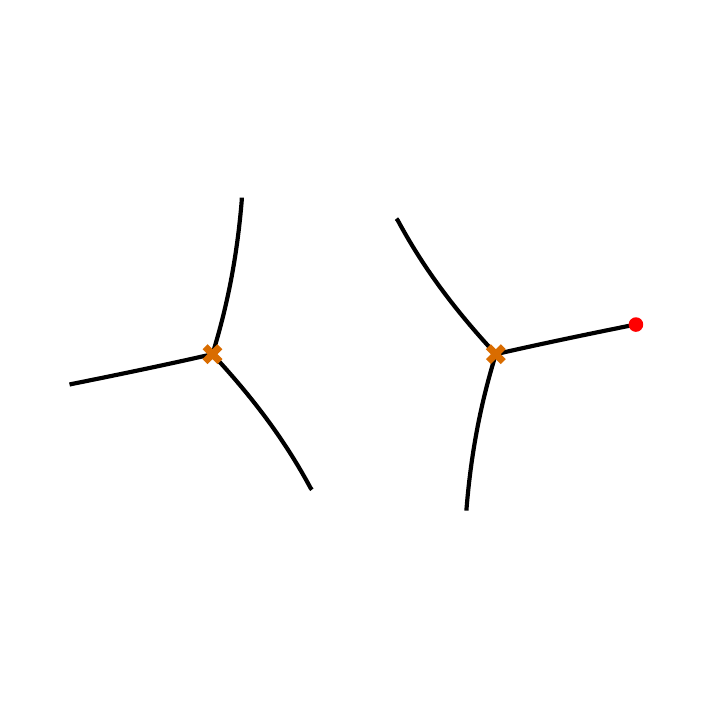}
         \vspace{-0.8cm}
         \caption{For fixed $\zeta=-4.155-1.457\ri$ we plot a cutoff version 
     of the Stokes graph at $\arg(\epsilon) = \arg(\zeta)$, 
     plotting the Stokes curves only up
     to $|2\int_{b_0}^{z} \lambda| = \abs{\zeta}$.
    As explained around \eqref{polesSol}, a cutoff Stokes curve runs exactly up to the point $z=4\, \e^{\frac{\ri\pi}{30}}$, which is plotted as a red dot.}
         \label{BPSraywk}
     \end{subfigure}
            \caption{The Borel plane for $\Upsilon_{\rm red}^{(1)}(z,\epsilon)$ and the corresponding soliton  at $z=4\, \e^{\frac{\ri\pi}{30}}$.}
        \label{4borelweber}
\end{figure}

\begin{figure}
     \centering
     \begin{subfigure}[c]{0.32\textwidth}
         \centering
         \vspace{0.5cm}
         \includegraphics[width=0.7\textwidth]{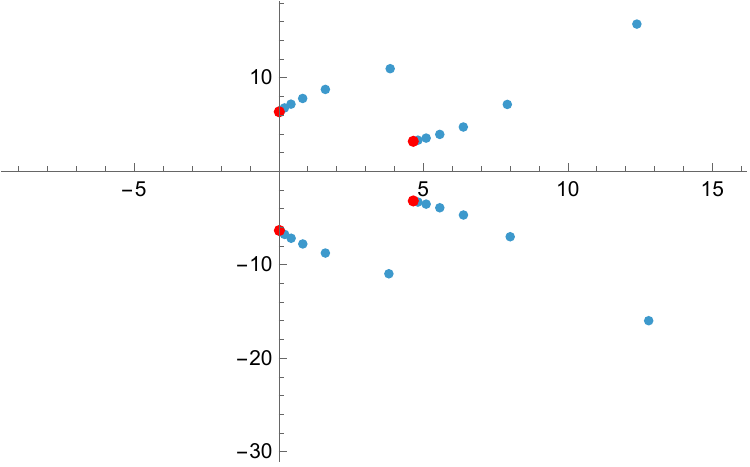}
         \caption{Borel plane of $\Upsilon_{\rm red}^{(1)}(z,\epsilon)$ at $z=2\ri$ with 60 terms. Only the first pole in each series of poles corresponds to the actual singularity of $\hat{\Upsilon}_{\rm red}^{(1)}(z,\zeta)$ that has physical meaning and we show it in red. The numerical values of singularities are at $\zeta=4.655+3.189\ri, 4.655-3.19\ri, 0.019-6.347\ri, 0.023+6.35\ri$.}
         \label{BPSrayst}
     \end{subfigure}
     \hfill
          \begin{subfigure}[c]{0.32\textwidth}
         \centering
         \vspace{-0.1cm}
         \includegraphics[width=0.7\textwidth]{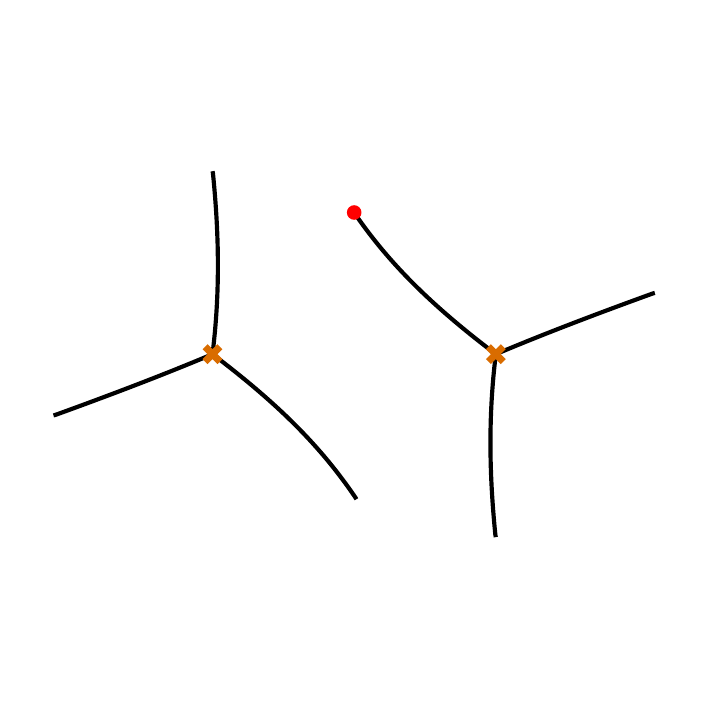}
         \vspace{-0.7cm}
         \caption{For fixed $\zeta=4.655+3.189\ri$ we plot a cutoff version 
     of the Stokes graph with $\arg(\epsilon) = \arg(\zeta)$, 
     plotting the Stokes curves only up
     to $|2\int_{b_0}^{z} \lambda| = \abs{\zeta}$.
    As explained around \eqref{polesSol}, a cutoff Stokes curve runs exactly up to the point $z=2\ri$, which is plotted as a red dot.}
         \label{BPSraywk}
     \end{subfigure}
         \hfill
          \begin{subfigure}[c]{0.32\textwidth}
         \centering
         \vspace{-0.1cm}
         \includegraphics[width=0.7\textwidth]{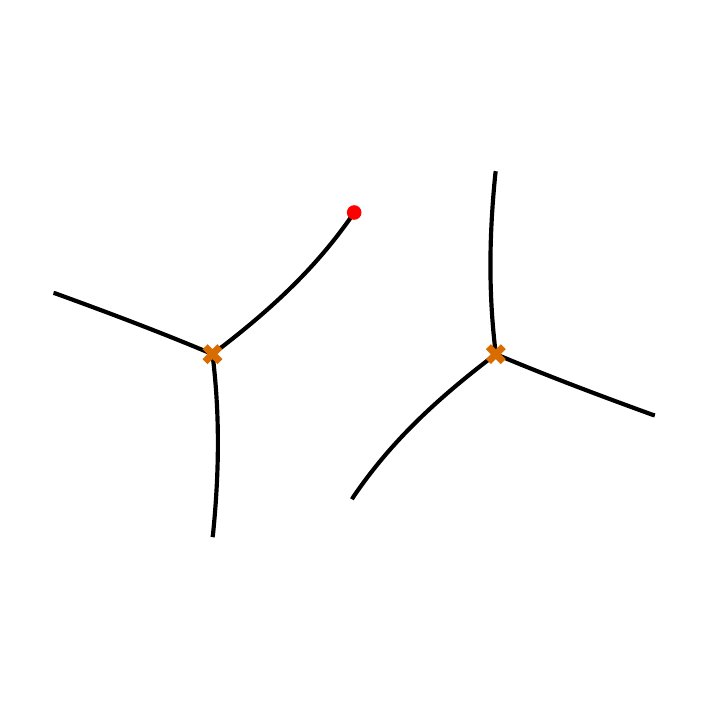}
         \vspace{-0.7cm}
         \caption{For fixed $\zeta=4.655-3.19\ri$ we plot a cutoff version 
     of the Stokes graph with $\arg(\epsilon) = \arg(\zeta)$, 
     plotting the Stokes curves only up
     to $|2\int_{b_0}^{z} \lambda| = \abs{\zeta}$.
    As explained around \eqref{polesSol}, a cutoff Stokes curve runs exactly up to the point $z=2\ri$, which is plotted as a red dot.}
         \label{BPSraywk}
     \end{subfigure}
            \caption{The Borel plane of $\Upsilon_{\rm red}^{(1)}(z,\epsilon)$  and solitons at $z=2\ri$.}
        \label{2iborel}
\end{figure}

We observe that on the Borel plane, there are indeed two types of poles for $\hat{\Upsilon}_{\rm red}^{(1)}(z,\zeta)$ as described in \autoref{padeborelsol} and \autoref{sec:borel4d}:
\begin{itemize}
\item Poles at the minus central charges of the 4d BPS states $-Z_{\pm \gamma_f}$. 
Technically, the Pad\'{e} approximation is more sensitive to poles with smaller central charges\footnote{We thank Maximilian Schwick for explaining this point to the authors.}. For example, the pole at $- Z_{\pm \gamma_f}$ is more obvious in  \autoref{4borelweber} than in \autoref{3borel} because in \autoref{4borelweber} $|Z_{\gamma_f}|$ and $|Z_{\gamma_{12}}|$ are more close. 
\item Poles at the minus central charges  of BPS solitons supported at $z$, $-Z_{\gamma_{21}}$.
Note that the number of BPS solitons depends on the position of $z$ on the Riemann surface $C$. There are two regions \circled{1} and \circled{2} (and \circled{1'} and \circled{2'} which are related to \circled{1} and \circled{2} by symmetry) as shown in the picture with boundaries given by the walls in the degenerate spectral network $\cW$ at $\arg(\epsilon)=\pm\frac{\pi}{2}$. Those walls are known as the walls of marginal stability\cite{Gaiotto:2011tf,Longhi:2012mj}.
\begin{center}
\begin{tikzpicture}[scale=0.5]
\draw plot [smooth] coordinates {(0.5, 0)  (0.93592, -0.615053)  (1.18603, -0.913777) (1.38783, -1.14376)  (1.5622, -1.33749)  (1.71812, -1.50792) (1.86051, -1.66178)  (1.99241, -1.80308)  (2.11587, -1.93445) };
\draw plot [smooth] coordinates {(0.5, 0)  (0.93592, 0.615053)  (1.18603, 0.913777)  (1.38783, 1.14376)  (1.5622, 1.33749)  (1.71812, 1.50792)  (1.86051, 1.66178)  (1.99241, 1.80308)  (2.11587, 1.93445)};
\draw plot [smooth] coordinates {(-0.5, 2.16506*10^-9) (-0.93592, 0.615053) (-1.18603, 0.913777) (-1.38783, 1.14376) (-1.5622, 1.33749) (-1.71812, 1.50792) (-1.86051, 1.66178) (-1.99241, 1.80308) (-2.11587, 1.93445) };
\draw plot [smooth] coordinates {(-0.5, -2.16506*10^-9) (-0.93592, -0.615053) (-1.18603, -0.913777) (-1.38783, -1.14376) (-1.5622, -1.33749) (-1.71812, -1.50792) (-1.86051, -1.66178) (-1.99241, -1.80308) (-2.11587, -1.93445) (-2.23234, -2.05771)};
\draw (-0.5, 0) to (0.5, 0);
\draw [line width=0.2cm, white] (0,0) circle (3cm);
\draw[dashed,gray] (0,0) circle (2.82cm);
\drawbranchpointmarker{-0.5, 0};
\drawbranchpointmarker{0.5, 0};
\node at (-0.8,1.5) [right] {$\circled{2}$};
\node at (-0.8,-1.5) [right] {$\circled{2'}$};
\node at (1.25,0) [right] {$\circled{1}$};
\node at (-2.75,0) [right] {$\circled{1'}$};

\end{tikzpicture}
\end{center}

When $z\in\circled{1}$, there is only one BPS soliton supported at $z$ with the charge $\gamma_{21_+}$ surrounding $b_0=2$, some examples are shown in \autoref{3borel}, \autoref{4borelweber}.
In region \circled{2}, there are two BPS solitons with charges $\gamma_{21_+}$ surrounding $b_0=2$ and $\gamma_{21_-}$ surrounding $b_0=-2$ respectively, an example can be found in  \autoref{2iborel}.

\end{itemize}

According to \eqref{sec2:jumppi} the ${\rm disc}(\frac{\Upsilon_{\rm red}^{(1)}}{\epsilon})(z,\epsilon)$ at $\arg(\epsilon)=\arg( Z_{\pm\gamma_f})=\pm \frac{\pi}{2}$ is 
\begin{equation}\label{jumppi} 
{\rm disc}(\frac{\Upsilon_{\rm red}^{(1)}}{\epsilon})(z,\epsilon)=-\frac{1}{2}\log(1+\cX_{\gamma_f}(\epsilon))=-\frac{1}{2}\log(1+\E^{-2\pi\ri/\epsilon}), \text{ for }\epsilon\in \E^{\ri\,\arg( -Z_{\gamma_f})}\mathbb{R}_+,
\end{equation}
\begin{equation}\label{jumpni} 
{\rm disc}(\frac{\Upsilon_{\rm red}^{(1)}}{\epsilon})(z,\epsilon)=\frac{1}{2}\log(1+\cX_{-\gamma_f}(\epsilon))=\frac{1}{2}\log(1+\E^{2\pi\ri/\epsilon}), \text{ for }\epsilon\in \E^{\ri\,\arg(-Z_{-\gamma_f})}\mathbb{R}_+.
\end{equation}
We show numerical checks for \eqref{jumppi} and  \eqref{jumpni} in \autoref{discflavor}\footnote{For simplicity, we neglect the imaginary part which is smaller than the stable digits. We will keep this convention in this paper.} at $z=8\e^{\frac{\ri\pi}{30}}\in\circled{1}$ and $z=6\ri\in\circled{2}$ respectively.

\begin{table}[h]
    \centering
    \begin{tabular}{| c | c c  c| }
    \hline
     $\epsilon$ & ${\rm disc}(\frac{\Upsilon_{\rm red}^{(1)}}{\epsilon})(8\e^{\frac{\ri\pi}{30}},\epsilon)$ &  ${\rm disc}(\frac{\Upsilon_{\rm red}^{(1)}}{\epsilon})(6\ri,\epsilon)$   & $-\frac{1}{2}\log(1+\cX_{\gamma_f}(\epsilon))$\\

    \hline 

 $\frac{1}{\sqrt{7}}\ri$     & $-\underline{3.0155666}88\times 10^{-8}$ & $-\underline{3.0155666}87\times 10^{-8}$ & $-3.015566689\times 10^{-8}$ \\

 $\frac{\pi}{3}\ri$       & $-\underline{0.0012378}44$ & $\underline{-0.0012378}49$ & $-0.001237843$ \\

 $\sqrt{2}\ri$     & $-\underline{0.005846}65$  & $-\underline{0.005846}382$ & $-0.005846673$  \\
 \hline
  $\epsilon$ & $\disc(\frac{\Upsilon_{\rm red}^{(1)}}{\epsilon})(8\e^{\frac{\ri\pi}{30}},\epsilon)$ &  ${\rm disc}(\frac{\Upsilon_{\rm red}^{(1)}}{\epsilon})(6\ri,\epsilon)$   & $\frac{1}{2}\log(1+\cX_{-\gamma_f}(\epsilon))$\\

    \hline 
 $-\frac{1}{\sqrt{7}}\ri$     & $\underline{3.0155666}87\times 10^{-8}$ & $\underline{3.0155666}87\times10^{-8}$ & $3.015566689\times 10^{-8}$ \\
 $-\frac{\pi}{3}\ri$       & $\underline{0.0012378}51$ &\underline{0.0012378}49 & $0.001237843$ \\
 $-\sqrt{2}\ri$       & $\underline{0.005846}252$ & $\underline{0.005846}382$ & $0.005846673$ \\
          \hline
    \end{tabular}
    \caption{To check the discontinuities corresponding to $-Z_{\pm\gamma_f}$, it is convenient to choose $|Z_{\gamma_{21}}
   |\gg |Z_{\gamma_f}|$ so that most poles in the Pad\'{e} approximation concentrate around the poles at  $-Z_{\pm\gamma_f}$ on the Borel plane. Thus we show the result at $z=8\e^{\frac{\ri\pi}{30}}$ and $z=6\ri$ as examples. In the numerical calculation, we use 60 terms. The underlines show stable digits of the Pad\'{e}-Borel summation.}
    \label{discflavor}
\end{table}

We now come to the discontinuities at minus BPS soliton central charges. Recall that  $-\gamma_{21_+}$ is the charge surrounding $b_0=2$ and $-\gamma_{21_-}$ is the charge surrounding $b_0=-2$. The discontinuities of $\Upsilon_{\rm red}(z,\epsilon)$ defined in \eqref{defUpsilonred} at
\begin{align}
\arg(\epsilon)=&
\arg\left( -Z_{\gamma_{21_+}}\right)\\
\nonumber=&\arg\left(\frac{1}{2} \left(-\sqrt{z^2-4} z+\log \left(z^2-4\right)+2 \log \left(\frac{z}{\sqrt{z^2-4}}+1\right)-2 \log \left(z-\sqrt{z^2-4}\right)\right)\right),\\
\arg(\epsilon)=&
\arg\left( -Z_{\gamma_{21_-}}\right)\\
\nonumber=&\arg\left(\frac{1}{2} \left(-\sqrt{z^2-4} z+\log \left(z^2-4\right)+2 \log \left(\frac{z}{\sqrt{z^2-4}}+1\right)-2 \log \left(z-\sqrt{z^2-4}\right)\right)-2\pi\ri\right),
\end{align}
are proposed by \eqref{sec2:jumpsol2} as 
\begin{equation}\label{weberjumpsol}
 {\rm disc}(\Psi^{(1)}_{\rm red})(z,\epsilon)=\ri\Psi^{(1)}_{\rm red}(z,-\epsilon)\e^{\frac{Z_{\gamma_{21_\pm}}}{\epsilon }}.
 \end{equation}
We check numerical the matching of the RHS and LHS of \eqref{weberjumpsol} and the comparison is shown in \autoref{discsol1}, where we choose $z=\frac5 2+\frac\ri {100}\in\circled{1}$ and \autoref{discsol2} where we choose $z=\frac{\ri}{5}\in\circled{2}$.

\begin{table}[h]
    \centering
    \begin{tabular}{| c | c  c| }
    \hline
     $\epsilon$ & ${\rm disc}(\Psi^{(1)}_{\rm red})(z,\epsilon)$  & $\ri\Psi^{(1)}_{\rm red}(z,-\epsilon)\e^{\frac{Z_{\gamma_{21_+}}}{\epsilon }}$\\
    \hline 
     $\frac{1}{3\pi}\E^{\arg(-Z_{\gamma_{21_+}})\ri}$     & $-\underline{0.01124491286}561984 $  & $-\underline{0.01124491286537348}$  \\
 & $+\underline{0.00006197024}802560  \ri$ &  $ +\underline{0.00006197024802813}  \ri$ \\

 $\frac{1}{6\sqrt{3}}\E^{\arg(-Z_{\gamma_{21_+}})\ri}$     & $\underline{-0.00702090325}918099 $  & $\underline{-0.00702090325920407}$  \\
 & $+\underline{0.00003872141}506306  \ri$ &  $ +\underline{0.00003872141506112}  \ri$ \\
     \hline
    \end{tabular}
    \caption{To detect the discontinuity at $-Z_{\gamma_{21_+}}$, it is convenient to choose $|Z_{\gamma_{21_+}}
   |\gg |Z_{\gamma_f}|$ so that most poles in the Pad\'{e} approximation concentrate around the pole at  $-Z_{\gamma_{21_+}}$ on the Borel plane. Thus we show the result at $z=\frac5 2+\frac\ri {100}$ as an example where most of poles form a series of poles at $-Z_{\gamma_{21_+}}(z)$. In the numerical calculation, we use 60 terms.}
    \label{discsol1}
\end{table}

\begin{table}[h]
    \centering
    \begin{tabular}{| c | c  c| }
    \hline
     $\epsilon$ & ${\rm disc}(\Psi^{(1)}_{\rm red})(z,\epsilon)$  & $\ri\Psi^{(1)}_{\rm red}(z,-\epsilon)\e^{\frac{Z_{\gamma_{21_\pm}}}{\epsilon }}$\\
    \hline 
 $\frac{1}{6\sqrt{3}}\E^{\arg(-Z_{\gamma_{21_+}})\ri}$     & $-\underline{3.58134936}546739\times 10^{-15}$  & $-\underline{3.581349365494}69\times10^{-15} $ \\
 & $+\underline{3.59033913}497574\times 10^{-15}\ri$ &  $ -\underline{3.590339134881}24\times10^{-15} \ri$ \\
 \hline
 $\frac{1}{6\sqrt{3}}\E^{\arg(-Z_{\gamma_{21_-}})\ri}$     & $-\underline{3.59033913}497574\times10^{-15}$ & $-\underline{3.5903391348812}4\times10^{-15}$ \\
 & $ +\underline{3.58134936}546739\times10^{-15}\ri$ &  $ +\underline{3.5813493654946}9 \times10^{-15}\ri$ \\
    \hline
    \end{tabular}
    \caption{To detect the discontinuity at $-Z_{\gamma_{21_\pm}}$, it is convenient to choose $|Z_{\gamma_{21_\pm}}
   |\gg |Z_{\gamma_f}|$ so that most poles in the Pad\'{e} approximation concentrate around the pole at  $-Z_{\gamma_{21_\pm}}$ on the Borel plane. Thus we show the result at $z=\frac{\ri}{5}$ as an example  where most of poles form a series of poles at $-Z_{\gamma_{21_\pm}}(z)$. In the numerical calculation, we use 60 terms.}
    \label{discsol2}
\end{table}

We check numerically that $\Psi^{(1)}(z,\epsilon)$ is indeed a solution. We discuss its relation to the  more familiar solution given by the parabolic cylinder D function in \autoref{appen:matchwkbD}.

\subsection{Open TBA equations}

Using the discontinuities of the solution $\Psi^{(1)}_{\rm red}$ -- \eqref{jumppi},\eqref{jumpni} and \eqref{weberjumpsol}, together with the leading asymptotic $\Psi^{(1)}_{\rm red}\sim\frac{1}{(-\sqrt{P(z,0)})^{\frac12}}$, $\Psi^{(1)}_{\rm red}$ at $z\in\circled{1}$ satisfies an open TBA equation  
\begin{align}\label{Weber1soltba}
\nonumber\Psi^{(1)}_{\rm red}(z,\epsilon)=&\E^{-\frac{1}{2\pi\ri}\mathcal{I}_{\gamma_f}(\epsilon)}\Bigg(\frac{1}{\left(-\sqrt{P(z,0)}\right)^{\frac{1}{2}}}\\
&+\frac{1}{2\pi}\int\limits_{\ell_{\gamma_{21_+}}}\frac{\epsilon}{\epsilon'}\frac{1}{\epsilon'-\epsilon}\Psi^{(1)}_{\rm red}(z,-\epsilon')\mathbb{X}_{\gamma_{21_+}}(z,\epsilon')\rd\epsilon'\Bigg),
 \end{align}
as proposed in \eqref{natTBA}. Here we define
\begin{equation}\label{mathfrakX}
\mathbb{X}_{\gamma_{21_\pm}}(z,\epsilon)=\e^{\frac{Z_{\gamma_{21_\pm}}}{\epsilon }+\frac1 {2\pi\ri}\mathcal{I}_{\gamma_f}(\epsilon)}
\end{equation}
and $\mathcal{I}_{\gamma}(\epsilon)$ is defined in \eqref{Idefm}.
Using the relation \eqref{relateTBAborel} which in this example is
\begin{equation}\label{GMNwkbnorm}
g_1(z,\epsilon)=N_{\rm GMN}(\epsilon)\Psi^{(1)}_{\rm red}(z,\epsilon),
\end{equation}
where 
\begin{equation}\label{nweber}
N_{\rm GMN}(\epsilon)=\E^{\frac{1}{2\pi\ri}\mathcal{I}_{\gamma_f}(\epsilon)},
\end{equation}
$g_1(z,\epsilon)$ satisfies the GMN open TBA equation 
\begin{align}\label{weberGMN}
g_1(z,\epsilon)=&\frac{1}{\left(-\sqrt{P(z,0)}\right)^{\frac{1}{2}}}+\frac{1}{2\pi}\int\limits_{\ell_{\gamma_{21_+}}}\frac{\epsilon}{\epsilon'}\frac{1}{\epsilon'-\epsilon} g_1(z,-\epsilon')X_{\gamma_{21_+}}(z,\epsilon')\rd\epsilon',
 \end{align}
 where 
\begin{equation}
X_{\gamma_{21_\pm}}(z,\epsilon)=\e^{\frac{Z_{\gamma_{21_\pm}}}{\epsilon }+\frac1 {\pi\ri}\mathcal{I}_{\gamma_f}(\epsilon)}.
\end{equation}

Similarly, when $z\in\circled{2}$ we have
\begin{align}\label{Weber2soltba}
\nonumber\Psi^{(1)}_{\rm red}(z,\epsilon)=&\E^{-\frac{1}{2\pi\ri}\mathcal{I}_{\gamma_f}(\epsilon)}\Bigg(\frac{1}{\left(-\sqrt{P(z,0)}\right)^{\frac{1}{2}}}\\
&+\frac{1}{2\pi}\int\limits_{\ell_{\gamma_{21_+}}}\frac{\epsilon}{\epsilon'}\frac{1}{\epsilon'-\epsilon}\Psi^{(1)}_{\rm red}(z,-\epsilon')\mathbb{X}_{\gamma_{21_+}}(z,\epsilon')\rd\epsilon'\\
&+\frac{1}{2\pi}\int\limits_{\ell_{\gamma_{21_-}}}\frac{\epsilon}{\epsilon'}\frac{1}{\epsilon'-\epsilon}\Psi^{(1)}_{\rm red}(z,-\epsilon')\mathbb{X}_{\gamma_{21_-}}(z,\epsilon')\rd\epsilon'\Bigg),
 \end{align}
 Or using \eqref{GMNwkbnorm}, the corresponding GMN open TBA reads
 \begin{align}\label{WeberGMN2}
\nonumber g_1(z,\epsilon)=&\frac{1}{\left(-\sqrt{P(z,0)}\right)^{\frac{1}{2}}}+\frac{1}{2\pi}\int\limits_{\ell_{\gamma_{21_+}}}\frac{\epsilon}{\epsilon'}\frac{1}{\epsilon'-\epsilon}g_1(z,-\epsilon')X_{\gamma_{21_+}}(z,\epsilon')\rd\epsilon'\\
&+\frac{1}{2\pi}\int\limits_{\ell_{\gamma_{21_-}}}\frac{\epsilon}{\epsilon'}\frac{1}{\epsilon'-\epsilon} g_1(z,-\epsilon')X_{\gamma_{21_-}}(z,\epsilon')\rd\epsilon'\Bigg).
 \end{align}

We compare the results of open TBA with the ones of Pad\'{e}-Borel summation at $z=\e^{\frac{\pi\ri}{30}}\in\circled{1}$ in \autoref{openTBA3} using \eqref{Weber1soltba} and at $z=2\ri \in\circled{2}$ in \autoref{openTBA2i} using \eqref{WeberGMN2}. We see that both TBA equations of $\Psi^{(1)}_{\rm red}(z,\epsilon)$ and $g_1(z,\epsilon)$ work and in practice, it’s easier to work with the GMN open TBA. So in future calculations, we will only use the GMN open TBA.
\begin{table}[h]
\begin{tabular}{ |p{2.8cm}||p{3.8cm}|p{3.8cm}|p{3.8cm}| }
 \hline 
 \hline
 \multicolumn{4}{|c|}{$z=3\e^{\frac{\pi\ri}{30}}\in\circled{1}$} \\
 \hline
 \hline
$\epsilon$ & $\frac1{2\sqrt{2}}$ & $\frac3{2\sqrt{2}}$ & $\frac5{2\sqrt{2}}$\\
\hline
   $\Psi^{(1), {\rm TBA}}_{\rm red}(z,\epsilon)$, s=$\frac1{3\sqrt{2}}$, c=200 & 0.0747650099 $+0.9165798994$ i &0.0617787405 $+0.8926859184$ i  & 0.0542550859 $+0.8760679146$ i  \\
\hline
   $\Psi^{(1), {\rm TBA}}_{\rm red}(z,\epsilon)$, s=$\frac1{3\sqrt{2}}$, c=250 & 0.0747547267 $+0.9165655498$ i &0.0617449795 $+0.8926402374$ i  & 0.0541953665 $+0.8759881749$ i  \\
\hline
   $\Psi^{(1), {\rm TBA}}_{\rm red}(z,\epsilon)$, s=$\frac1{3\sqrt{2}}$, c=300 & 0.0747483924 $+0.9165567612$ i &0.0617241191 $+0.8926122227$ i  & 0.0541583729 $+0.8759392145$ i  \\
  \hline
  \hline
   $\Psi^{(1), {\rm PB}}_{\rm red}(z,\epsilon)$ & 0.0747385428 $+0.916546668$i & 0.0617030584 $+0.8925871368$ i  & 0.05413127 $+0.87591105$ i \\
\hline
\hline
\end{tabular}
\caption{Numerical results for  $\Psi^{(1)}_{\rm red}(z,\epsilon)$. PB stands for Pad\'{e}-Borel summation method and TBA stands for the numerical result of \eqref{Weber1soltba} obtained by iteration method. We only show the stable digits of $\Psi^{(1), {\rm PB}}_{\rm red}(z,\epsilon)$. $s$ and $c$ represent the step size and cutoff of the integral in the numerical discretization. We see that the TBA results converge to the Pad\'{e}-Borel result.}
\label{openTBA3}
\end{table}

\begin{table}[h]
\begin{tabular}{ |p{4cm}||p{3.7cm}|p{3.7cm}|p{3.7cm}| }
 \hline 
 \hline
 \multicolumn{4}{|c|}{$z=2\ri\in\circled{2}$} \\
 \hline
 \hline
$\epsilon$ & $\frac1{2\sqrt{2}}$ & $\frac3{2\sqrt{2}}$ & $\frac5{2\sqrt{2}}$\\
\hline
$g_1(z,\epsilon)/N_{\rm GMN}(\epsilon)$, s=$\frac1{\sqrt{2}}$, c=200 & 0.6055840041 $+0.6055840041$ i & 0.6259036077 $+0.6259036077$ i  & 0.6391055749 $+0.6391055749 $ i \\
\hline
$g_1(z,\epsilon)/N_{\rm GMN}(\epsilon)$, s=$\frac1{3\sqrt{2}}$, c=200 & 0.6055785433 $+0.6055785433$ i & 0.6259749263 $+0.6259749263$ i  &0.639057921 $+0.639057921 $ i \\
\hline
$g_1(z,\epsilon)/N_{\rm GMN}(\epsilon)$, s=$\frac1{3\sqrt{2}}$, c=250 & 0.6055882339 $+0.6055882339$ i & 0.6260035042 $+0.6260035042$ i  & 0.6391028374 $+0.6391028374 $ i \\
\hline
$g_1(z,\epsilon)/N_{\rm GMN}(\epsilon)$, s=$\frac1{3\sqrt{2}}$, c=300 & 0.6055941351 $+0.6055941351$ i & 0.6260208870 $+0.6260208870$ i  & 0.6391315362 $+0.6391315362 $ i \\
  \hline
  \hline
   $\Psi^{(1), {\rm PB}}_{\rm red}(z,\epsilon)$ & 0.6056134206 $+0.6056134206$ i &0.626 $+0.626$ i  & 0.64 $+0.64 $ i \\
\hline
\hline
\end{tabular}
\caption{Numerical iteration computation for $g_1(z,\epsilon)$ and $\Psi^{(1)}_{\rm red}(z,\epsilon)$. PB stands for Pad\'{e}-Borel summation method, $g_1(z,\epsilon)$ stands for the result using \eqref{WeberGMN2} and $N_{\rm GMN}(\epsilon)$ is defined in \eqref{nweber}. We only show the stable digits of $\Psi^{(1), {\rm PB}}_{\rm red}(z,\epsilon)$. $s$ and $c$ represent the step size and cutoff of the integral. We see that the TBA results converge to the Pad\'{e}-Borel result.}
\label{openTBA2i}
\end{table}

\section{Pure $SU(2)$ at $u=0$}\label{sec:su2}
A richer example to study is the pure $SU(2)$ theory which corresponds to
\begin{equation}\label{su2inz}
(- \epsilon^2 \partial_z^2 +  \frac{\Lambda}{z} + \frac{u-\epsilon^2/4}{z^2} + \frac{\Lambda}{z^3})\psi(z)=0.
\end{equation}
This equation has an alternative expression obtained using the change of variable $z=\E^x$ and $\psi(z)=z^{1/2}\phi\big(x(z)\big)$, 
\begin{equation}\label{su2inx}
-\epsilon^2\phi''(x)+(u+2\Lambda\cosh(x))\phi(x)=0,
\end{equation}
which is known as the modified Mathieu equation. 
We will concern ourselves with the former representation.

Without loss of generality, we set $\Lambda=1$ following the convention of \cite{Gaiotto:2014bza}. We focus on the strong coupling region in this paper and choose the most symmetrical case when $u=0$ for simplicity. The case for arbitrary $u$ in the strong coupling region can be studied similarly. 

When $u=0$ the two branch points are at $b_0=\pm \ri$. We fix the convention of sheet label by
\begin{equation}
Y^{0,(1)}(z)=-\sqrt{\frac{1}{z}+\frac{1}{z^3}}.
\end{equation}

\begin{figure}
      \begin{subfigure}[c]{0.45\textwidth}
         \centering
         \includegraphics[width=0.6\textwidth]{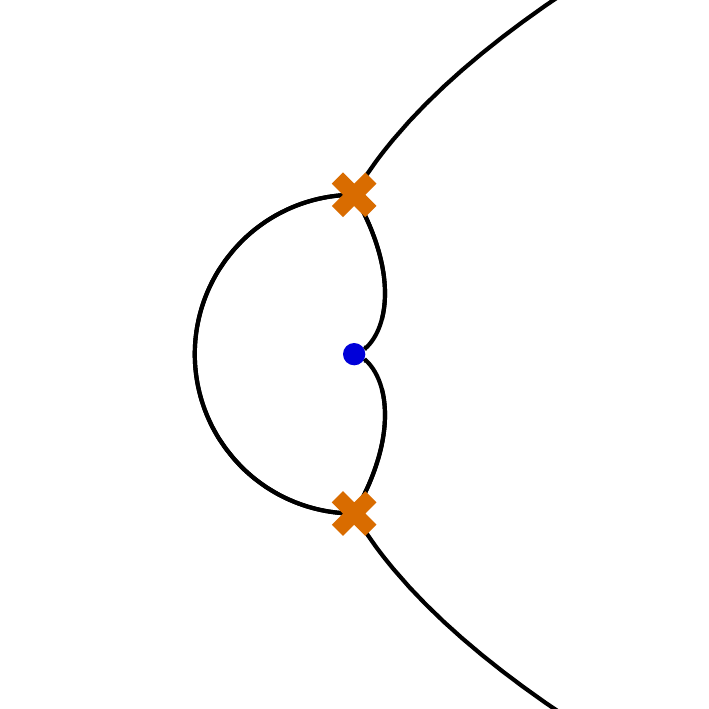}
         \caption{The degenerate spectral network at $\arg(\epsilon)=\arg(\pm Z_{\gamma_d})=0\text{ or }\pi$.}
         \label{su2deg1}
      \end{subfigure}
          \hspace{1cm}
            \begin{subfigure}[c]{0.45\textwidth}
         \centering
         \includegraphics[width=0.6\textwidth]{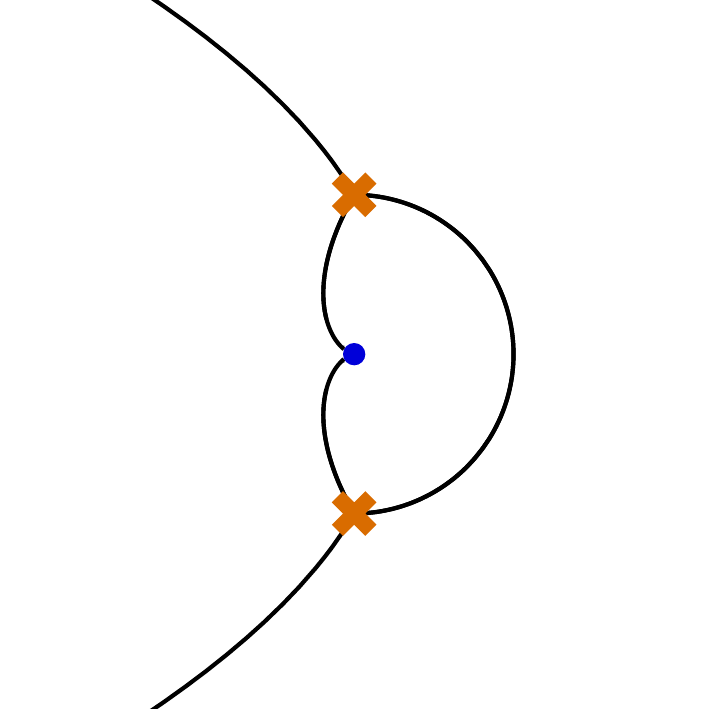}
         \caption{The degenerate spectral network at $\arg(\epsilon)=\arg(\pm Z_{\gamma_m})=\mp\frac{\pi}{2}$.}
         \label{su2deg2}
      \end{subfigure}
             \caption{The degenerate spectral networks of the pure $SU(2)$ theory at $u=0$.}
        \label{degsu2}
\end{figure}

The pure SU(2) theory has 4 BPS states in the strong coupling region, the monopole with charge $\gamma_m$ and the dyon with charge $\gamma_{d}$ and their antiparticles, detected by the degenerate spectral networks in \autoref{degsu2}.
\begin{center}
\begin{tikzpicture}[scale=0.5]
\draw (0,0) circle (3cm);
\draw (0,0) circle (2cm);
\draw (0,2.5) circle (0.5cm);
\draw (0,-2.5) circle (0.5cm);
\drawbranchpointmarker{0,2.5};
\drawbranchpointmarker{0,-2.5};
\drawbranchpointmarker{0,0};
\draw[snake it, orange] (0,2.5) to (0,0);
\draw[snake it, orange] (0,-2.5) to (0,-4);
\filldraw [blue] (0,0) circle (1.5pt);
\node at (3,1) [right] {$\gamma_m$};
\node at (3,0) [right] {$1$};
\draw[->] (3,-0.2) -- (3,0.2);
\node at (-4.5,1) [right] {$\gamma_d$};
\node at (-4,0) [right] {$1$};
\draw[<-] (-3,-0.2) -- (-3,0.2);
\end{tikzpicture}
\end{center}
Their central charges are
\begin{align}
\label{Zgammad}Z_{\gamma_{d}}=&8 \sqrt{2 \Lambda - u}\, \textbf{E}\left(\frac{4 \Lambda }{2 \Lambda - u}\right)+8 \ri \sqrt{2 \Lambda - u} \left(\textbf{K}\left(\frac{- u-2 \Lambda }{2
   \Lambda - u}\right)-\textbf{E}\left(\frac{- u-2 \Lambda}{2 \Lambda - u}\right)\right),\\
\label{Zgammam}Z_{\gamma_{m}}=&8 \ri \sqrt{2 \Lambda - u} \left(\textbf{K}\left(\frac{- u-2 \Lambda }{2 \Lambda - u}\right)-\textbf{E}\left(\frac{- u-2 \Lambda }{2 \Lambda -
   u}\right)\right),
\end{align}
where \textbf{E} and \textbf{K} are the complete elliptic integrals and the corresponding BPS indices are
\begin{equation}
\Omega(\pm \gamma_m)=\Omega(\pm \gamma_d)=1.
\end{equation}
The quantum periods $\cX_{\pm\gamma_m}$ and $\cX_{\pm\gamma_d}$ have $\mathbb{Z}_4$ symmetry at the $u=0$ point  and they are related by
\begin{align}\label{z4sym}
\cX_{\gamma}(\epsilon)=&\cX_{-\gamma}(-\epsilon),\\
\cX_{\gamma_d}(\epsilon)=&\cX_{\gamma_m}(-\ri\epsilon).
\end{align}
In this example, the BPS soliton degeneracies $\mu(\gamma_{21})$ are all 1.

For a more pedagogical exposition the reader may consult the previous Weber example in \autoref{sec:weber}, whereas in the present example we will attempt to be more concise.

\subsection{Pad\'{e}-Borel summation}

The wall of marginal stability divides the Riemann surface into 4 regions and their symmetric parts under $z\rightarrow -z$ and $\epsilon\rightarrow \ri\epsilon$. By further using the symmetry $z\rightarrow \frac1z$, we can relate $\circled{1}$ to $\circled{2}$ and $\circled{3}$ to $\circled{4}$. So it is enough to study $\circled{1}$ to $\circled{2}$. In this part, we will focus mostly on region $\circled{1}$ which has two BPS solitons corresponding to the walls emanated from branch points $b_0=\pm \ri$ respectively. We discuss $\circled{3}$ which has 3 BPS solitons with less numerical checks for individual discontinuities since the Pad\'{e}-Borel summation has limited resolution to deal with the three singularities in this example which are always different by $Z_{\gamma_m}$ and $Z_{-\gamma_d}$\footnote{\label{diff3rays}This is because the Pad\'{e}-Borel summation is not working very well when the singularities are too close to each other or for a singularity always have relatively larger absolute value, so it is hard to discuss them separately in this example.}. But we directly check  the numerical matching of the GMN open TBA method and the Pad\'{e}-Borel summation method and this supports the validity of the proposals in \autoref{sec:disc}.

\begin{center}
\begin{tikzpicture}[scale=0.5]
\draw plot [smooth] coordinates {(4.999999999*10^-11, -0.9999999999133974596)
(0.14218913919420198909, -0.6848694535423721891)
(0.17892816862663017427, -0.5381235755819253629)
(0.19165306212939956301, -0.4354347328217642541)
(0.19308285997670873771, -0.3582181247728311461)
(0.18858589242316844830, -0.2981445018636956629)
(0.18094132180352489125, -0.2504450877101520722)
(0.17172405421320848597, -0.2120350328738031612)
(0.16186441762438360469, -0.18077255147807083249)
(0.15191900769675342215, -0.1551034260401074439)
(0.14221739612063630987, -0.13386645275303664749)
(0.13294764121496449320, -0.11617636359554414900)
(0.12420857269463011733, -0.10134825329404550105)
(0.11604275259358488927, -0.08884621149964734952)
(0.10845764199560586004, -0.07824706229186059579)
(0.10143932823758217834, -0.06921405062548066576)
(0.09496145446352134856, -0.06147735557469268377)
(0.08899101149006435363, -0.05481943777543297573)
(0.08349206050990491976, -0.04906389181985071502)
(0.07842808600880972687, -0.04406688488756151359)
(0.07376344169455547417, -0.03971053008187809179)
(0, 0)
};
\draw plot [smooth] coordinates {(-1.0000000000*10^-10, -1.0000000000000000000)
(-0.14075351439071713687, -0.9900446698063377371)
(-0.22246204144489686305, -0.9749413521401348501)
(-0.28999491974814070962, -0.9570281849067075370)
(-0.3492500679546641518, -0.9370295568979267980)
(-0.4026736729297001306, -0.9153436039485417125)
(-0.45157891718038885135, -0.8922311816521436371)
(-0.4967732724651013837, -0.8678803584066988672)
(-0.5387975140108788741, -0.8424353033435769857)
(-0.5780358138172545157, -0.8160113957814883960)
(-0.6147731520006936323, -0.7887039837039920517)(-0.6147731520006936323, -0.7887039837039920517)
(-0.7672020986583027385, -0.6414054402739741423)
(-0.8768037980049633896, -0.4808483117139152168)
(-0.9502010100032840380, -0.3116376735911002467)
(-0.9905346356807437957, -0.1372630089036478310)
(-0.9992277878327505615, 0.0392915741336469611)
(-0.9765638944950855534, 0.2152276937480743439)
(-0.9217873869031693989, 0.3876957740105655436)
(-0.8328249101511311970, 0.5535365176268740729)
};
\draw plot [smooth] coordinates {(5.000000001*10^-11, -1.0000000000866025404)
(0.29061848405964639945, -1.3997955413600150137)
(0.5563820517287352472, -1.6733100332965960641)
(0.8467698367376610975, -1.9238565577431718089)
(1.1659507364896639268, -2.1631370478028968782)
(1.5152964701189916581, -2.3956050155358626984)
};
\draw plot [smooth] coordinates {(4.999999999*10^-11, 0.9999999999133974596)
(0.14218913919420198909, 0.6848694535423721891)
(0.17892816862663017427, 0.5381235755819253629)
(0.19165306212939956301, 0.4354347328217642541)
(0.19308285997670873771, 0.3582181247728311461)
(0.18858589242316844830, 0.2981445018636956629)
(0.18094132180352489125, 0.2504450877101520722)
(0.17172405421320848597, 0.2120350328738031612)
(0.16186441762438360469, 0.18077255147807083249)
(0.15191900769675342215, 0.1551034260401074439)
(0.14221739612063630987, 0.13386645275303664749)
(0.13294764121496449320, 0.11617636359554414900)
(0.12420857269463011733, 0.10134825329404550105)
(0.11604275259358488927, 0.08884621149964734952)
(0.10845764199560586004, 0.07824706229186059579)
(0.10143932823758217834, 0.06921405062548066576)
(0.09496145446352134856, 0.06147735557469268377)
(0.08899101149006435363, 0.05481943777543297573)
(0.08349206050990491976, 0.04906389181985071502)
(0.07842808600880972687, 0.04406688488756151359)
(0.07376344169455547417, 0.03971053008187809179)
(0,0)
};
\draw plot [smooth] coordinates {(5.000000001*10^-11, 1.0000000000866025404)
(0.29061848405964639945, 1.3997955413600150137)
(0.5563820517287352472, 1.6733100332965960641)
(0.8467698367376610975, 1.9238565577431718089)
(1.1659507364896639268, 2.1631370478028968782)
(1.5152964701189916581, 2.3956050155358626984)
};
\draw plot [smooth] coordinates {(-1.0000000000*10^-10, 1.0000000000000000000)
(-0.14075351439071713687, 0.9900446698063377371)
(-0.22246204144489686305, 0.9749413521401348501)
(-0.28999491974814070962, 0.9570281849067075370)
(-0.3492500679546641518, 0.9370295568979267980)
(-0.4026736729297001306, 0.9153436039485417125)
(-0.45157891718038885135, 0.8922311816521436371)
(-0.4967732724651013837, 0.8678803584066988672)
(-0.5387975140108788741, 0.8424353033435769857)
(-0.5780358138172545157, 0.8160113957814883960)
(-0.6147731520006936323, 0.7887039837039920517)
(-0.6147731520006936323, 0.7887039837039920517)
(-0.7672020986583027385, 0.6414054402739741423)
(-0.8768037980049633896, 0.4808483117139152168)
(-0.9502010100032840380, 0.3116376735911002467)
(-0.9905346356807437957, 0.1372630089036478310)
(-0.9992277878327505615, -0.0392915741336469611)
(-0.9765638944950855534, -0.2152276937480743439)
(-0.9217873869031693989, -0.3876957740105655436)
(-0.8328249101511311970, -0.5535365176268740729)
};
\draw plot [smooth] coordinates {(-5.000000001*10^-11, 0.9999999999133974596)
(-0.14218913919420198909, 0.6848694535423721891)
(-0.17892816862663017427, 0.5381235755819253629)
(-0.19165306212939956301, 0.4354347328217642541)
(-0.19308285997670873771, 0.3582181247728311461)
(-0.18858589242316844830, 0.2981445018636956629)
(-0.18094132180352489125, 0.2504450877101520722)
(-0.17172405421320848597, 0.2120350328738031612)
(-0.16186441762438360469, 0.18077255147807083249)
(-0.15191900769675342215, 0.1551034260401074439)
(-0.14221739612063630987, 0.13386645275303664749)
(-0.13294764121496449320, 0.11617636359554414900)
(-0.12420857269463011733, 0.10134825329404550105)
(-0.11604275259358488927, 0.08884621149964734952)
(-0.10845764199560586004, 0.07824706229186059579)
(-0.10143932823758217834, 0.06921405062548066576)
(-0.09496145446352134856, 0.06147735557469268377)
(-0.08899101149006435363, 0.05481943777543297573)
(-0.08349206050990491976, 0.04906389181985071502)
(-0.07842808600880972687, 0.04406688488756151359)
(-0.07376344169455547417, 0.03971053008187809179)
(0,0)
};
\draw plot [smooth] coordinates {(-4.999999999*10^-11, 1.0000000000866025404)
(-0.29061848405964639945, 1.3997955413600150137)
(-0.5563820517287352472, 1.6733100332965960641)
(-0.8467698367376610975, 1.9238565577431718089)
(-1.1659507364896639268, 2.1631370478028968782)
(-1.5152964701189916581, 2.3956050155358626984)};
\draw plot [smooth] coordinates {(1.0000000000*10^-10, 0.99999999999999999999)
(0.14075351439071713687, 0.9900446698063377371)
(0.22246204144489686305, 0.9749413521401348501)
(0.28999491974814070962, 0.9570281849067075370)
(0.3492500679546641518, 0.9370295568979267980)
(0.4026736729297001306, 0.9153436039485417125)
(0.45157891718038885135, 0.8922311816521436371)
(0.4967732724651013837, 0.8678803584066988672)
(0.5387975140108788741, 0.8424353033435769857)
(0.5780358138172545157, 0.8160113957814883960)
(0.6147731520006936323, 0.7887039837039920517)
(0.6147731520006936323, 0.7887039837039920517)
(0.7672020986583027385, 0.6414054402739741423)
(0.8768037980049633896, 0.4808483117139152168)
(0.9502010100032840380, 0.3116376735911002467)
(0.9905346356807437957, 0.1372630089036478310)
(0.9992277878327505615, -0.0392915741336469611)
(0.9765638944950855534, -0.2152276937480743439)
(0.9217873869031693989, -0.3876957740105655436)
(0.8328249101511311970, -0.5535365176268740729)
};
\draw plot [smooth] coordinates {(1.0000000000*10^-10, -0.99999999999999999999)
(0.14075351439071713687, -0.9900446698063377371)
(0.22246204144489686305, -0.9749413521401348501)
(0.28999491974814070962, -0.9570281849067075370)
(0.3492500679546641518, -0.9370295568979267980)
(0.4026736729297001306, -0.9153436039485417125)
(0.45157891718038885135, -0.8922311816521436371)
(0.4967732724651013837, -0.8678803584066988672)
(0.5387975140108788741, -0.8424353033435769857)
(0.5780358138172545157, -0.8160113957814883960)
(0.6147731520006936323, -0.7887039837039920517)
(0.6147731520006936323, -0.7887039837039920517)
(0.7672020986583027385, -0.6414054402739741423)
(0.8768037980049633896, -0.4808483117139152168)
(0.9502010100032840380, -0.3116376735911002467)
(0.9905346356807437957, -0.1372630089036478310)
(0.9992277878327505615, 0.0392915741336469611)
(0.9765638944950855534, 0.2152276937480743439)
(0.9217873869031693989, 0.3876957740105655436)
(0.8328249101511311970, 0.5535365176268740729)
};
\draw plot [smooth] coordinates {(-4.999999999*10^-11, -1.0000000000866025404)
(-0.29061848405964639945, -1.3997955413600150137)
(-0.5563820517287352472, -1.6733100332965960641)
(-0.8467698367376610975, -1.9238565577431718089)
(-1.1659507364896639268, -2.1631370478028968782)
(-1.5152964701189916581, -2.3956050155358626984)
};
\draw plot [smooth] coordinates {(-5.000000001*10^-11, -0.9999999999133974596)
(-0.14218913919420198909, -0.6848694535423721891)
(-0.17892816862663017427, -0.5381235755819253629)
(-0.19165306212939956301, -0.4354347328217642541)
(-0.19308285997670873771, -0.3582181247728311461)
(-0.18858589242316844830, -0.2981445018636956629)
(-0.18094132180352489125, -0.2504450877101520722)
(-0.17172405421320848597, -0.2120350328738031612)
(-0.16186441762438360469, -0.18077255147807083249)
(-0.15191900769675342215, -0.1551034260401074439)
(-0.14221739612063630987, -0.13386645275303664749)
(-0.13294764121496449320, -0.11617636359554414900)
(-0.12420857269463011733, -0.10134825329404550105)
(-0.11604275259358488927, -0.08884621149964734952)
(-0.10845764199560586004, -0.07824706229186059579)
(-0.10143932823758217834, -0.06921405062548066576)
(-0.09496145446352134856, -0.06147735557469268377)
(-0.08899101149006435363, -0.05481943777543297573)
(-0.08349206050990491976, -0.04906389181985071502)
(-0.07842808600880972687, -0.04406688488756151359)
(-0.07376344169455547417, -0.03971053008187809179)
(0,0)
};
\draw [line width=0.2cm, white] (0,0) circle (3cm);
\draw[dashed,gray] (0,0) circle (2.82cm);
\drawbranchpointmarker{0, 1};
\drawbranchpointmarker{0,-1};
\node[scale=0.2] at (-0.15,0.5) [right] {$\circled{3}$};
\node[scale=0.2] at (-0.17,-0.5) [right] {$\circled{3'}$};
\node[scale=0.3] at (0.35,0) [right] {$\circled{2}$};
\node[scale=0.3] at (-0.85,0) [right] {$\circled{2'}$};
\node[scale=0.8] at (1.2,0) [right] {$\circled{1}$};
\node[scale=0.8] at (-2.6,0) [right] {$\circled{1'}$};
\node[scale=0.8] at (-0.65,2) [right] {$\circled{4}$};
\node[scale=0.8] at (-0.7,-2) [right] {$\circled{4'}$};
\drawbranchpointmarker{0,0};
\filldraw [blue] (0,0) circle (1.5pt);
\end{tikzpicture}
\end{center}

In region \circled{1}, there are two natural normalizations we can use 
\begin{equation}\label{defPsi1su2}
\Psi_{\rm red}^{\text{norm}_1,(1)}=\e^{\frac{\Upsilon_{\rm red}^{\text{norm}_1,(1)}(z,\epsilon)}{\epsilon}},
\end{equation}
\begin{equation}\label{defPsi2su2}
\Psi_{\rm red}^{\text{norm}_2,(1)}=\e^{\frac{\Upsilon_{\rm red}^{\text{norm}_2,(1)}(z,\epsilon)}{\epsilon}},
\end{equation}
depending on which branch point and which soliton we use in \eqref{Upsilonallorder}. To be more precise, norm$_{1,2}$ correspond to using $b_0=\pm \ri$ and $-\gamma_{21_{1,2}}$ for regularization respectively.
\begin{center}
\begin{tikzpicture}[scale=1]
\def\theta{16}
\draw plot [smooth] coordinates {(2,0) (0,0.75)};
\draw plot [smooth] coordinates {(2,0) (0.21,1.1)};
\draw [domain=170:380] plot ( {0.1+0.2*cos(\x)*sin(\theta)+0.3*sin(\x)*cos(\theta)},{0.93+0.2*cos(\x)*cos(\theta)-0.4*sin(\x)*sin(\theta)});
\def\theta{90+35+35}
\draw plot [smooth] coordinates {(2,0) (0,-0.75)};
\draw plot [smooth] coordinates {(2,0) (0.21,-1.1)};
\draw [domain=-175:-370] plot ( {0.1+0.2*cos(\x)*sin(\theta)+0.3*sin(\x)*cos(\theta)},{-0.93+0.2*cos(\x)*cos(\theta)-0.4*sin(\x)*sin(\theta)});
\draw[snake it, orange] (0,1) to (0,0);
\draw[snake it, orange] (0,-1) to (0,-2);
\drawbranchpointmarker{0,-1};
\drawbranchpointmarker{0,1};
\drawbranchpointmarker{0,0};
\filldraw [blue] (0,0) circle (1.5pt);
\node at (1.2,0.5) [right] {$-\gamma_{21_1}$};
\node at (1.2,-0.5) [right] {$-\gamma_{21_2}$};
\node at (2,0) [right] {$z$};
\draw[<-] (0.4,-0.99) --(0.2,-1.1);
\node at (0.2,-1.24) [right]  {$1$};
\draw[<-] (0.4,0.61) --(0.2,0.67);
\node at (0.2,0.36) [right] {$1$};
\end{tikzpicture}
\end{center}

As proposed in \autoref{padeborelsol} and \autoref{sec:borel4d}, we find that the singularities on the Borel plane are at:
\begin{itemize}
\item minus 4d BPS state central charge 
\begin{equation}-Z_{\pm\gamma_d}=\eqref{Zgammad}\sim \mp 6.7777,\end{equation}
and
\begin{equation}-Z_{\pm\gamma_m}=\eqref{Zgammam}\sim \pm 6.7777\ri.\end{equation}
\item minus BPS soliton central charge
\begin{equation}
- Z_{\gamma_{21_1}}=4 \sqrt{\frac{1}{z^3}} z \, _2F_1\left(-\frac{1}{2},-\frac{1}{4};\frac{3}{4};-z^2\right)-\frac{(1-\ri) \sqrt{2 \pi } \Gamma \left(\frac{3}{4}\right)}{\Gamma
   \left(\frac{5}{4}\right)},
\end{equation}
and
\begin{equation}
 -Z_{\gamma_{21_2}}=4 \sqrt{\frac{1}{z^3}+\frac{1}{z}} z \left(z^2+1\right) \, _2F_1\left(1,\frac{5}{4};\frac{3}{4};-z^2\right)-\frac{(1+\ri) \sqrt{2 \pi } \Gamma
   \left(\frac{3}{4}\right)}{\Gamma \left(\frac{5}{4}\right)}.
\end{equation}
\end{itemize}

We summarize  the numerically checked discontinuities  corresponding to BPS soliton central charges in \autoref{su2jump2d4d} and the discontinuities corresponding to 4d BPS state central charges in \autoref{su2jump4d}. They all match the proposals in \eqref{sec2:jumpsol2} and \eqref{sec2:jumppi}. 
We show an example of the numerical checks for \autoref{su2jump4d} in \autoref{disc4d} and some examples of the numerical checks for
\autoref{su2jump2d4d} in \autoref{disc2d4d}, where
 we use
 \begin{align}
\label{defjumpmn1}
 \bX_{\gamma_{21_1}}^{m,n}(z,\epsilon)&=\E^{\frac{Z_{\gamma_{21_1}}}{\epsilon}+m\frac{1}{\pi\ri}\mathcal{I}_{\gamma_m}(\epsilon)+n\frac{1}{\pi\ri}\mathcal{I}_{\gamma_d}(\epsilon)},\\
\label{defjumpmn2}\bX_{\gamma_{21_2}}^{m,n}(z,\epsilon)
&=\E^{\frac{Z_{\gamma_{21_2}}}{\epsilon}+m\frac{1}{\pi\ri}\mathcal{I}_{\gamma_m}(\epsilon)+n\frac{1}{\pi\ri}\mathcal{I}_{\gamma_d}(\epsilon)},
\end{align}
which are labeled by $m,n=\pm 1,\pm2$ and $\mathcal{I}_{\gamma}(\epsilon)$ is defined in \eqref{Idefm}.

\begin{figure}
     \centering
         \includegraphics[width=0.5\textwidth]{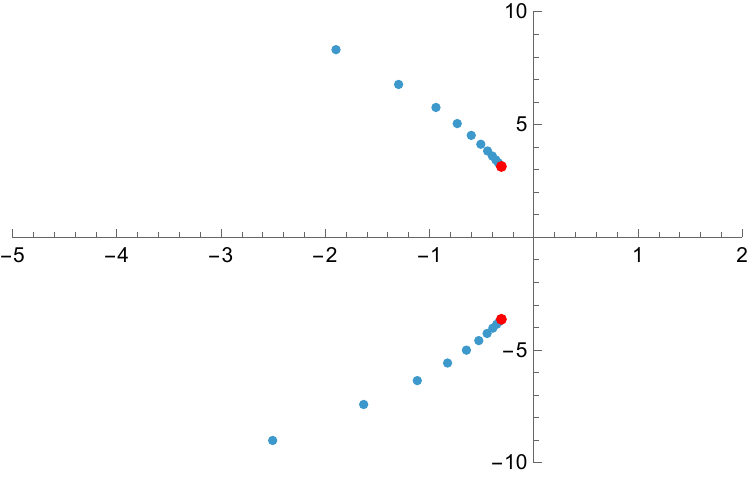}
         \caption{Borel plane of $\Upsilon_{\rm red}^{\text{norm}_{\rm tot},(1)}(\epsilon)$ at $z=\frac{10}{9}+\frac{\ri}{10}$ with 60 terms. Only the first pole in each series of poles corresponds to the actual singularity that has physical meaning. The numerical values of the singularities match with $- Z_{\gamma_{21_1}}$ and $- Z_{\gamma_{21_2}}$, which we show as the red dots. The corresponding solitons are the same as the ones in \autoref{su2reg1sol1} and \autoref{su2reg1sol2}. When varying $z$, we don't see any singularity on the Borel plane of $\hat{\Upsilon}_{\rm red}^{\text{norm}_{\rm tot},(1)}$ at $- Z_{\pm\gamma_{d}}$ and $- Z_{\pm\gamma_{m}}$ as predicted.}
        \label{su22borel}
\end{figure}

Just as the discussion in \autoref{sec:matchTBAborel}, for the purpose of matching with the GMN open TBA solutions, we need to multiply \eqref{defPsi1su2} or \eqref{defPsi2su2} by the factor shown in \eqref{relateTBAborel}. 
To be more precise, 
we define
\begin{align}\label{normfac1su2}
\Upsilon_{\rm red}^{\text{norm}_{\rm tot},(1)}(z,\epsilon)=\Upsilon_{\rm red}^{\text{norm}_1,(1)}(z,\epsilon)-\frac{1}{2\pi\ri}\mathcal{I}_{\gamma_d}(\epsilon)-\frac{1}{2\pi\ri}\mathcal{I}_{\gamma_m}(\epsilon),
\end{align}
which can be obtained similarly from $\Upsilon_{\rm red}^{\text{norm}_2,(1)}(z,\epsilon)$. More details about how to obtain the corresponding WKB series can be found in \autoref{appen:obtainnormtot}. We check that $\Upsilon_{\rm red}^{\text{norm}_{\rm tot},(1)}(z,\epsilon)$ indeed doesn't have any discontinuity corresponding to $Z_{\pm\gamma_d}$ or $Z_{\pm\gamma_m}$ and the discontinuities corresponding to $Z_{\gamma_{21_1}}(z)$ and $Z_{\gamma_{21_2}}(z)$ are as expected for the GMN open TBA equations as shown in \autoref{su2jump4d} and \autoref{su2jump2d4d}. We show some numerical checks of \autoref{su2jump2d4d} in \autoref{disc2d4dtot}. From now on we will focus on  $\Upsilon_{\rm red}^{\text{norm}_{\rm tot},(1)}(z,\epsilon)$ since it corresponds to the GMN open TBA solutions. We also check numerically that $\Psi_{\rm red}^{\text{norm}_{\rm tot},(1)}(z,\epsilon)$ indeed solves \eqref{su2inz}.

\begin{table}[h]
\begin{tabular}{ |p{4cm}|p{6cm}|p{6 cm}| }
\hline
$\arg(\epsilon)$ &  $\arg(-Z_{\gamma_{21_1}})$ &  $\arg(-Z_{\gamma_{21_2}})$ \\
\hline
${\rm disc}(\Psi^{\text{norm}_{1},(1)}_{\rm red})(z,\epsilon)$ & $\ri\Psi^{\text{norm}_{1},(1)}_{\rm red}(z,-\epsilon)\bX_{\gamma_{21_1}}^{0,0}(z,\epsilon)$ & $\ri\Psi^{\text{norm}_{1},(1)}_{\rm red}(z,-\epsilon)\bX_{\gamma_{21_2}}^{0,2}(z,\epsilon)$\\
\hline
${\rm disc}(\Psi^{\text{norm}_{2},(1)}_{\rm red})(z,\epsilon)$ & $\ri\Psi^{\text{norm}_{2},(1)}_{\rm red}(z,-\epsilon)\bX_{\gamma_{21_1}}^{0,-2}(z,\epsilon)$ & $\ri\Psi^{\text{norm}_{2},(1)}_{\rm red}(z,-\epsilon)\bX_{\gamma_{21_2}}^{0,0}(z,\epsilon)$\\
\hline
${\rm disc}(\Psi^{\text{norm}_{\rm tot},(1)}_{\rm red})(z,\epsilon)$ & $\ri\Psi^{\text{norm}_{\rm tot},(1)}_{\rm red}(z,-\epsilon)\bX_{\gamma_{21_1}}^{-1,-1}(z,\epsilon)$ & $\ri\Psi^{\text{norm}_{\rm tot},(1)}_{\rm red}(z,-\epsilon)\bX_{\gamma_{21_2}}^{-1,1}(z,\epsilon)$\\
\hline
\end{tabular}
\caption{Discontinuities corresponding to BPS soliton central charges in region \textcircled{1}. }
\label{su2jump2d4d}
\end{table}

\begin{table}[h]
\begin{tabular}{ |p{2.6cm}|p{3cm}|p{3.5 cm}|p{3 cm}|p{3.5cm}| }
\hline
$\arg(\epsilon)$ &  $\arg(-Z_{\gamma_m})$ &  $\arg(-Z_{-\gamma_m})$ &  $\arg(-Z_{\gamma_d})$ &  $\arg(-Z_{-\gamma_d})$ \\
 \hline
$\disc(\frac{\Upsilon^{\text{norm}_1,(1)}_{\rm red}}{\epsilon})$ & $\frac{1}{2}\log(1+\cX_{\gamma_{m}}(\epsilon))$ &  $-\frac{1}{2}\log(1+\cX_{-\gamma_{m}}(\epsilon))$ & $\frac{1}{2}\log(1+\cX_{\gamma_{d}}(\epsilon))$ & $-\frac{1}{2}\log(1+\cX_{-\gamma_{d}}(\epsilon))$ \\
\hline
$\disc(\frac{\Upsilon^{\text{norm}_2,(1)}_{\rm red}}{\epsilon})$ &  $\frac{1}{2}\log(1+\cX_{\gamma_{m}}(\epsilon))$ & $-\frac{1}{2}\log(1+\cX_{-\gamma_{m}}(\epsilon))$ &$-\frac{1}{2}\log(1+\cX_{\gamma_{d}}(\epsilon))$ &$\frac{1}{2}\log(1+\cX_{-\gamma_{d}}(\epsilon))$ \\
\hline
$\disc(\frac{\Upsilon^{\text{norm}_{\rm tot},(1)}_{\rm red}}{\epsilon})$ & 0 & 0 & 0 & 0\\
\hline
\end{tabular}
\caption{Discontinuities corresponding to 4d BPS state central charges in region \textcircled{1}. }
\label{su2jump4d}
\end{table}

In region \circled{3}, we use the same $\Upsilon^{\text{norm}_{\tot},(1)}_{\rm red}$  as in \circled{1}. It has 3 discontinuities at minus BPS soliton central charges
\begin{equation}
- Z_{\gamma_{21_1}}=-4 \sqrt{\frac{1}{z}}  \, _2F_1\left(-\frac{1}{2},-\frac{1}{4};\frac{3}{4};-z^2\right)+\frac{(8+24 \ri) \pi ^{3/2}}{\Gamma \left(\frac{1}{4}\right)^2}-8 \sqrt{2} \textbf{E} (2),
\end{equation}
\begin{equation}
-Z_{\gamma_{21_2}}=- Z_{\gamma_{21_1}}+Z_{\gamma_m},
\end{equation}
and
\begin{equation}
-Z_{\gamma_{21_3}}=- Z_{\gamma_{21_1}}+Z_{-\gamma_d}.
\end{equation}

\begin{center}
\begin{tikzpicture}[scale=1]
\def\theta{0}
\draw [domain=0+90-40:0+270-90] plot ( {1*cos(\x)*sin(\theta)+1.2*sin(\x)*cos(\theta)},{0.2+1*cos(\x)*cos(\theta)-1.2*sin(\x)*sin(\theta)});
\draw [domain=0+90-40:0+270-90] plot ( {1.4*cos(\x)*sin(\theta)+1.6*sin(\x)*cos(\theta)},{0.2+1.4*cos(\x)*cos(\theta)-1.6*sin(\x)*sin(\theta)});
\draw [domain=90:270] plot ( {0+0.2*cos(\x)},{-1+0.2*sin(\x)});
\draw plot [smooth] coordinates {( {1*cos(0+90-40)*sin(\theta)+1.2*sin(0+90-40)*cos(\theta)},{0.2+1*cos(0+90-40)*cos(\theta)-1.2*sin(0+90-40)*sin(\theta)}) (0,2)};
\draw plot [smooth] coordinates {( {1.4*cos(0+90-40)*sin(\theta)+1.6*sin(0+90-40)*cos(\theta)},{0.2+1.4*cos(0+90-40)*cos(\theta)-1.6*sin(0+90-40)*sin(\theta)}) (0,2)};

\draw [domain=0+90-40:0+270-90] plot ( {-1*cos(\x)*sin(\theta)-1.2*sin(\x)*cos(\theta)},{0.2+1*cos(\x)*cos(\theta)-1.2*sin(\x)*sin(\theta)});
\draw [domain=0+90-40:0+270-90] plot ( {-1.4*cos(\x)*sin(\theta)-1.6*sin(\x)*cos(\theta)},{0.2+1.4*cos(\x)*cos(\theta)-1.6*sin(\x)*sin(\theta)});
\draw [domain=90:270] plot ( {0-0.2*cos(\x)},{-1+0.2*sin(\x)});
\draw plot [smooth] coordinates {( {-1*cos(0+90-40)*sin(\theta)-1.2*sin(0+90-40)*cos(\theta)},{0.2+1*cos(0+90-40)*cos(\theta)-1.2*sin(0+90-40)*sin(\theta)}) (0,2)};
\draw plot [smooth] coordinates {( {-1.4*cos(0+90-40)*sin(\theta)-1.6*sin(0+90-40)*cos(\theta)},{0.2+1.4*cos(0+90-40)*cos(\theta)-1.6*sin(0+90-40)*sin(\theta)}) (0,2)};

\draw [domain=180:360] plot ( {0+0.2*cos(\x)},{1+0.2*sin(\x)});
\draw plot [smooth] coordinates {(0.2,1) (0,2)};
\draw plot [smooth] coordinates {(-0.2,1) (0,2)};

\draw[snake it, orange] (0,1) to (0,0);
\draw[snake it, orange] (0,-1) to (0,-2);
\drawbranchpointmarker{0,-1};
\drawbranchpointmarker{0,1};
\drawbranchpointmarker{0,0};
\filldraw [blue] (0,0) circle (1.5pt);

\node[scale=0.8]  at (0,0.65) [right] {$-\gamma_{21_1}$};
\node[scale=0.8]  at (-1.4,-0.5) [left] {$-\gamma_{21_3}$};
\node[scale=0.8]  at (1.4,-0.5) [right] {$-\gamma_{21_2}$};
\node at (0,2) [above] {$z$};
\draw[->] (0.13,1.3) --(0.12,1.4);
\node at (0.1,1.3) [right] {$1$};
\draw[<-] (1.6,0.2) --(1.6,0.2);
\node at (1.6,0.2)  [right]  {$1$};
\draw[<-] (-1.2,0.2) --(-1.2,0.2);
\node at (-1.2,0.2)  [right]  {$1$};
\end{tikzpicture}
\end{center}

\begin{figure}
     \centering
     \begin{subfigure}[c]{0.47\textwidth}
         \centering
         \includegraphics[width=0.7\textwidth]{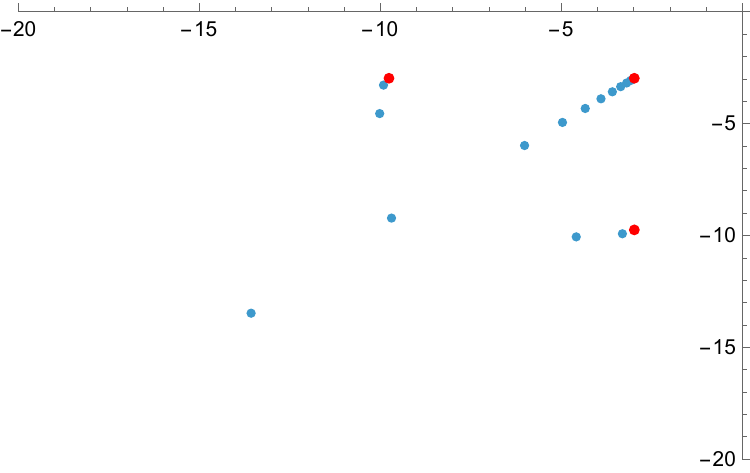}
         \caption{Borel plane of $\Upsilon_{\rm red}^{\text{norm}_{\rm tot},(1)}(\epsilon)$ at $z=5\ri$ with 40 terms. Only the first pole in each series of poles corresponds to the actual singularity that has physical meaning. The numerical values of singularities match with $- Z_{\gamma_{21_1}}$, $- Z_{\gamma_{21_2}}$  and  $- Z_{\gamma_{21_3}}$, which we show as the red dots. }
         \label{BPSrayst2}
     \end{subfigure}
     \\
      \begin{subfigure}[c]{0.25\textwidth}
         \centering
          \vspace{0.5cm}
         \includegraphics[width=0.8\textwidth]{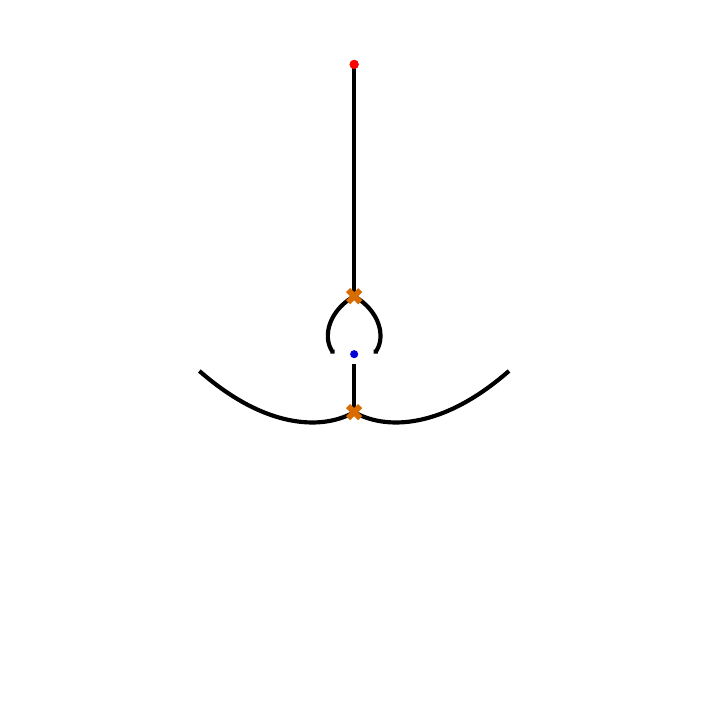}
         \vspace{-1cm}
         \caption{The cutoff version 
     of the Stokes graph with $\arg(\epsilon) = \arg(-Z_{\gamma_{21_1}})$, where we 
     plot the Stokes curves only up
     to $|2\int_{b_0}^{z} \lambda| = Z_{\gamma_{21_1}}$.
    $z=5\ri$ is plotted as a red dot.}
      \end{subfigure}
          \hspace{1cm}
            \begin{subfigure}[c]{0.25\textwidth}
         \centering
         \vspace{0.5cm}
         \includegraphics[width=0.8\textwidth]{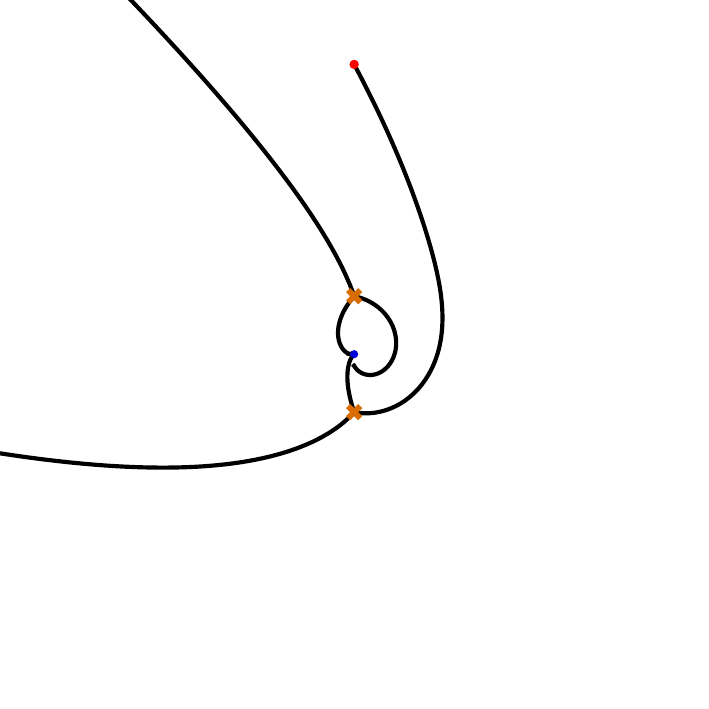}
           \vspace{-1cm}
         \caption{The cutoff version 
     of the Stokes graph with $\arg(\epsilon)  = \arg(-Z_{\gamma_{21_2}})$, where we 
     plot the Stokes curves only up
     to $|2\int_{b_0}^{z} \lambda| = Z_{\gamma_{21_2}}$.
    $z=5\ri$ is plotted as a red dot.}
      \end{subfigure}
      \hspace{1cm}
      \begin{subfigure}[c]{0.25\textwidth}
         \centering
          \vspace{0.5cm}
         \includegraphics[width=0.8\textwidth]{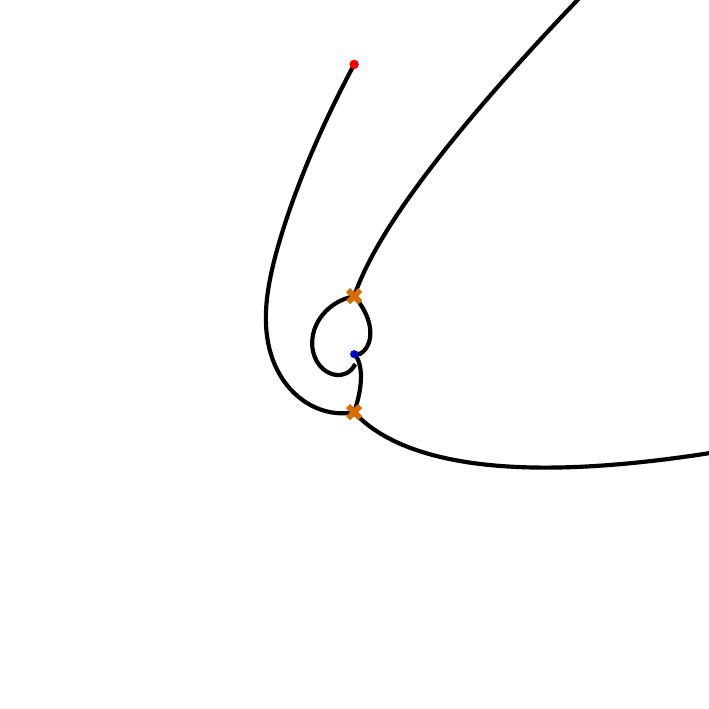}
                       \vspace{-1cm}
         \caption{The cutoff version 
     of the Stokes graph with $\arg(\epsilon)  = \arg(-Z_{\gamma_{21_3}})$, where we 
     plot the Stokes curves only up
     to $|2\int_{b_0}^{z} \lambda| = Z_{\gamma_{21_3}}$.
    $z=5\ri$ is plotted as a red dot.}
      \end{subfigure}
           \caption{The Borel plane for pure $SU(2)$ theory in region \textcircled{3} and corresponding solitons.}
        \label{su23borel}
\end{figure}

We show an example of the Borel plane in \autoref{su23borel}. The predicted (see \autoref{sec:disc} and \autoref{sec:matchTBAborel}) discontinuities corresponding to the singularities are shown in \autoref{su2jump2d4d2} which agree with the ones of GMN open TBA. Instead of numerically checking \autoref{su2jump2d4d2} because of the difficulties described around \autoref{diff3rays}, we directly check the matching of Pad\'{e}-Borel summation result and GMN open TBA result in \autoref{sec:su2match}.

\begin{table}[h]
\begin{centering}
\begin{tabular}{ |p{4cm}|p{6cm}| }
\hline
$\arg(\epsilon)$ & ${\rm disc}(\Psi^{\text{norm}_{\rm tot},(1)}_{\rm red})(z,\epsilon)$\\
\hline
$\arg(-Z_{\gamma_{21_1}})$ &  $\ri\Psi^{\text{norm}_{\rm tot},(1)}_{\rm red}(z,-\epsilon)\bX_{\gamma_{21_1}}^{-1,-1}(z,\epsilon)$\\
\hline
$\arg(-Z_{\gamma_{21_2}})$ & $\ri\Psi^{\text{norm}_{\rm tot},(1)}_{\rm red}(z,-\epsilon)\bX_{\gamma_{21_2}}^{-1,1}(z,\epsilon)$\\
\hline
$\arg(-Z_{\gamma_{21_3}})$ & $\ri\Psi^{\text{norm}_{\rm tot},(1)}_{\rm red}(z,-\epsilon)\bX_{\gamma_{21_3}}^{1,-1}(z,\epsilon)$\\
\hline
\end{tabular}
\caption{Discontinuities corresponding to BPS soliton central charges in region \textcircled{3}. }
\label{su2jump2d4d2}
\end{centering}
\end{table}

\subsection{Closed TBA equations}\label{sec:su2closetba}
Using the $\mathbb{Z}_4$ symmetry \eqref{z4sym}, the GMN closed TBA equations are simplified to
\begin{equation}
\log\cX_{\gamma_d}(\epsilon)=\frac{Z_{\gamma_d}}{\epsilon}+\frac{1}{\pi}\int\limits_{\ell_{\gamma_d}}\frac{\epsilon}{\epsilon'^2+\epsilon^2}\log(1+\cX_{\gamma_d}(\epsilon'))\rd\epsilon',
\end{equation}
\begin{equation}
\log\cX_{\gamma_m}(\epsilon)=\frac{Z_{\gamma_m}}{\epsilon}+\frac{1}{\pi}\int\limits_{\ell_{\gamma_m}}\frac{\epsilon}{\epsilon'^2+\epsilon^2}\log(1+\cX_{\gamma_m}(\epsilon'))\rd\epsilon'.
\end{equation}
A thorough discussion and numerical calculation of the closed TBA equations in the pure $SU(2)$ theory strong coupling region can be found in \cite{Grassi:2019coc}.

\subsection{Open TBA equations}\label{sec:su2match}
In \circled{1}, we check numerically that $\Psi^{\text{norm}_{\rm tot},(1)}_{\rm red}(z,\epsilon)$ calculated by Pad\'{e}-Borel summation indeed matches the solution of the GMN open TBA equations \eqref{eq:int-22}:
\begin{align}\label{su2tba}
\nonumber g_1(z,\epsilon)=&\frac{1}{(-\sqrt{\frac1z+\frac1{z^3}})^{\frac12}}+\frac{1}{2\pi}\int\limits_{\ell_{\gamma_{21_1}}}\frac{\epsilon}{\epsilon'}\frac{1}{\epsilon'-\epsilon} g_1(z,-\epsilon')\bX_{\gamma_{21_1}}^{-1,-1}(z,\epsilon')\rd\epsilon'\\
&+\frac{1}{2\pi}\int\limits_{\ell_{\gamma_{21_2}}}\frac{\epsilon}{\epsilon'}\frac{1}{\epsilon'-\epsilon}g_1(z,-\epsilon')\bX_{\gamma_{21_2}}^{-1,1}(z,\epsilon')\rd\epsilon'\Bigg).
 \end{align}

The numerical result is shown in \autoref{openTBAsu2}. Note that in the pure $SU(2)$ example, the data of quantum periods are generated by the closed TBA equations in \autoref{sec:su2closetba}. 

\begin{table}[h]
\begin{tabular}{ |p{3cm}||p{4.3cm}|p{4cm}|p{4cm}| }
 \hline 
 \hline
 \multicolumn{4}{|c|}{$z=2\in\circled{1}$} \\
 \hline
 \hline
$\epsilon$ & $\frac1{2}$ & $\frac5{2}$ & $\frac9{2}$\\
\hline
$g_1(z,\epsilon)$, s=$\frac13$,  c=100& $-1.115885915$i & $-1.0693870250$i & $-1.0283991468$i\\
\hline
$g_1(z,\epsilon)$, s=$\frac13$, c=150 & $-1.1159157529$i  & $-1.0695100811$i & $-1.0285764157$i  \\
\hline
$g_1(z,\epsilon)$, s=$\frac13$, c=200 & $-1.1159291043$i  & $-1.0695688246$i & $-1.0286683017$i  \\
\hline
\hline
$\Psi^{\text{norm}_{\rm tot},(1), {\rm PB}}_{\rm red}(z,\epsilon)$ & $-1.1159534113$i & $-1.0696877893$i & $-1.02887231$i\\
\hline
\hline
\end{tabular}
\caption{Numerical iteration computation for $g_1(z,\epsilon)$. PB stands for Pad\'{e}-Borel summation method and $g_1(z,\epsilon)$ stands for the result of open TBA \eqref{su2tba}. We only show the stable digits of $\Psi^{(1), {\rm PB}}_{\rm red}(z,\epsilon)$. $s$ and $c$ represent the step size and cutoff of the integral in the numerical discretization respectively. We see that the open TBA results converge to the Pad\'{e}-Borel result.}
\label{openTBAsu2}
\end{table}

Analogously, in \circled{3}, the GMN TBA equation reads
\begin{equation}\label{su2tba2}
\begin{aligned}
 g_1(z,\epsilon)=&\frac{1}{(-\sqrt{\frac1z+\frac1{z^3}})^{\frac12}}+\frac{1}{2\pi}\int\limits_{\ell_{\gamma_{21_1}}}\frac{\epsilon}{\epsilon'}\frac{1}{\epsilon'-\epsilon} g_1(z,-\epsilon')\bX_{\gamma_{21_1}}^{-1,-1}(z,\epsilon')\rd\epsilon'\\
&+\frac{1}{2\pi}\int\limits_{\ell_{\gamma_{21_2}}}\frac{\epsilon}{\epsilon'}\frac{1}{\epsilon'-\epsilon} g_1(z,-\epsilon')\bX_{\gamma_{21_2}}^{-1,1}(z,\epsilon')\rd\epsilon'\\
&+\frac{1}{2\pi}\int\limits_{\ell_{\gamma_{21_3}}}\frac{\epsilon}{\epsilon'}\frac{1}{\epsilon'-\epsilon}g_1(z,-\epsilon')\bX_{\gamma_{21_3}}^{1,-1}(z,\epsilon')\rd\epsilon'
 \end{aligned}
 \end{equation}
 We check numerically that the result of open TBA equation \eqref{su2tba2} matches with the result of Pad\'{e}-Borel method; the numerics are shown in \autoref{openTBAsu22}.

\begin{table}[h]
\begin{tabular}{ |p{3cm}||p{4.3cm}|p{4cm}|p{4cm}| }
 \hline 
 \hline
 \multicolumn{4}{|c|}{$z=\frac 4 3\ri\in\circled{3}$} \\
 \hline
 \hline
$\epsilon$ & $\frac1{4}$ & $\frac5{4}$ & $\frac9{4}$\\
\hline
\hline
$g_1(z,\epsilon)$, s=$\frac1{10}$, & $0.5180536456$    & $0.4970744744$ & $0.4817195233$ \\
 c=50  & $-1.1298287041$i  & $-0.9978281667 $i  & $-0.9374825040$i \\
\hline
$g_1(z,\epsilon)$, s=$\frac1{10}$, & $0.5180815897$    & $0.4972625820$ & $0.4821120637$ \\
 c=100  & $-1.1297711378 $i  & $-0.9974950743 $i  & $-0.9368303124$i \\
\hline
$g_1(z,\epsilon)$, s=$\frac1{14}$, & $0.5179337086$    & $0.4970021414$ & $0.4818450046$ \\
 c=100  & $-1.1291826503 $i  & $-0.9969153436 $i  & $-0.9362791446 $i \\
 \hline
\hline
$\Psi^{\text{norm}_{\rm tot},(1), {\rm PB}}_{\rm red}(z,\epsilon)$ & 0.5178997882  & 0.4970543& $0.482$ \\
& $-1.1290698449$i & $-0.996694$i & $-0.9359$i\\
\hline
\hline
\end{tabular}
\caption{Numerical iteration computation for $g_1(z,\epsilon)$. PB stands for Pad\'{e}-Borel summation method and $g_1(z,\epsilon)$ stands for the result of open TBA  \eqref{su2tba2}. We only show the stable digits of $\Psi^{(1), {\rm PB}}_{\rm red}(z,\epsilon)$. $s$ and $c$ represent the step size and cutoff of the integral in the numerical discretization respectively. We see that the open TBA results converge to the Pad\'{e}-Borel result.}
\label{openTBAsu22}
\end{table}

\appendix
\section{The WKB solution of Weber equation and parabolic cylinder D function}\label{appen:matchwkbD}

The Weber equation \eqref{eqn:weber} is known to be solved by the parabolic cylinder D functions which can be further expressed in terms of the generalized Hermite polynomials
\begin{align}\label{harmpoly}
&D_{\frac{1}{\epsilon}-\frac{1}{2}}(\frac{ z}{\sqrt{\epsilon}})=2^{-\frac{1}{2\epsilon}+\frac1 4}\e^{-\frac{z^2}{4\epsilon}}H_{\frac{1}{\epsilon}-\frac{1}{2}}(\frac{ z}{\sqrt{2}\sqrt{\epsilon}}),\\
&D_{-\frac{1}{\epsilon}-\frac{1}{2}}(\frac{\ri z}{\sqrt{\epsilon}})=2^{\frac{1}{2\epsilon}+\frac1 4}\e^{\frac{z^2}{4\epsilon}}H_{-\frac{1}{\epsilon}-\frac{1}{2}}(\frac{\ri z}{\sqrt{2}\sqrt{\epsilon}}).
\end{align}

The WKB solution obtained using Borel summation satisfies
\begin{equation}\label{matchwkbD}
\Psi^{(1)}(z,\epsilon)=N(\epsilon) D_{\frac{1}{\epsilon}-\frac{1}{2}}(\frac{ z}{\sqrt{\epsilon}})
\end{equation}
in a certain region of the Riemann surface up to a normalization by 
\begin{equation}\label{normDfcn}
N(\epsilon)=N^0(\epsilon)(1+\mathcal{O}(\epsilon))={\ri}\frac{\sqrt[4]{\pi } 2^{\frac{5}{4}} \sqrt[4]{\frac{2}{\epsilon }-1}}{\sqrt{\Gamma \left(\frac{1}{2}+\frac{1}{\epsilon }\right)}}(1+\mathcal{O}(\epsilon)),
\end{equation} 
which is a function depending only on $\epsilon$.

We further define
\begin{align}
&D^{\rm red}_{\frac{1}{\epsilon}-\frac{1}{2}}(\frac{ z}{\sqrt{\epsilon}})\\
=&D_{\frac{1}{\epsilon}-\frac{1}{2}}(\frac{ z}{\sqrt{\epsilon}})\exp \left(-\frac{-\sqrt{z^2-4} z+\log \left(z^2-4\right)+2 \log \left(\frac{z}{\sqrt{z^2-4}}+1\right)-2 \log
   \left(z-\sqrt{z^2-4}\right)}{4 \epsilon }\right).
\end{align}

In principle, \eqref{normDfcn} can be obtained by a careful analysis of the expansion of the parabolic cylinder D function at $\epsilon\rightarrow 0$. We provide numerical evidence that $N(\epsilon)$ doesn't depend on $z$.
We show the result 
at $z=3\e^{\frac{\pi\ri}{30}}$ in \autoref{PB3} and the result at $z=4\e^{\frac{\pi\ri}{30}}$ in \autoref{PB4}; it is clear that the series in \eqref{normDfcn} is $z$ independent. 
\begin{table}[h]
    \centering
    \begin{tabular}{| c | c  c| }
    \hline
     $\epsilon$ & $\Psi^{(1)}_{\rm red}(3\,\e^{\frac{\pi\ri}{30}},\epsilon)$  & $\Psi^{(1)}_{\rm red}(3\,\e^{\frac{\pi\ri}{30}},\epsilon)/(N^0(\epsilon) D^{\rm red}_{\frac{1}{\epsilon}-\frac{1}{2}}(\frac{ 3\,\e^{\frac{\pi\ri}{30}}}{\sqrt{\epsilon}}))$\\

    \hline 

 $\frac{1}{10}$     & $0.0827590186 + 0.9287849458 $ i  & 1.012905895   \\

 $\frac{1}{5}$       & $0.0792142413 + 0.9236103398$ i &  1.0266900961   \\

 $\frac{3}{10}$     & $0.0761906876 + 0.9189022272 $ i  &1.0414664128 \\

 $\frac{2}{5}$       & $0.0735583096 + 0.9145873397$ i  & 1.0573712634  \\
         \hline
    \end{tabular}
    \caption{Pad\'{e}-Borel summation of $\Psi^{(1)}_{\rm red}$ at $z=3\,\e^{\frac{\pi\ri}{30}}$ and its comparison to the parabolic cylinder D function.}
    \label{PB3}
\end{table}

\begin{table}[h]
    \centering
    \begin{tabular}{| c | c  c| }
    \hline
     $\epsilon$ & $\Psi^{(1)}_{\rm red}(4\,\e^{\frac{\pi\ri}{30}},\epsilon)$  & $\Psi^{(1)}_{\rm red}(4\,\e^{\frac{\pi\ri}{30}},\epsilon)/(N^0(\epsilon) D^{\rm red}_{\frac{1}{\epsilon}-\frac{1}{2}}(\frac{ 4\,\e^{\frac{\pi\ri}{30}}}{\sqrt{\epsilon}}))$\\

    \hline 

 $\frac{1}{10}$     & $0.0517743510 + 0.7552467066$ i  & 1.012905895  \\

 $\frac{1}{5}$       & $0.0509522447 + 0.7543909226$ i&  1.0266900961  \\

 $\frac{3}{10}$     & $0.0501881251 + 0.7535997488$ i  & 1.0414664128\\

 $\frac{2}{5}$       & $0.0494739715 + 0.7528627368$ i & 1.0573712634\\
         \hline
    \end{tabular}
    \caption{Pad\'{e}-Borel summation of $\Psi^{(1)}_{\rm red}$ at $z=4\,\e^{\frac{\pi\ri}{30}}$ and its comparison to the parabolic cylinder D function.}
    \label{PB4}
\end{table}

\section{Numerical checks of the discontinuities in the pure $SU(2)$ theory}
We show an example of the numerical checks for \autoref{su2jump4d} in \autoref{disc4d} and some examples of the numerical checks for
\autoref{su2jump2d4d} in \autoref{disc2d4d} and \autoref{disc2d4dtot}.
\begin{table}[h!]
\begin{tabular}{ |p{7cm}|p{8cm}| }
 \hline
${\rm disc}(\frac{\Upsilon^{\text{norm}_1,(1)}_{\rm red}(8+9\ri,\frac{\ri}{10})}{\frac{\ri}{10}})(8+9\ri,\frac{\ri}{10})$   & $\frac{1}{2}\log(1+\cX_{\gamma_{m}}(\frac{\ri}{10}))$\\
 \hline 
$\underline{1.8641}50101\times 10^{-30}$ & $\underline{1.864147689}\times 10^{-30}$\\
 \hline
 ${\rm disc}(\frac{\Upsilon^{\text{norm}_1,(1)}_{\rm red}(8+9\ri,-\frac{\ri}{10})}{-\frac{\ri}{10}})(8+9\ri,-\frac{\ri}{10})$   & $-\frac{1}{2}\log(1+\cX_{-\gamma_{m}}(-\frac{\ri}{10}))$\\
 \hline 
$\underline{-1.864}146098\times 10^{-30}$ & $-\underline{1.864147689}\times 10^{-30}$\\
 \hline
  ${\rm disc}(\frac{\Upsilon^{\text{norm}_1,(1)}_{\rm red}(8+9\ri,\frac{1}{10})}{\frac{1}{10}})(8+9\ri,\frac{1}{10})$   & $-\frac{1}{2}\log(1+\cX_{-\gamma_{d}}(\frac{1}{10}))$\\
 \hline 
$-\underline{1.864}149949\times 10^{-30}$ & $-\underline{1.864147689}\times 10^{-30}$\\
 \hline
   ${\rm disc}(\frac{\Upsilon^{\text{norm}_1,(1)}_{\rm red}(8+9\ri,-\frac{1}{10})}{-\frac{1}{10}})(8+9\ri,-\frac{1}{10})$   & $\frac{1}{2}\log(1+\cX_{\gamma_{d}}(-\frac{1}{10}))$\\
 \hline 
$\underline{1.86}4147403\times 10^{-30}$ & $\underline{1.864147689}\times 10^{-30}$\\
\hline
\end{tabular}
\caption{Numerical computation for the discontinuities of $\frac{\Upsilon^{\text{norm}_1,(1)}_{\rm red}(z,\epsilon)}{\epsilon}$ corresponding to the 4d BPS central charges in region \textcircled{1}.}
\label{disc4d}
\end{table}

\begin{table}[h]
\begin{tabular}{ |p{6cm}|p{10cm}| }
\hline
${\rm disc}(\Psi^{\text{norm}_1,(1)}_{\rm red})(z,\frac{1}{10}\E^{\arg(-Z_{\gamma_{21_1}})\ri})$  & $\ri\Psi^{\text{norm}_1,(1)}_{\rm red}(z,-\frac{1}{10}\E^{\arg(Z_{-\gamma_{21_1}})\ri})\bX_{\gamma_{21_1}}^{0,0}(z,\frac{1}{10}\E^{\arg(-Z_{\gamma_{21_1}})\ri})$\\
\hline
$\underline{1.865429542}8\times10^{-14} $ & $ \underline{1.865429543}\times10^{-14} $ \\ 
$+\underline{8.1667454}013\times10^{-16}\ri$ & $+ \underline{8.1667453753}\times10^{-16}  \ri$\\
 \hline 
${\rm disc}(\Psi^{\text{norm}_1,(1)}_{\rm red})(z,\frac{1}{10}\E^{\arg(-Z_{\gamma_{21_2}})\ri})$  & $\ri\Psi^{\text{norm}_1,(1)}_{\rm red}(z,-\frac{1}{10}\E^{\arg(Z_{-\gamma_{21_2}})\ri})\bX_{\gamma_{21_2}}^{0,2}(z,\frac{1}{10}\E^{\arg(-Z_{\gamma_{21_2}})\ri})$\\
\hline
$\underline{1.18526198}37\times10^{-16} $ & $ \underline{1.18526198}35\times10^{-16} $ \\ 
$+\underline{4.784501}8563\times10^{-18}\ri$ & $+ \underline{4.784501}7427\times10^{-18}  \ri$\\
 \hline 
 ${\rm disc}(\Psi^{\text{norm}_2,(1)}_{\rm red})(z,\frac{1}{10}\E^{\arg(Z_{-\gamma_{21_1}})\ri})$  & $\ri\Psi^{\text{norm}_2,(1)}_{\rm red}(z,-\frac{1}{10}\E^{\arg(-Z_{\gamma_{21_1}})\ri})\bX_{\gamma_{21_1}}^{0,-2}(z,\frac{1}{10}\E^{\arg(-Z_{\gamma_{21_1}})\ri})$\\
\hline
$\underline{1.879758452}7\times10^{-14} $ & $\underline{1.8797584527}\times10^{-14}$\\
$+ \underline{8.3721193}63\times10^{-16} \ri$ & $+ \underline{8.3721193752}\times10^{-16}  \ri$\\
\hline
${\rm disc}(\Psi^{\text{norm}_2,(1)}_{\rm red})(z,\frac{1}{10}\E^{\arg(Z_{-\gamma_{21_2}})\ri})$  & $\ri\Psi^{\text{norm}_2,(1)}_{\rm red}(z,-\frac{1}{10}\E^{\arg(-Z_{\gamma_{21_2}})\ri})\bX_{\gamma_{21_2}}^{0,0}(z,\frac{1}{10}\E^{\arg(-Z_{\gamma_{21_2}})\ri})$\\
\hline
$\underline{1.17614517}85\times10^{-16} $ & $\underline{1.176145176}9\times10^{-16}$\\
$+ \underline{4.824556}5394\times10^{-18} \ri$ & $+ \underline{4.8245566}777\times10^{-18}  \ri$\\
\hline
\end{tabular}
\caption{Numerical computation for the discontinuities of $\Psi_{\rm red}^{\text{norm}_1,(1)}(z,\epsilon)$ and $\Psi_{\rm red}^{\text{norm}_2,(1)}(z,\epsilon)$ at $\arg(\epsilon)=\arg(-Z_{\gamma_{21_1}})$ and $\arg(\epsilon)=\arg(-Z_{\gamma_{21_2}})$, respectively. We choose $z=\frac{10}{9}+\frac{\ri}{10}\in$\textcircled{1}.}
\label{disc2d4d}
\end{table}

\begin{table}[h]
\begin{tabular}{ |p{6cm}|p{10cm}| }
 \hline
${\rm disc}(\Psi^{\text{norm}_{\rm tot},(1)}_{\rm red})(z,\frac{1}{10}\e^{\arg(-Z_{\gamma_{21_1}})\ri})$   & $\ri\Psi^{\text{norm}_{\rm tot},(1)}_{\rm red}(z,-\frac{1}{10}\e^{\arg(-Z_{\gamma_{21_1}})\ri})\bX_{\gamma_{21_1}}^{-1,1}(z,\frac{1}{10}\e^{\arg(-Z_{\gamma_{21_1}})\ri})$\\
 \hline 
$\underline{1.8715382607}\times10^{-14} $ & $ \underline{1.8715382607}\times10^{-14} $ \\ 
$+\underline{8.98569524}24\times10^{-16}\ri$ & $+ \underline{8.985695242}\times10^{-16}  \ri$\\
 \hline
 ${\rm disc}(\Psi^{\text{norm}_{\rm tot},(1)}_{\rm red})(z,\frac{1}{10}\e^{\arg(-Z_{\gamma_{21_2}})\ri})$   & $\ri\Psi^{\text{norm}_{\rm tot},(1)}_{\rm red}(z,-\frac{1}{10}\e^{\arg(-Z_{\gamma_{21_1}})\ri})\bX_{\gamma_{21_2}}^{-1,-1}(z,\frac{1}{10}\e^{\arg(-Z_{\gamma_{21_2}})\ri})$\\
 \hline
$\underline{1.180485135}5\times10^{-16} $ & $\underline{1.1804851355}\times10^{-16}$\\
$+ \underline{4.3485987}027\times10^{-18} \ri$ & $+ \underline{4.3485987022}\times10^{-18}  \ri$\\
\hline
\end{tabular}
\caption{Numerical computation for the discontinuities of $\Psi^{\text{norm}_{\rm tot},(1)}_{\rm red}$ corresponding to the BPS soliton central charges when $\arg(\epsilon)=\arg(-Z_{\gamma_{21_1}})$ and $\arg(\epsilon)=\arg(-Z_{\gamma_{21_2}})$, respectively.  We choose $z=\frac{\ri}{10}+\frac{10}{9}\in$\textcircled{1}}
\label{disc2d4dtot}
\end{table}

\section{WKB series for \texorpdfstring{$\Upsilon_{\rm red}^{{\rm norm}_{\rm tot},(1)}(z,\epsilon)$}{Upsilon_red^(norm_tot,1)(z,epsilon)}}
\label{appen:obtainnormtot}
In general, we can use the exact WKB methods to obtain the quantum periods $\cX_{\gamma_d}(\epsilon)$ and $\cX_{\gamma_m}(\epsilon)$ and then substitute the $\cI_{\gamma_d}(\epsilon)$ and $\cI_{\gamma_m}(\epsilon)$ in \eqref{normfac1su2}. An example of the exact WKB methods is the GMN closed TBA equations in \autoref{sec:su2closetba}. However, in the $u=0$ case, there is a simpler procedure to obtain \eqref{normfac1su2} directly at the WKB series level using \eqref{rationnorm}.
More precisely, we define
\begin{equation}\label{su2f1}
f_1(\epsilon)\equiv\frac{1}{2\pi\ri}\int\limits_{\ell_{\gamma_d}}\frac{\epsilon}{\epsilon'^2-\epsilon^2}\log(1+\cX_{\gamma_d}(\epsilon'))=-\frac{1}{2\pi}\int\limits_{\ell_{\gamma_m}}\frac{\epsilon}{\epsilon'^2+\epsilon^2}\log(1+\cX_{\gamma_m}(\epsilon')).
\end{equation}
$f_1(\epsilon)$ can be obtained by
\begin{equation}
f_1(\epsilon)=\frac{1}{2}\Upsilon^{\text{norm}_1,(1)}_{\rm red}(z=b_0=-\ri,\epsilon).
\end{equation}
Thus
\begin{equation}
\begin{aligned}
&\quad\ \Upsilon_{\rm red}^{\text{norm}_{\rm tot},(1)}(z,\epsilon)\\
&=\Upsilon_{\rm red}^{\text{norm}_1,(1)}(z,\epsilon)-\frac{1}{2\pi\ri}\int\limits_{\ell_{\gamma_d}}\frac{\epsilon}{\epsilon'^2-\epsilon^2}\log(1+\cX_{\gamma_d}(\epsilon'))-\frac{1}{2\pi\ri}\int\limits_{\ell_{\gamma_m}}\frac{\epsilon}{\epsilon'^2-\epsilon^2}\log(1+\cX_{\gamma_m}(\epsilon'))\\
&=\Upsilon_{\rm red}^{\text{norm}_1,(1)}(z,\epsilon)-f_1(\epsilon)-f_1(\ri\epsilon).
\end{aligned}
\end{equation}
Similarly, $\Upsilon^{\text{norm}_{\tot},(1)}_{\rm red}$ can also be obtained as
\begin{align}\label{su2psitot}
\Upsilon_{\rm red}^{\text{norm}_{\rm tot},(1)}(z,\epsilon)=\Upsilon_{\rm red}^{\text{norm}_2,(1)}(z,\epsilon)+f_1(\epsilon)- f_1(\ri\epsilon).
\end{align}

\bibliography{open-tba}

\end{document}